%% file: cf_a496.tex
\documentclass[useAMS,usenatbib]{mn2e}

\usepackage{lscape}
\usepackage{graphicx}

\newcommand{\Sub}{_{\mathrm{sub}}}

\newcommand{\Max}{_{\mathrm{max}}}
\newcommand{\Min}{_{\mathrm{min}}}

\newcommand{\ICM}{_{\mathrm{ICM}}}
\newcommand{\Hot}{_{\mathrm{hot}}}
\newcommand{\Cold}{_{\mathrm{cold}}}

\newcommand{\Proj}{_{\mathrm{proj}}}
\newcommand{\Sun}{_{\sun}}

\newcommand{\Fe}{_{\mathrm{Fe}}}

\newcommand{\degree}{^o}

\newcommand{\K}{\,\textrm{K}}
\newcommand{\KeV}{\,\textrm{keV}}
\newcommand{\Kpc}{\,\textrm{kpc}}
\newcommand{\Mpc}{\,\textrm{Mpc}}
\newcommand{\PC}{\,\textrm{pc}}

\newcommand{\Gyr}{\,\textrm{Gyr}}
\newcommand{\Kms}{\,\textrm{km}\,\textrm{s}^{-1}}

\newcommand{\ccm}{\,\textrm{cm}^{-3}}
\newcommand{\gccm}{\,\textrm{g}\,\textrm{cm}^{-3}}

\title[Sloshing cold fronts in A496]{Gas sloshing, cold fronts, Kelvin-Helmholtz instabilities and the merger history of  the cluster of galaxies Abell 496}
\author[Roediger et al.]{E. Roediger$^{1}$\thanks{E-mail:
email@address (ER)}, 
L. Lovisari$^{2,3}$,
R. Dupke$^{4,5,6,7}$,  
S. Ghizzardi$^{8}$, 
M. Br\"uggen$^{1}$, \newauthor
R. P. Kraft$^{9}$ and 
M. E. Machacek$^{9}$\\
$^{1}$Jacobs University Bremen, PO Box 750 561, 28725 Bremen, Germany\\
$^{2}$Argelander-Institut f\"ur Astronomie, Rheinische Friedrich-Wilhelms-Universit\"at Bonn, Auf dem H\"ugel 71, D-53121 Bonn, Germany\\
$^{3}$Institut f\"ur Astro- und Teilchenphysik, Universit\"at Innsbruck, Technikerstr. 25, A-6020 Innsbruck, Austria\\
$^{4}$Department of Astronomy, University of Michigan, 930 Dennison Bldg., Ann Arbor, MI 48109-1090, USA\\
$^{5}$Eureka Scientific Inc., Oakland, CA 94602-3017\\
$^{6}$Observatorio Nacional, Rua Gal. Jose Cristino, 20921-400, Rio de Janeiro, Brazil\\
$^{7}$University of Alabama, Tuscaloosa, AL 35487-0324\\
$^{8}$INAF, Istituto di Astrofisica Spaziale e Fisica Cosmica, via E. Bassini 15, 20133 Milano, Italy\\
$^{9}$ Harvard/Smithsonian Center for Astrophysics, 60 Garden Street, MS-4, Cambridge, MA 02138, USA
}
\begin{document}

\date{Accepted 1988 December 15. Received 1988 December 14; in original form 1988 October 11}

\pagerange{\pageref{firstpage}--\pageref{lastpage}} \pubyear{2011}

\maketitle

\label{firstpage}

\begin{abstract}
We investigate the origin and nature of the multiple sloshing cold fronts in the core of Abell 496 by direct comparison between observations and dedicated hydrodynamical simulations. Our simulations model a minor merger with a 4$\times 10^{13}M\Sun$ subcluster crossing A496 from the south-west to the north-north-east, passing the cluster core in the south-east at a pericentre distance 100 to a few 100 kpc about 0.6 to 0.8 Gyr ago. The gas sloshing triggered by the merger can reproduce almost all observed features, e.g.~the characteristic spiral-like brightness residual distribution in the cluster centre and its asymmetry out to 500 kpc, also the positions of and contrasts across the cold fronts. If the subcluster passes close ($100 \Kpc$) to the cluster core, the resulting shear flows are strong enough to trigger Kelvin-Helmholtz instabilities that in projection resemble the peculiar kinks in the cold fronts of Abell 496. Finally, we show that sloshing does not lead to a significant modification of the global ICM profiles but a mild oscillation around the initial profiles. 
\end{abstract}


%iiiiiiiiiiiiiiiii
\input{intro}

%iiiiiiiiiiiiiiiiiiiiiiiiiii
\input{comparison}

%iiiiiiiiiiiiiiiii
\input{discussion}

%**********************
\section{Summary}  \label{sec:summary}
%**********************
%
We have presented a consistent scenario of minor-merger triggered gas sloshing in the slightly elliptical  cluster of galaxies A496 by comparing observations to dedicated simulations. We have deduced the following scenario as most the likely: about 
0.6 to 0.8 Gyr
ago, a subcluster of about $4\times 10^{13}M\Sun$ passed 
between 100 and a few 100 kpc
SE of the cluster core, moving from the SW to the N-NE. The subcluster should now be located about 1 Mpc N-NE of the cluster centre, but is probably already dispersed by tidal disruption and thus maybe impossible to identify. 
No obvious subcluster is found in the galaxy distribution near the expected position.

Our proposed scenario can explain almost all observed features of the ICM in A496: the spiral-like surface brightness excess wrapped around the cluster core and the associated asymmetries in temperature and metallicity maps, the positions of the cold fronts along the outer edge of the brightness excess spiral, the profiles for X-ray brightness, temperature and metallicity inside 150 kpc, and the large-scale distribution of brightness excess and deficit within the observed field of view, which extends out to 500 kpc. Only the   outermost southern cold front is not a discontinuity in our simulations.

For a small pericentre distance of 100 kpc, our simulations even reproduce the kinks observed in the inner cold fronts. This suggests that they are Kelvin-Helmholtz instabilities arising from shear flows along the cold fronts. This puts an upper limit of roughly $10\,\mu$G or a magnetic pressure of 10\% of the thermal pressure on the magnetic field parallel to the cold front.

We have studied the effect of gas sloshing on fully azimuthally averaged profiles for X-ray brightness, projected temperature and metallicity and found deviations from the original profiles to be below 10 percent. Thus, only in very deep data, traces of sloshing can be detected in fully averaged profiles, and only then sloshing could introduce errors in the mass reconstruction by violating the hydrostatic equilibrium.

%**********************
\section*{Acknowledgments}
%**********************
E.R.~acknowledges the support of the Priority Programme 
ÓWitnesses of Cosmic HistoryÓ of the DFG (German Re- 
search Foundation),  the supercomputing grants NIC 
 3711 and 4368 at the John-Neumann Institut at the 
Forschungszentrum J\"ulich, a visiting scientist fellowship of the Smithsonian Astrophysical Observatory, and the hospitality of the Center for Astrophysics in Cambridge. L.L.~acknowledges support by the DFG through Heisenberg grant RE 1462/6, by the German Aerospace Agency (DLR) with funds from the Ministry of Economy and Technology (BMWi) through grant 50 OR 1102Ê and the Austrian Science Foundation (FWF) through grant  P19300-N16.
R.A.D acknowledges support from NASA Grant
NNH10CD19C.
M.B.~acknowledges support by the DFG Research Unit "Magnetisation of Interstellar and Intergalactic Media: The Prospects of Low-Frequency Radio Observations".
We thank Suresh Sivanandam for providing us with observational data and  Georgiana Ogrean, Annalisa Bonafede and Paul Nulsen for helpful discussions. The results presented were produced using the FLASH 
code, a product of the DOE ASC/Alliances-funded Center 
for Astrophysical Thermonuclear Flashes at the University 
of Chicago. 
This research has made use of Aladin. It  has also made use of the NASA/IPAC Extragalactic Database (NED) which is operated by the Jet  Propulsion Laboratory, California Institute of Technology, under contract with the National Aeronautics and Space Administration.

%*******************************************************************
%*************** R E F E R E N C E S *******************************
%*******************************************************************
%
\bibliographystyle{mn2e}
\bibliography{library}

\appendix

\input{appendix}

%\input{appendix2}

\bsp

\label{lastpage}

\end{document}

%% file: intro.tex
%*****************
\section{Introduction}
%*****************
%
During their lifetimes, galaxy clusters experience a number of major and minor mergers. Constraining the merger rate and understanding the properties of mergers are  important for cosmological studies, since cluster dynamics have a strong impact on cluster mass determination, which is widely used to set constraints to cosmological parameters (e.g.~\citealt{Rasia2006,Lau2009,Allen2011}). The investigation of individual clusters reveals insights into the details of the physics of galaxy clusters and their intra-cluster medium (ICM).   Depending on the mass ratio and impact parameter, mergers leave observable traces in the clusters, among them shocks and cold fronts (see review by \citealt{Markevitch2007}). Cold fronts (CFs) appear as discontinuities in X-ray images and temperature maps, where the brighter and denser side is also the cooler one. The direction of the temperature jump distinguishes them from shocks, where the  jump is in the opposite direction. 

CFs come in two varieties: the first are merger CFs, which are found in  merging clusters, e.g.~A3667 (\citealt{ Vikhlinin2001}),  the bullet cluster 1E 0657-56 (\citealt{Markevitch2002}), and A2146 (\citealt{Russell2010a2146,Russell2011a2146}). These CFs are  contact discontinuities between the gaseous atmospheres of the two merging clusters. More recently, also the shocks associated with mergers have been discovered in some of these clusters (see \citealt{Markevitch2010shocks} for a review).  

CFs of the second variety form arcs around the cool cores of apparently relaxed clusters and have more subtle temperature contrasts of  a factor of  around 2. \citet{Markevitch2001} proposed gas sloshing as the origin of these CFs: while a gas-free subcluster falls through the main galaxy cluster, its gravitational impact during pericentre passage offsets the main cluster's central ICM. Consequently, the offset ICM starts falling back towards the cluster centre and starts sloshing inside the main potential well. Usually, the subcluster passes the main cluster core at some distance, it transfers angular momentum to the ICM, and the sloshing takes on a spiral-like appearance. The details of the dynamics and offset mechanisms are described in \citet{Ascasibar2006}, who performed hydrodynamical SPH+$N$-body simulations confirming this scenario. \citet{ZuHone2010} presented similar simulations, but concentrated on the heating efficiency of the sloshing process. 

\citet{Roediger2011} (R11 hereafter) presented an extensive study of gas sloshing in the Virgo cluster. In a one-to-one comparison between simulations and observations, they showed that the sloshing mechanism can explain all  properties of the observed CFs also quantitatively, and have identified several new characteristic features. 

Here, we continue this approach by performing a similar study for the cluster Abell 496. To this end, we run a set of numerical simulations of minor-merger induced gas sloshing tailored to this cluster, produce synthetic observations and compare them in detail to the available observations. 

The article is structured as follows: in Sect.~\ref{sec:a496} we set the stage by giving a comprehensive synopsis of the observed sloshing signatures in A496, of course including the CFs, but also other features.   Sect.~\ref{sec:compare} briefly introduces the simulations and compares simulation results and observations in detail. Here, we derive the most likely merger scenario and discuss uncertainties and possible alternatives. In Sect.~\ref{sec:discussion} we discuss the (im)possible identification of the responsible subcluster, 
the origin of the distortions in A496's cold fronts
and further implications.  Finally, Sect.~\ref{sec:summary} summarises our findings.  For the sake of clarity, the description of the observational data sets and the description of the  simulation method and cluster model are collected in Appendices~\ref{sec:compare_datasets} and \ref{sec:method}, respectively.

%*************************************************
\section{Setting the stage: \newline sloshing signatures in A496} \label{sec:a496}
%*************************************************
%
%FFFFFFFFFF
\begin{figure*}
\hspace{0.9cm} Xray image \hspace{2.cm} brightness residuals (centre) \hspace{0.2cm} brightness residuals (large-scale) \hspace{0.2cm} projected temperature \hfill\hfill\phantom{x}
\newline
\rotatebox{90}{\hspace{2cm} observations}
\includegraphics[trim=100 0 80 0,clip,width=0.24\textwidth]{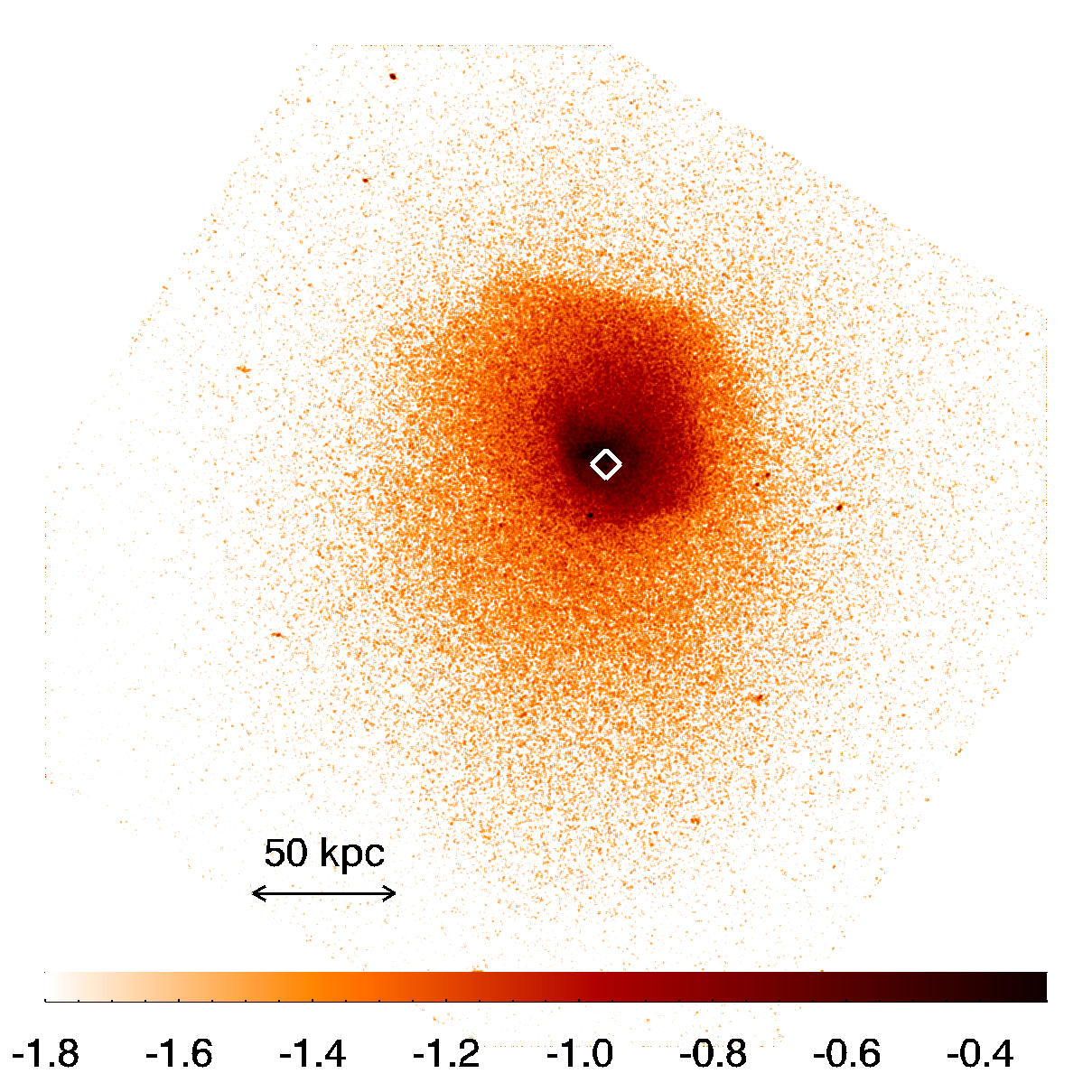}
\includegraphics[trim=100 0 80 0,clip,width=0.24\textwidth]{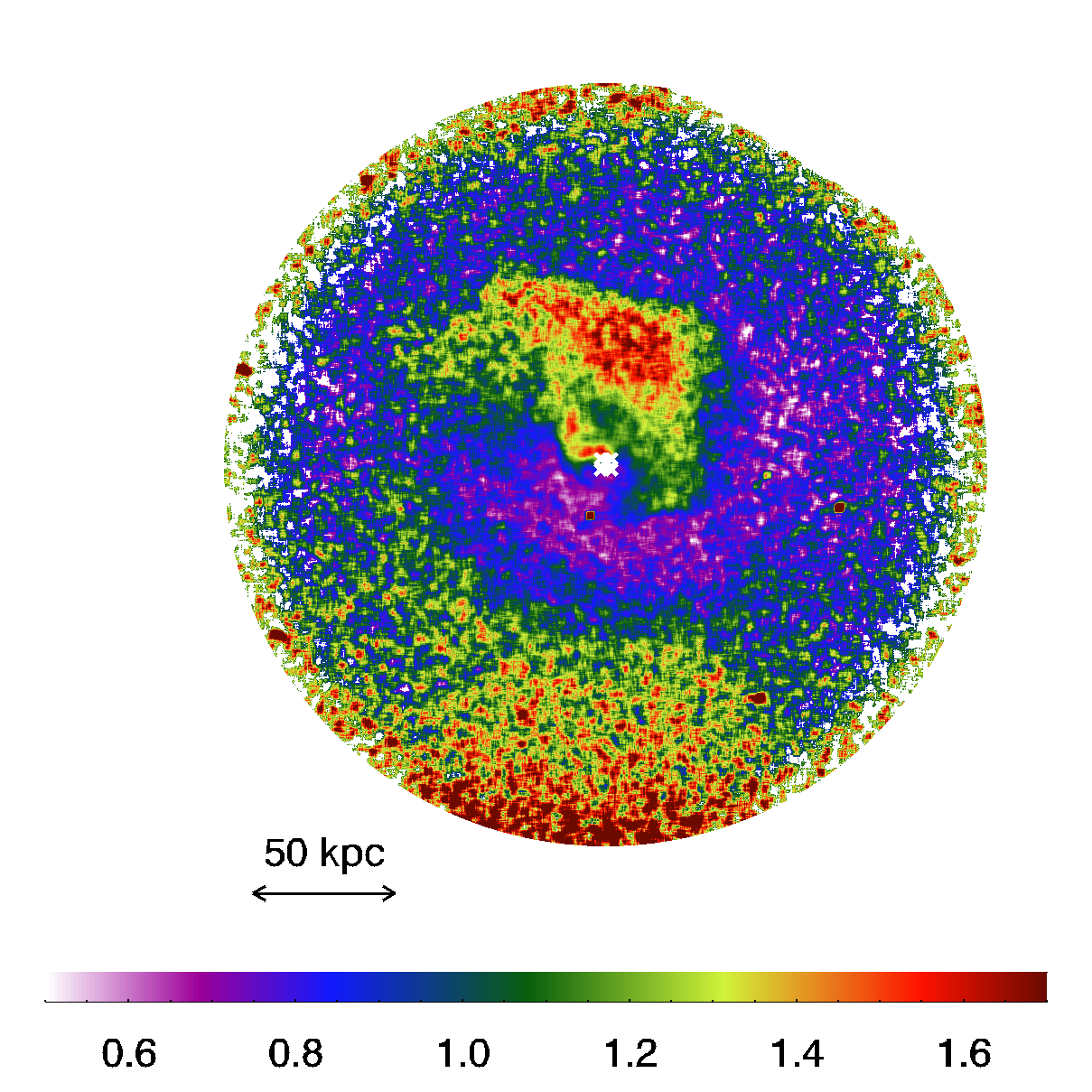}
\includegraphics[trim=10   0 10 0,clip,width=0.27\textwidth]{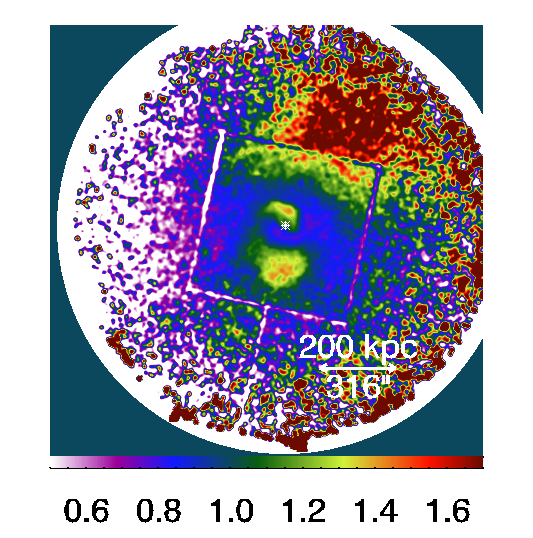}
\includegraphics[trim=0 0 0 0,clip,width=0.17\textwidth]{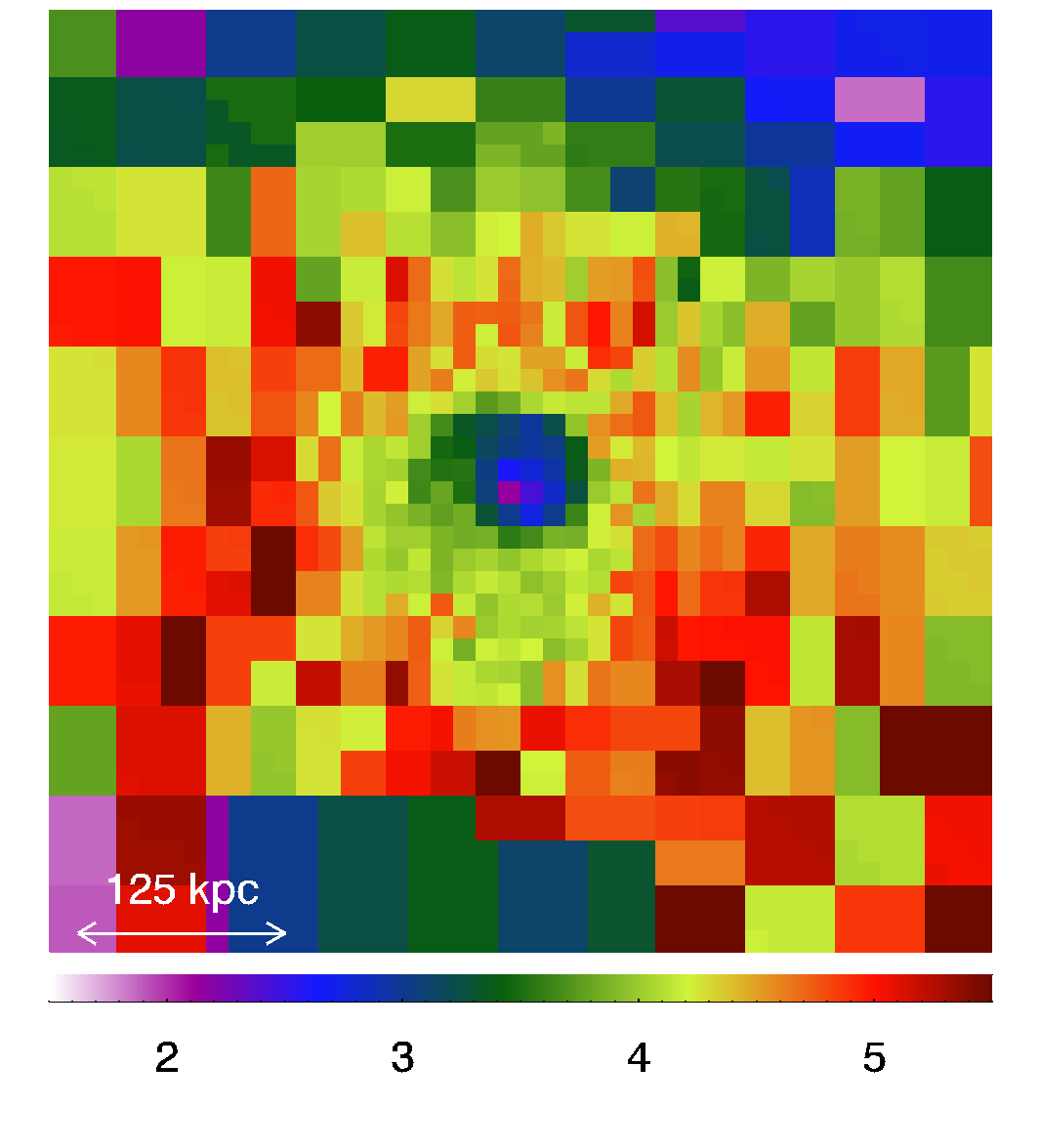}
\hfill\phantom{x}
\newline
\rotatebox{90}{fiducial simulation "close"}
\includegraphics[trim=30 0 300 100,clip,height=3.9cm,origin=c,angle=-15]{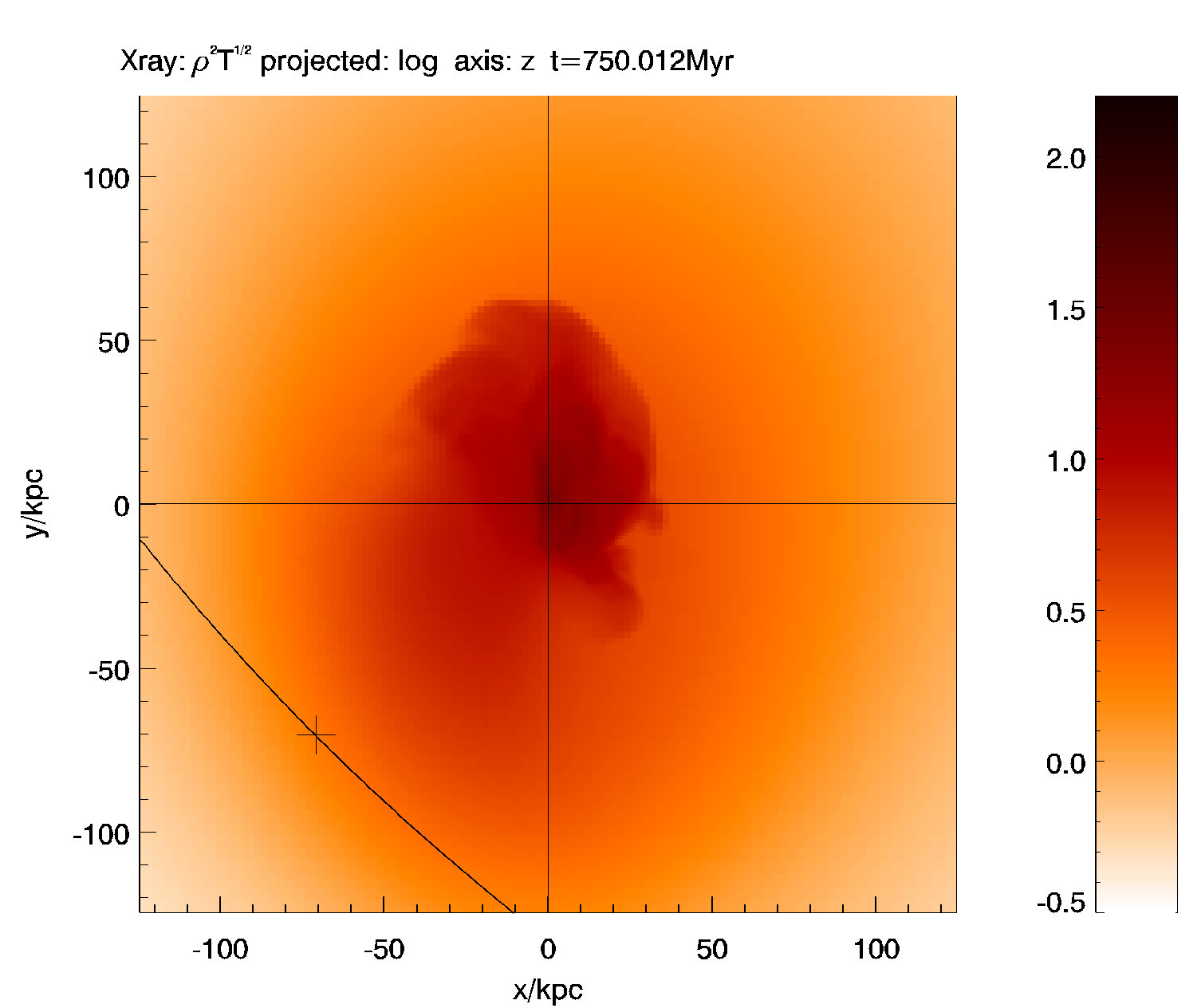}\hspace{-0.8cm}
\includegraphics[trim=30 0 300 100,clip,height=3.9cm,origin=c,angle=-15]{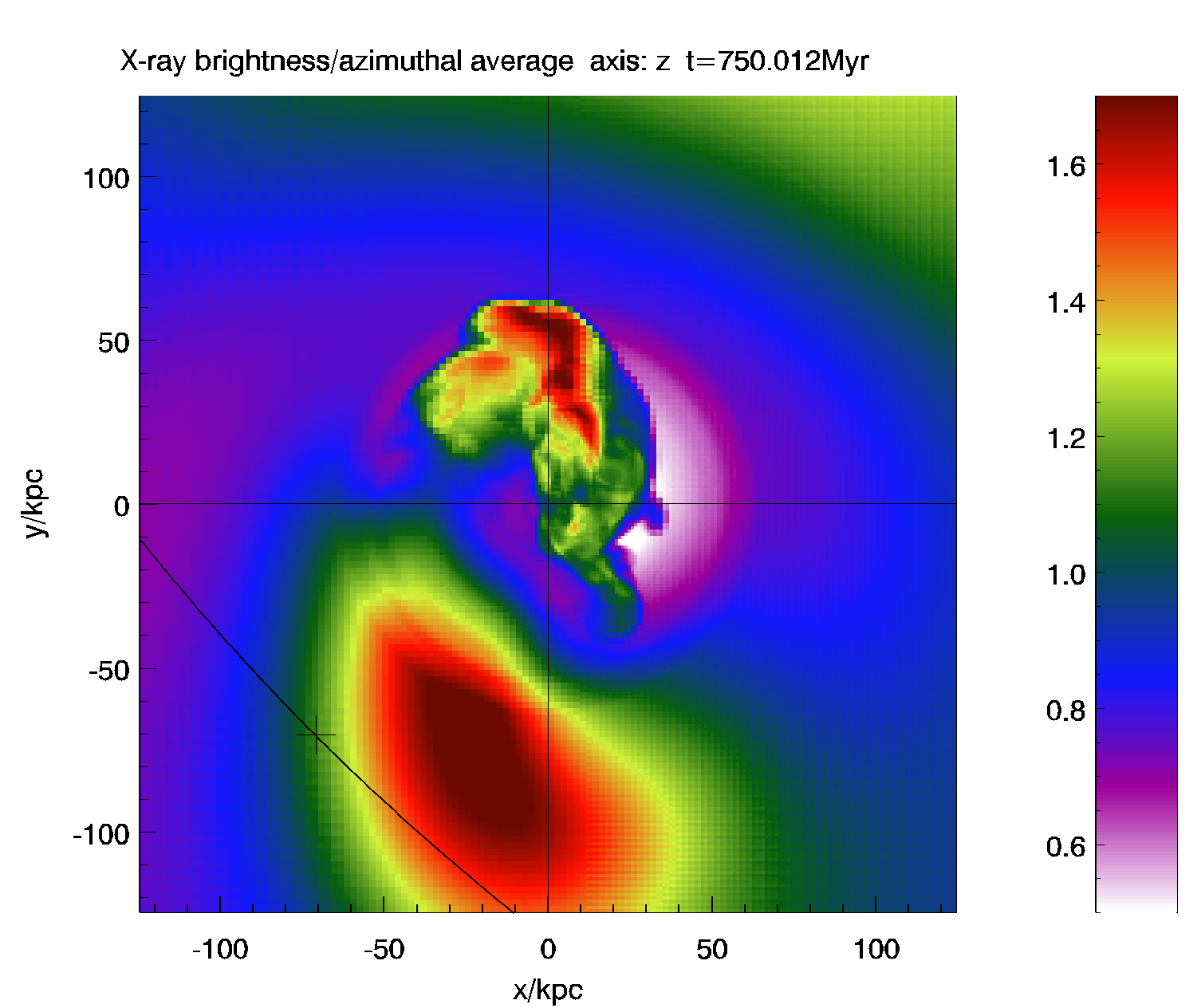}\hspace{-0.8cm}
\includegraphics[trim=30 0 300 100,clip,height=4.3cm,origin=c,angle=-15]{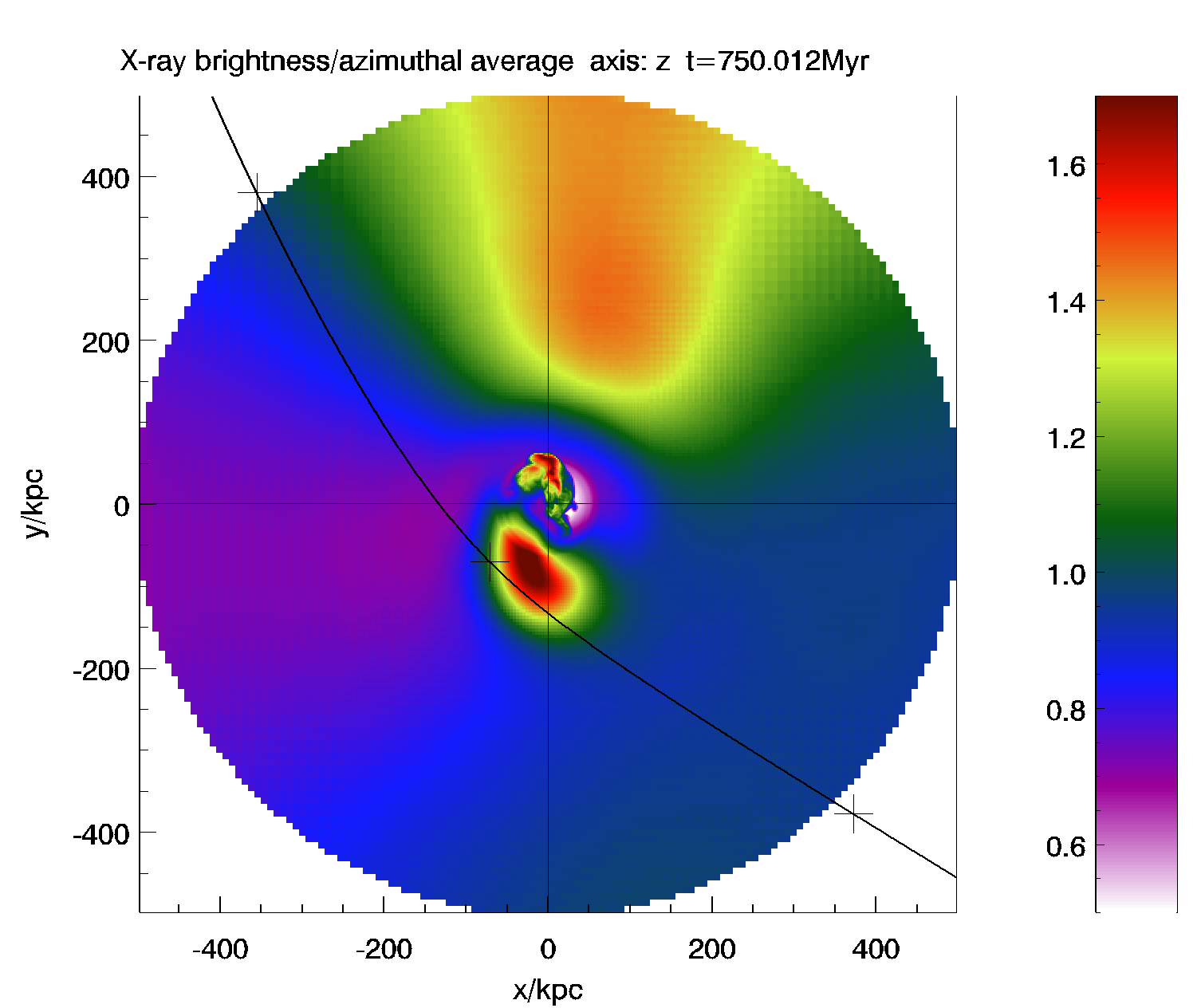}\hspace{-0.8cm}
\includegraphics[trim=30 0 0 100,clip,height=3.cm,origin=c,angle=-15]{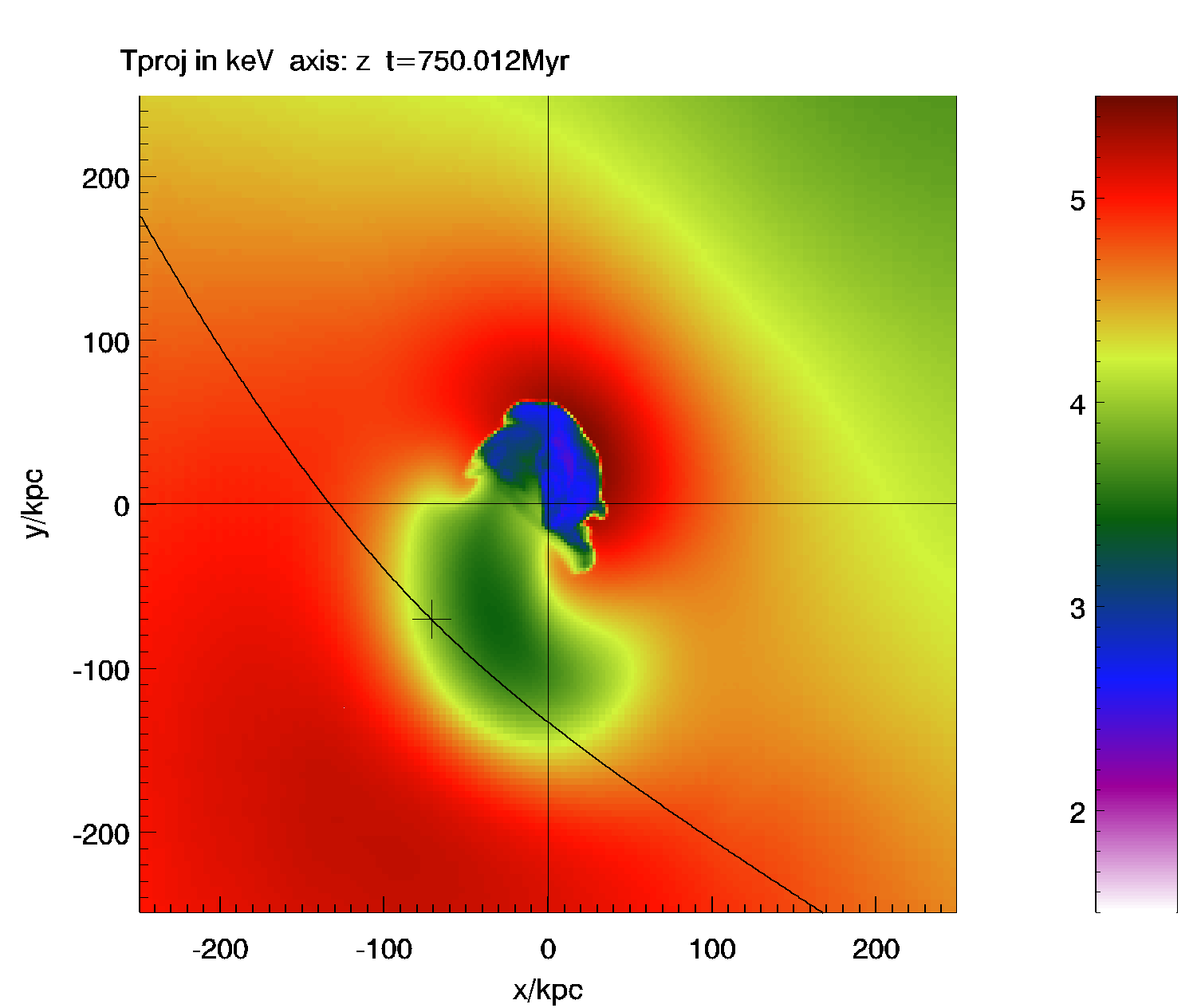}
\rotatebox{90}{fiducial simulation "distant"}
\includegraphics[trim=30 0 300 100,clip,height=3.9cm,origin=c,angle=-15]{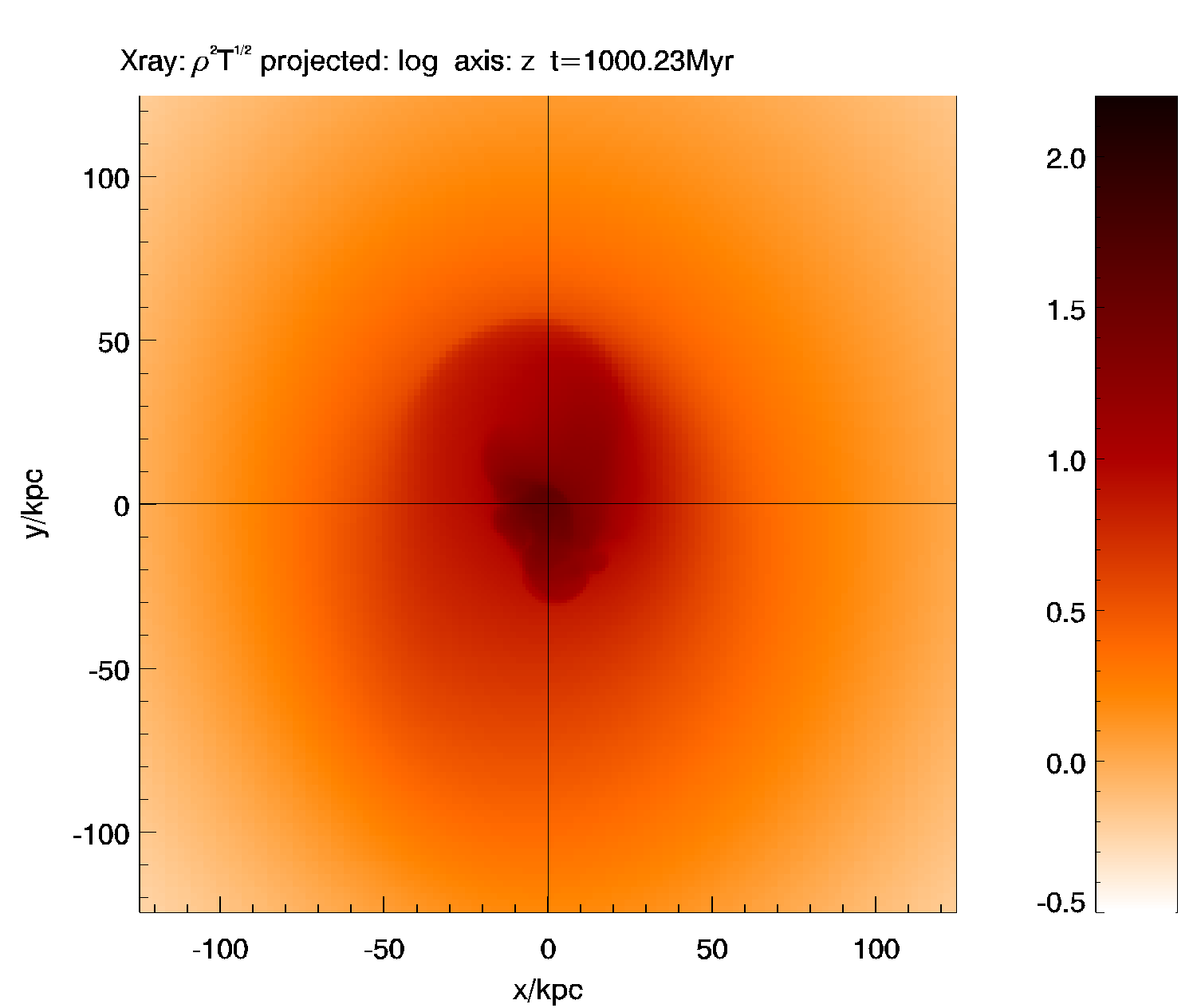}\hspace{-0.8cm}
\includegraphics[trim=30 0 300 100,clip,height=3.9cm,origin=c,angle=-15]{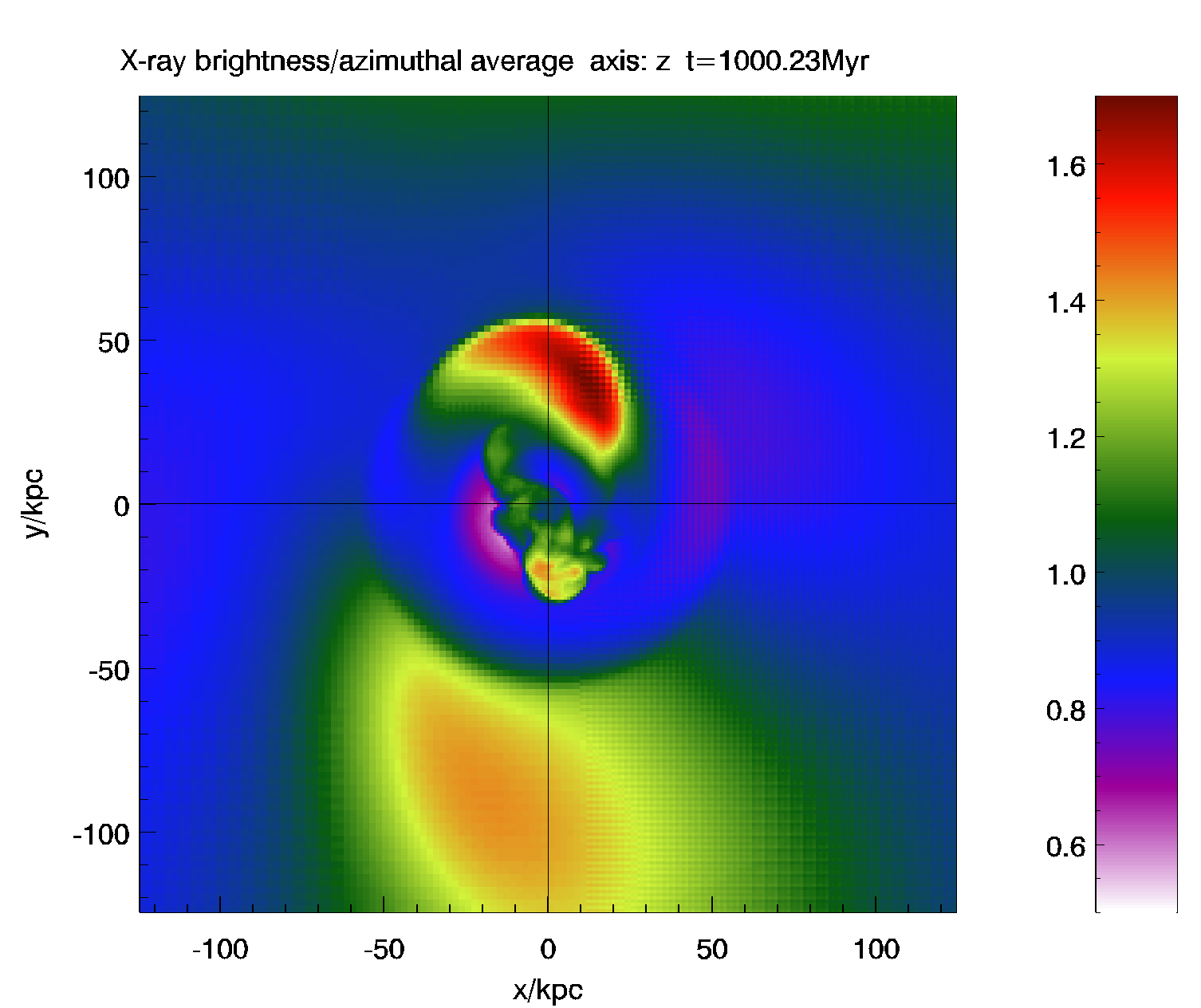}\hspace{-0.8cm}
\includegraphics[trim=30 0 300 100,clip,height=4.3cm,origin=c,angle=-15]{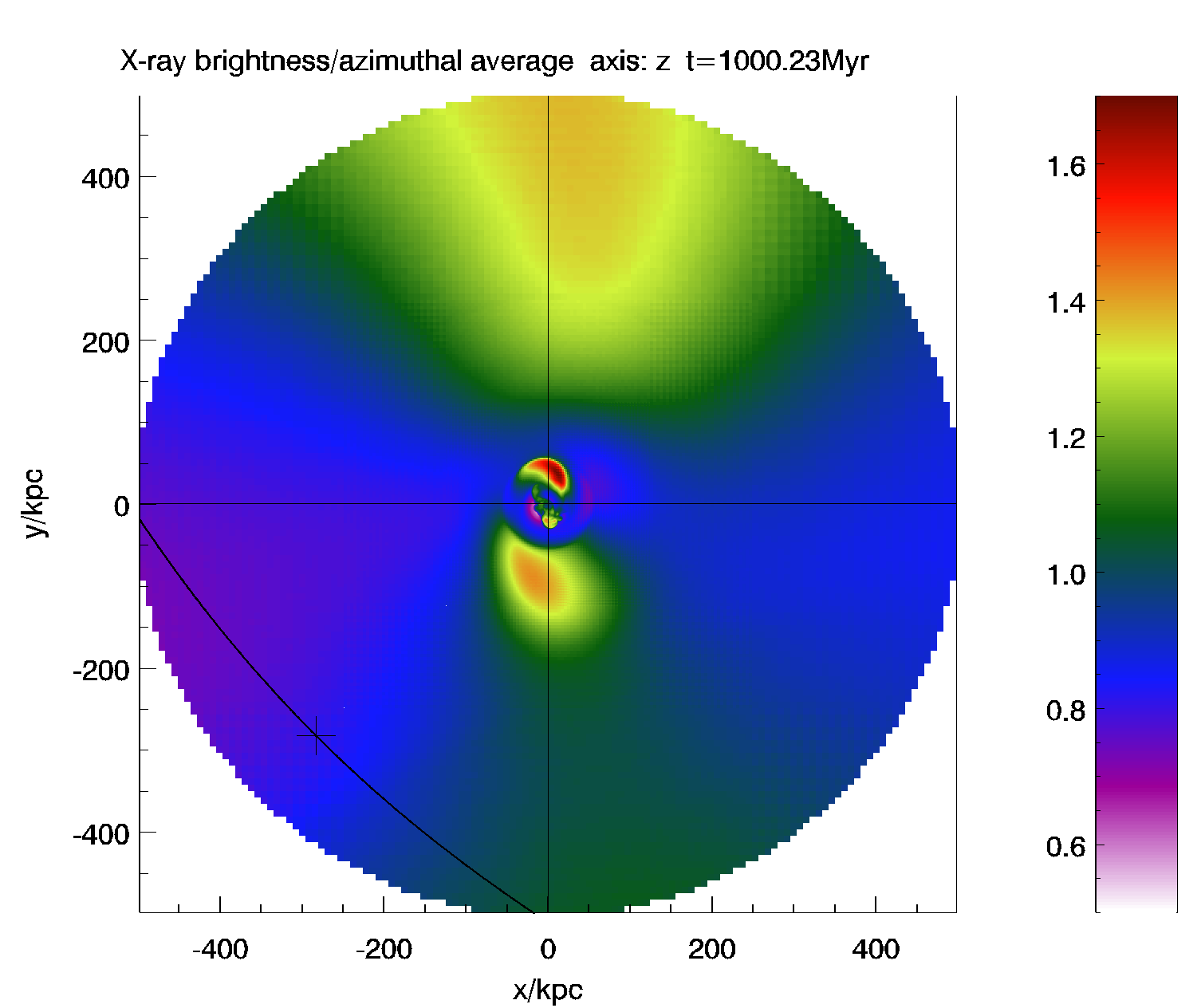}\hspace{-0.8cm}
\includegraphics[trim=30 0 0 100,clip,height=3.cm,origin=c,angle=-15]{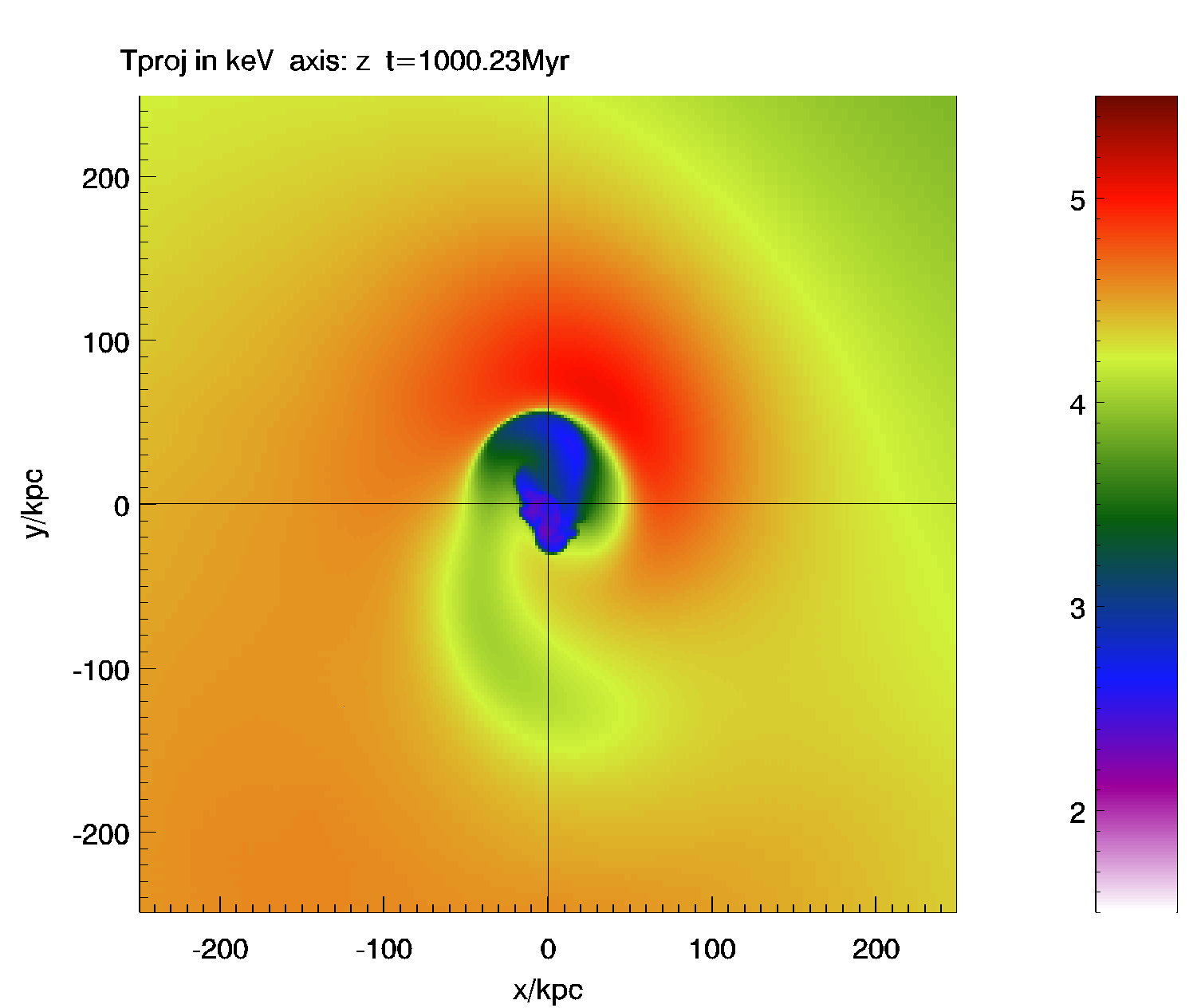}
\caption{Sloshing signatures in A496 -- qualitative comparison of observations and simulation results. Observations are shown in the top row, from left to right: logarithmic X-ray image for the central region from \citet{Dupke2007}, replotted in our colour scale; brightness residuals w.r.t. azimuthal average for the central region; brightness residuals out to 500 kpc from XMM archival data (see Table~\ref{tab:datasets}); projected temperature map out to 300 kpc, from \citet{Lovisari2011}, replotted in our colour scale. In the X-ray and residual images, we mark the cluster centre. The  middle and bottom rows display the same quantities in the same colour scales for our two fiducial simulations. In both runs, a subcluster with $4\times 10^{13}M\Sun$ and a scale radius of 100 kpc crossed the cluster on a diagonal orbit (see black lines in large-scale residual maps). In the middle row, the pericentre passage took place  at a distance of 100 kpc to the SE $0.75\Gyr$ ago, in the bottom row at 400 kpc $1\Gyr$ ago.
 The line-of-sight is perpendicular to the orbit. The cold fronts are the  outer edges of the brightness excess spiral. The fiducial runs reproduce nearly all features: 
the spiral pattern of the cold fronts and the brightness excess, the orientation of the brightness excess spiral, the large-scale brightness excess towards the N-NW, the strongest large-scale brightness deficit towards the E, and the appearance of the temperature map. These two best simulations differ in details. The larger impact parameter in the bottom row leads to a more regular structure of the cold fronts, and reproduces the cold front positions towards all directions simultaneously (see Fig.~\ref{fig:A496_profs_ell} and Sect.~\ref{sec:radii_age}). The smaller impact parameter in the middle row leads to stronger perturbation in the cluster core and higher shear velocities along the cold fronts. While this run does not match the cold front radii towards the East and West as well, it  resembles the disturbed shape of the observed cold fronts, i.e.~there are kinks of similar size as well as a double cold front towards the West.
}
\label{fig:A496_maps}
\end{figure*}
%FFFFFFFFFF

A496 is a nearby, X-ray bright galaxy cluster (redshift $z \approx 0.032$, \citealt{Dupke2007,Chilingarian2008} and refs.~therein). For a Hubble constant of $73 \Kms \Mpc^{-1}$,  the distance of A496 is $131\Mpc$, and one arcmin corresponds to  $38\Kpc$. Its X-ray contours are fairly regular, but the cluster is somewhat elongated in N-NW to S-SE direction.  Hence, this cluster is usually classified as a relaxed cool core cluster. Its X-ray peak coincides very well with central cD galaxy (\citealt{Dupke2007}). 

Apart from these global characteristics, all typical features of gas sloshing are found in A496. We summarise them in Fig.~\ref{fig:A496_maps} and give an overview of all sloshing signatures in the following subsections.   The comparison to the simulation results is discussed in Sect.~\ref{sec:compare}.

%***********
\subsection{Brightness edges in the X-ray image}
The \textit{Chandra} observations of \citet{Dupke2003} and \citet{Dupke2007} (D07 hereafter)  revealed a series of CFs wrapped around the cluster core in a spiral-like fashion (top left panel of Fig.~\ref{fig:A496_maps}). The major CF resides about 60 kpc north (N) of the cluster core and was also identified in XMM data (\citealt{Ghizzardi2010}, G10 hereafter). There is a secondary CF about 20 kpc towards the South (S) (D07, G10) although D07 split this CF into two, one towards S-SW and one towards SE. 
G10 and \citet{Tanaka2006} identified a further CF $\sim 150 \Kpc$ S of the cluster core, which is also evident in the Chandra image.

%***********
\subsection{Spiral-shaped brightness excess}
The  structure of the cluster becomes more evident in  brightness residual maps which show the ratio of the local brightness to the average value at each radius, thus   highlighting the deviation from circular symmetry. The second panel in the top row of Fig.~\ref{fig:A496_maps} displays the residual map for the cluster centre derived from the Chandra image of D07.%
\footnote{We note that the details of the residual map depend to some degree on the definition of the cluster centre as demonstrated in Appendix~\ref{sec:residual_center}. However, the overall structure including the spiral-shaped brightness excess, the northern CF and the brightness excess towards the S are robust features.}
The CFs are the outer edges of a spiral-shaped brightness excess structure. The secondary CF about 20 kpc south of the centre appears as a jump in the deficit.
 
 The residual map in the third panel is based on the background and vignetting corrected \textit{XMM-Newton} image and extends out to 500 kpc (archival data from 2007, 2008, see Table~\ref{tab:datasets}). It reveals the same structure as the one derived by  \citet{Tanaka2006} from an earlier and shorter XMM observation. Here we see the brightness excess spiral   extending out to $\sim 150 \Kpc$ towards the S. 
 Having detected the full extent of this central structure is helpful in constraining the merger geometry.   \citet{Lagana2010} find the same spiral-shaped morphology in their substructure map of A496, which highlights brightness excess w.r.t.~a $\beta$-model fit to the cluster.

%***********
\subsection{Large-scale and general asymmetry} \label{sec:asymmetry}
Furthermore, both, Tanaka's and our large-scale residual map clearly show an extended brightness excess outside of 300 kpc N-NW of the cluster core. R11 found a very similar feature in their  simulations for the Virgo cluster and identified it as a typical signature of gas sloshing. Such a large-scale brightness excess arises a few 100 kpc from the cluster centre towards the direction approximately opposite of the pericentre of the subcluster orbit because the cluster centre is displaced w.r.t.~the overall cluster. Consequently, this feature combined with the orientation of the brightness excess spiral indicates a subcluster orbit roughly from SW towards NE with the pericentre SE of the core. 

%
%FFFFFFFFFF
\begin{figure}
\centering\includegraphics[width=0.45\textwidth]{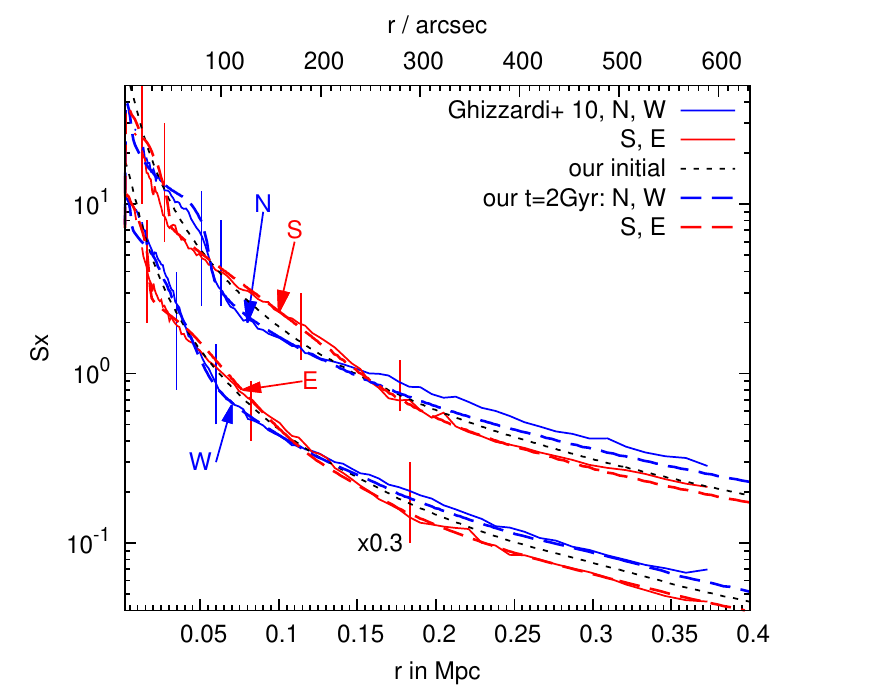}
\caption{Comparison between X-ray brightness profiles from opposite sides. The directions are colour-coded and labelled. For clarity, the profiles towards  E and W have been multiplied by 0.3. Solid lines are observed profiles (XMM-Newton, G10), dashed lines are simulation results for our fiducial run {"distant"}. Dotted lines indicate our initial profiles. Profiles from opposite sides oscillate around each other, switching over at the cold fronts. The vertical double lines bracket the cold fronts  towards the individual directions, using the same colour code as the profiles.}
\label{fig:profs_xray}
\end{figure}
%FFFFFFFFFF
%

The general asymmetry of the cluster is clearly evident when radial profiles towards opposite directions from the cluster centre are compared. As an example, we do so in Fig.~\ref{fig:profs_xray} for the X-ray brightness: N is compared to S and E to W. The profiles were averaged over  azimuthal ranges of  $30\degree$ as shown in Fig.~\ref{fig:sectors}. This comparison demonstrates a general feature typical of gas sloshing and reported already by R11: profiles from opposite sides oscillate around each other, switching over at the CFs. This is not only true for the X-ray brightness, but also for temperature and metallicity. The large-scale asymmetry appears as the split-up between opposite profiles at radii outside the outermost  CF.

 %***********
\subsection{Temperature and metallicity structure}
The XMM-Newton temperature map (top right panel in Fig.~\ref{fig:A496_maps}, \citealt{Lovisari2011},  L11 hereafter) confirms that brightness excess regions correspond to cooler regions, despite the fact that the  temperature map is very coarse. Also the metallicity map (L11)  shows a rough correspondence of brightness excess and higher metallicity, although it suffers from considerable systematic uncertainties (see  Appendix~\ref{sec:compare_datasets}). The less deep Chandra observation of D07 does not show significant structures in the metallicity map, but a clear cool region just inside the northern CF in also found in the Chandra data.

%***********
\subsection{Cold front positions from radial profiles} \label{sec:CFradii_obs}
%
%FFFFFFFFFF
\begin{figure*}
\includegraphics[width=0.49\textwidth]{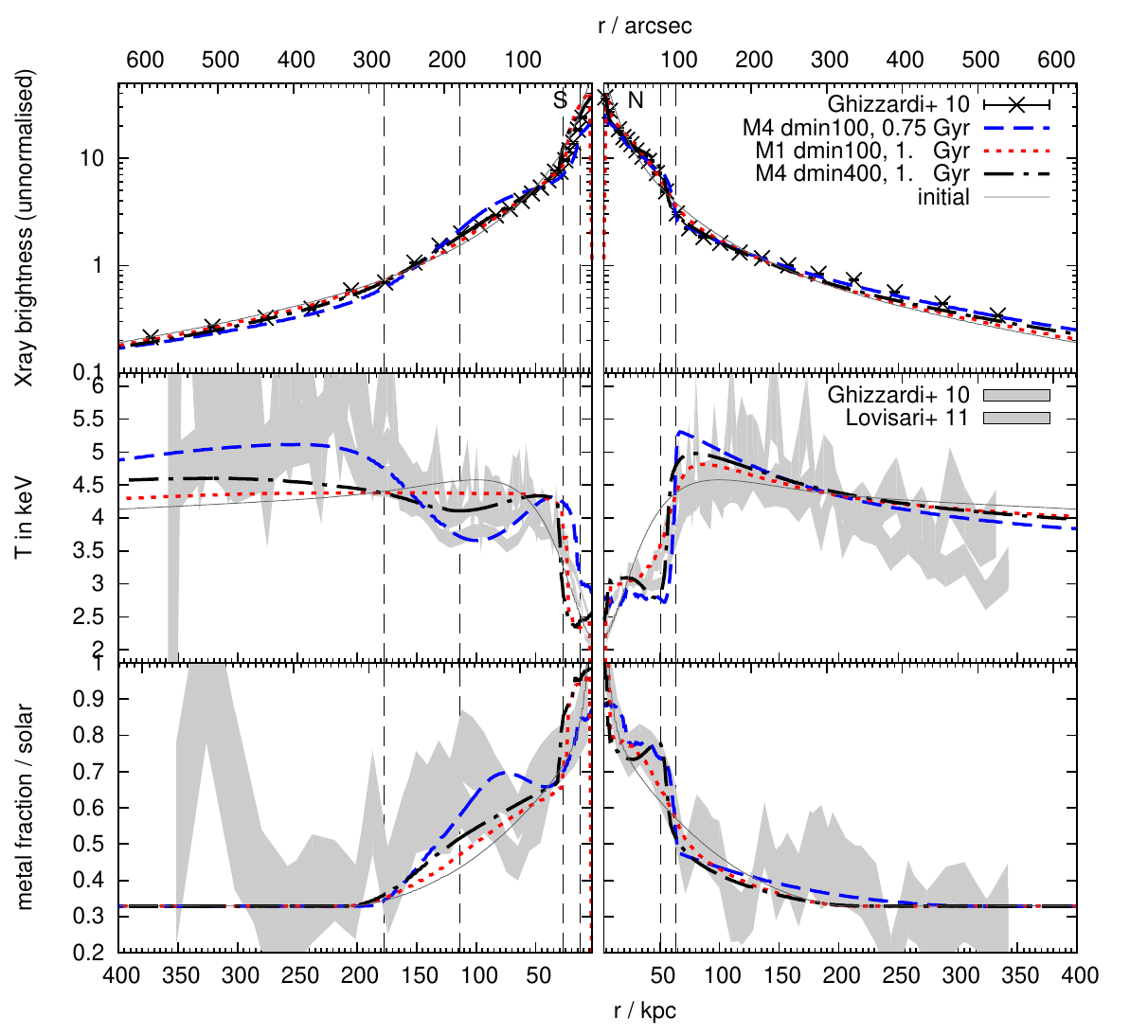}
\includegraphics[width=0.49\textwidth]{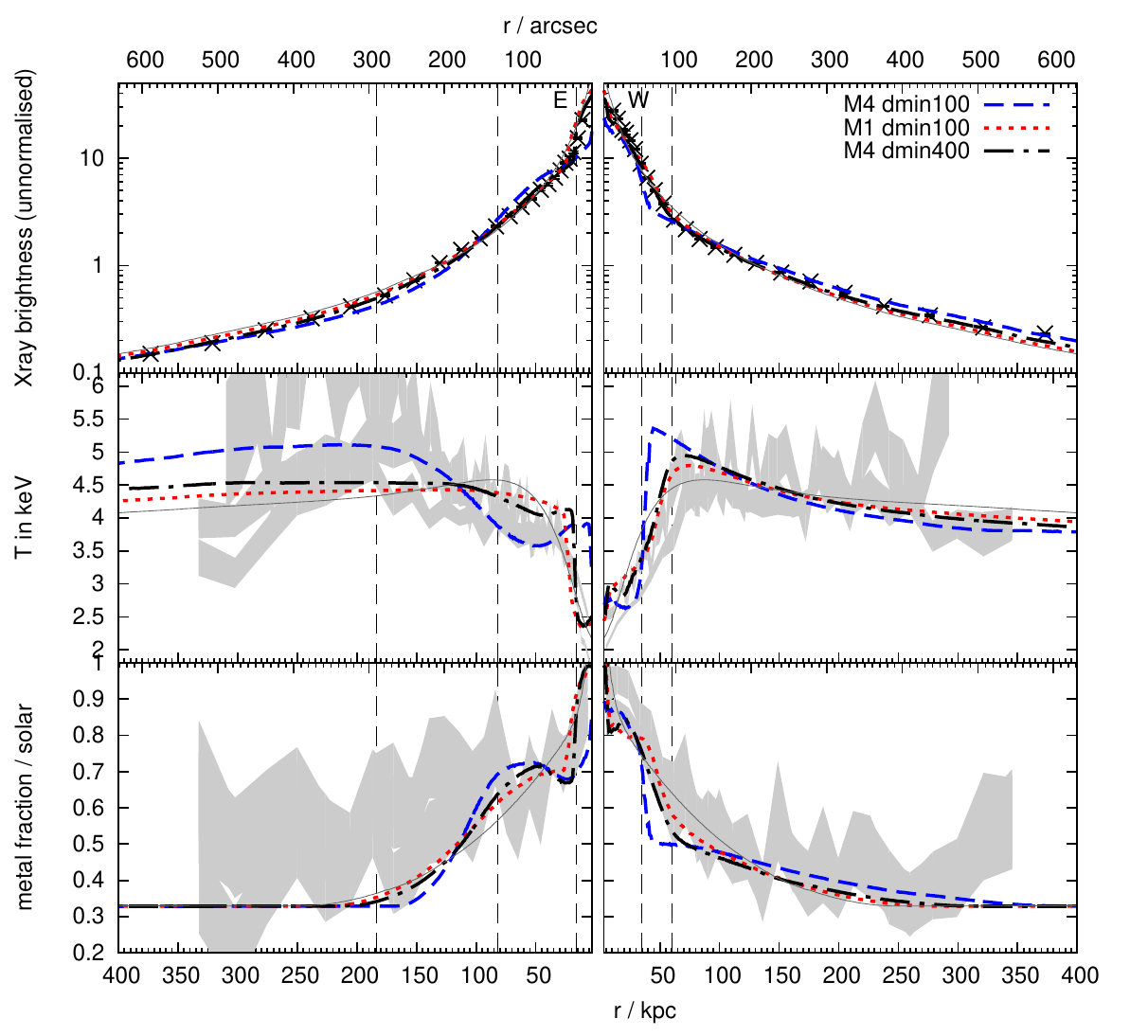}
\caption{Comparison of observed and synthetic profiles of X-ray brightness (top panels), projected temperature (middle panels) and metallicity (bottom panels) towards N and S (lhs figure), and E and W (rhs figure). For the observed X-ray profiles, we plot only every 4th available data point for the sake of clarity. The profiles are extracted from $30\degree$ azimuthal sectors as shown in Fig.~\ref{fig:sectors}. 
We show the following XMM-Newton observational data: X-ray brightness and temperature from G10,  and our re-calculation of temperature and metallicity profiles with different assumptions (APEC/VAPEC model, hydrogen column density fixed/free during fit; see Appendix~\ref{sec:compare_datasets}).  The different resulting temperature and metallicity profiles are shown as grey-shaded bands, together they indicate the amount of systematic error (see Fig.~\ref{fig:A496_profs_observed} for a distinction). 
We plot  three representative simulations: the {fiducial "distant" run, the fiducial "close" run}, and a low mass case with a subcluster of $10^{13}M\Sun$ and a pericentre distance of 100 kpc; see Table~\ref{tab:runs} for model parameters. 
The dashed vertical double lines mark the position of the cold fronts in the radial profiles (just single line for the inner eastern cold front).}
\label{fig:A496_profs_ell}
\end{figure*}
%FFFFFFFFFF
%
Radial profiles of X-ray brightness, temperature and metallicity are shown in Fig.~\ref{fig:A496_profs_ell} and are discussed in comparison to the simulations in Sect.~\ref{sec:mass_jumps}. Different observational datasets are available and are compared in Appendix~\ref{sec:compare_datasets}. Figure~\ref{fig:A496_profs_ell} shows a representative selection based on \textit{XMM-Newton} observations: the X-ray brightness and temperature profiles from G10,  and our re-calculation of temperature and metallicity profiles with three assumptions (APEC/VAPEC model, hydrogen column density fixed/free during fit; see Appendix~\ref{sec:compare_datasets}).  The different  temperature and metallicity profiles are shown as grey-shaded bands, together they indicate the amount of systematic error. In  Fig.~\ref{fig:A496_profs_observed} the different datasets can be distinguished. The profiles are averaged azimuthally over $30\degree$ as shown in Fig.~\ref{fig:sectors}.

In principle, CFs appear in radial profiles as downward jumps in X-ray brightness and metallicity, and as upward jumps in temperature. The position of these jumps identifies the CF position or CF radius. However, {projection and} the combination of azimuthal and radial binning reduces the intrinsic discontinuities to more or less steep gradients. 

In the radial profiles towards the N, the major, northern CF is clearly identified by such a steep gradient in all quantities. The other CFs are more subtle because they reside inside the cool core near the cluster centre. Here, not only the overall X-ray brightness and metallicity are strongly declining but, additionally, the overall temperature  increases with radius. This means that the general gradients of all quantities are in the same direction as typical for a CF. Thus, the gradient due to the CF blends with the overall gradient and the CF radius is hard to identify. As a consequence,  all but the major northern CF are mainly apparent as kinks in the observed X-ray brightness profiles, and their exact position is difficult to constrain {from the radial profiles}. 

Here, the oscillating behaviour of opposite profiles as described in Sect.~\ref{sec:asymmetry} and Fig.~\ref{fig:profs_xray} can help: the radius at which the profiles switch over identifies the position of the CF. Combining the information from each profile alone and from the comparison of opposite directions, we infer the positions of the CFs as listed in Table~\ref{tab:CFradii}.
%
%TTTTTTTTT
\begin{table}
\caption{Positions of cold fronts in A496 {derived from radial profiles}: we give a minimal extent in azimuth and the inner and outer edge of the cold front in radius in the sectors labelled in Fig.~\ref{fig:sectors}. Also for the azimuth we follow the convention from Fig.~\ref{fig:sectors}.}
\begin{tabular}{|l|l|r|r}
\hline
                          &                               & \multicolumn{2}{c}{radius in kpc} \\
                          &    azimuthal   & \multicolumn{2}{c}{(in arcsec)} \\
                          &   extent    & \multicolumn{2}{c}{in sectors} \\
\hline
major CF N-W: & $-10\degree$   to           &   N:  50-63&  W: 35-60 \\
                           &          $110\degree$                                                    &    (80-100)    &    (55-95)   \\
secondary CF E-S-SW: & $180\degree$ to &   S: 13-27 &  E: ?-15\\
                           &                  $300\degree$                              &    (20-43)    &    (?-25)   \\
outer CF E-S-SW: & $180\degree$ to &   S: 114-177 &  E: 82-184 \\
                           &        $300\degree$                    &    (180-280)    &  (130-290)  \\
\hline
\end{tabular}
\label{tab:CFradii}
\end{table}
%TTTTTTTTTTTT
%
For each CF we state its azimuthal range determined from the maps in Fig.~\ref{fig:A496_maps} and a radial range which encompasses the switch-over region as seen in Fig.~\ref{fig:profs_xray}.  For the secondary CF towards the E, the azimuthal profiles do not allow to reliably constrain the inner edge of the CF. {We measure the CF width more accurately in the next subsection}.

Our results mostly agree with the ones from D07 and G10. We have extended the major (northern) CF towards the W and the outermost southern CF towards the E.  We find the {outer southern CF} at the same radius as G10.  We have also interpreted the two inner southern CFs from D07 as one secondary CF. The same CF was reported by G10 at a slightly larger radius and  smaller azimuthal extent. 

We  mark these CF  positions in all relevant plots.

%***********************
\subsection{Detailed structure and width of the cold fronts} 
\begin{figure*}
\includegraphics[trim=0 100 0 0,clip,width=0.5\textwidth]{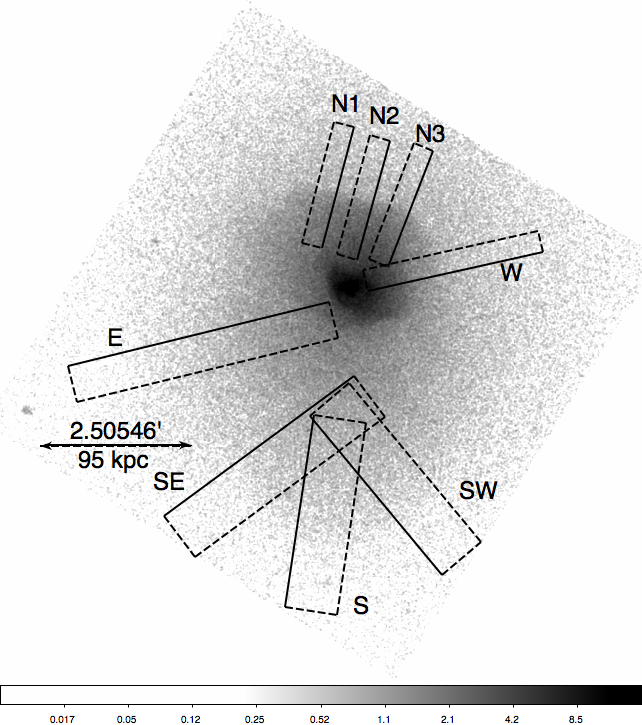}
\includegraphics[width=0.45\textwidth]{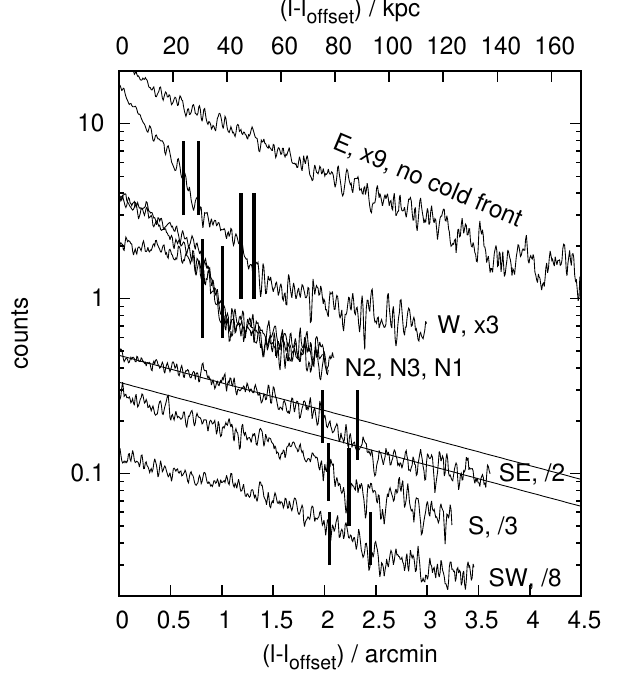}
\caption{{Width of cold fronts. Left: profile extraction regions perpendicular across the cold fronts. Right: Profiles along the regions shown on the left. The length coordinate along each profile is set off arbitrarily, it does not represent a distance from the cluster centre.  In order to avoid confusion, we have multiplied/divided each profile as noted in the legend. The double vertical lines denote the position and width of each cold front, or more precisely an upper limit on its width. We determine the width by fitting exponential functions to each profile inside and outside of the front as shown for the SE profile. The steeper transition region between the radial ranges that are well-fitted is the cold front. We clearly identify the N cold front in profiles N1, N2 and N3, and two fronts in the W profile. The fronts are more subtle in the three southern profiles, S, SE and SW, but clearly detectable. For comparison, the profile to the E does not contain a front.
}
}
\label{fig:A496_newprofs}
\end{figure*}

{
The inner CFs in A496 are peculiar because of their boxy morphology. Instead of smooth arcs as e.g.~in Virgo (\citealt{Simionescu2010}) or A2142 (\citealt{Markevitch2000,Owers2009hifid}) they show several kinks.  We discuss their  possible origin in Sect.~\ref{sec:KH}. The CFs being non-circular, radial azimuthally averaged profiles are not optimal for measuring their widths. Hence, we derived additional surface brightness profiles across the N and W cold front and the outer southern front, taking particular care to place the profile directions locally perpendicular to the fronts. For this reason, our profiles are not extracted from annuli but the box-like regions shown in the left-hand part of Fig.~\ref{fig:A496_newprofs}. The right-hand-side of this figure compares the resulting brightness profiles along the regions. 
}
%FFFFFFFFFF
%FFFFFFFFFF
{
All CFs appear as radial regions of steeper slope than ambient radial ranges. We determine their widths by fitting exponential functions to each profile inside and outside of the front as shown for the SE profile in Fig.~\ref{fig:A496_newprofs}. The steeper transition region between the radial ranges that are well-fitted is the cold front.  We mark them by bold double lines in Fig.~\ref{fig:A496_newprofs}. The CF in the N is clearly detected in all three northern profiles. Towards the W, there is a double CF. The outer southern CF is more subtle, but clearly detectable in all three southern profiles. The CF positions and widths for each profile are summarised in Table~\ref{tab:CFradii2}. 
}
%TTTTTTTTT
\begin{table}
\caption{{Positions and widths of cold fronts in A496 in profiles along regions marked in Fig.~\ref{fig:A496_newprofs}. }}
\begin{tabular}{|lcc}
\hline
                                                       & distance of cold   & upper limit on \\
                                                       & front to center      &  cold front  width \\
profile as labelled                           & in kpc                 &  in kpc  \\
 in Fig.~\ref{fig:A496_newprofs} & (in arcsec)         & (in arcsec) \\
\hline
N1          &            & 7.6 (12)\\
N2          & 57 (90)& 7.6 (12)\\
N3          &             & 7.6 (12)\\
W, outer & 57 (90)& 5 (8)\\
W, inner& 35 (56) & 5.7 (9)\\
S            & 158 (250) & 7.6 (12)\\
SW        && 13 (20)\\
SE         && 15 (24)\\
\hline
\end{tabular}
\label{tab:CFradii2}
\end{table}
%TTTTTTTTTTTT
%
{These widths are upper limits on the true  front width, because they contain projection effects. \citet{Vikhlinin2001} derived the X-ray brightness profile of an ellipsoidal gas cloud with a sharp outer edge. They showed that the brightness profile near the edge is approximately $\propto \sqrt{d}$, where $d$ is the distance from the edge. The radial ranges we have identified as CFs and marked by double vertical lines in the profiles in Fig.~\ref{fig:A496_newprofs} can be fitted by this function,  suggesting that also the CFs in A496 are discontinuities in ICM density. }

%***********************
\subsection{Previous suggestions regarding the origin of the cold fronts in A496} 
The observations of A496 have been interpreted differently by different authors.  D07 argued that the spiral pattern of the cold fronts and their multiplicity are unlikely to be caused by a head-on merging remnant core and favoured the sloshing scenario. Comparing qualitatively to the simulations of \citet{Ascasibar2006}, they derived a subcluster orbit roughly from south to north.  They speculated that the subcluster found in the  outskirts of A496 towards the N-NW  (\citealt{Flin2006}) could have triggered the sloshing.

Contrary, \citet{Tanaka2006} suggested that the non-uniform large-scale temperature and brightness residual distribution is unlikely to be caused by minor-merger induced gas sloshing. Hence, they interpreted  their southern CF as leading edge of a subcluster moving subsonically, and the northern one due to an oscillating central galaxy (see also \citealt{Tittley2005}). 

In this paper, we  qualitatively and quantitatively verify gas sloshing as the origin of  all observed features in A496 and infer an orbit similar to the prediction given in Sect.~\ref{sec:asymmetry} on the basis of the position of the extended large-scale brightness excess in the N-NW.

%% file: comparison.tex
%*************************************************
\section{Comparison between fiducial model and observations} \label{sec:compare}
%*************************************************
%
In this section, we present the most likely merger scenario derived from comparing between simulations and observations. We also discuss uncertainties and alternatives.

%**********************
\subsection{The  simulations} 
For the sake of clarity, here we only briefly summarise our simulation method and fiducial simulations. A detailed account of the simulation method, the initial cluster model, the subcluster model and orbit, and the synthetic observations are given in  Appendix~\ref{sec:method}. 

We follow the approach of R11  and perform  idealised simulations of a minor merger between a main cluster, here A496, and a smaller, gas-free subcluster. The ICM gas physics are described by the hydrodynamical equations. Additionally, the ICM is subject to the gravitational forces due to the main cluster and the subcluster. These are modelled by the adapted rigid potential approximation described  and verified by \citet{Roediger2011fastslosh}. Instead of modelling the DM distributions  by the N-body method, we assume static potentials for both clusters, which speeds up the simulations considerably. The subcluster potential is shifted through the main cluster along a test particle orbit. The simulations are run in the rest frame of the main cluster. As this is not an inertial frame, the resulting inertial accelerations are taken into account.
 
The main cluster, here A496, is  assumed to be initially in hydrostatic equilibrium and is set up according to observed ICM density and temperature profiles. Like A496, our model cluster  is slightly elliptical and elongated along the grid $y$-axis. 

The subclusters are described as  Hernquist halos (\citealt{Hernquist1990}).

%********************************
\subsection{Summary of fiducial scenario} 
%
%
%FFFFFFFF
\begin{figure}
\centering\includegraphics[origin=c,angle=-15,width=0.4\textwidth]{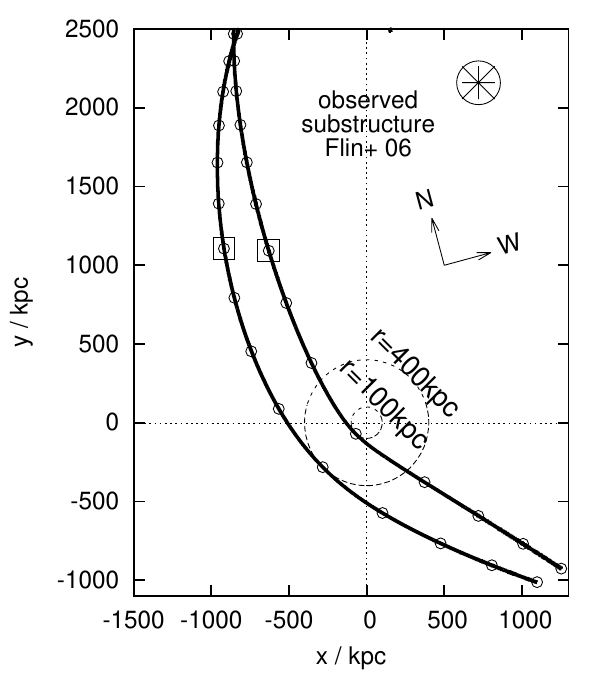}
\caption{Fiducial subcluster orbits in the $xy$-plane. Like the simulated maps in Fig.~\ref{fig:A496_maps}, we have rotated this plot by $15\degree$ clockwise, such that N is up and W is right.  We mark the subcluster position in steps of 250 Myr as predicted from the test particle orbit and highlight its position at {0.75 and $t=1\Gyr$ for the "close" and "distant" subcluster orbit, i.e.~with pericentre distances of 100 kpc and 400 kpc, respectively.} We note that the restricted potential approximation used here slightly over-predicts the cold front age and that the test-particle orbit neglects the subcluster's deceleration by dynamical friction, which is significant after pericentre passage. Hence, we  expect a current subcluster position considerably closer to the cluster centre, approximately 1 Mpc N-NW of the centre. We also mark the position of the subcluster found by \citet{Flin2006}.}
\label{fig:orbit_fiducial}
\end{figure}
%FFFFFFFFF

Our comparison of simulation results and observations leads us to the following {two fiducial runs that bracket the merger scenario}:
\begin{itemize}
\item Our fiducial subcluster has a mass of  $4\times 10^{13}M\Sun$ and a scale radius of 100 kpc.
\item It crosses the main cluster along a diagonal orbit from the SW to the N-NE, passing the cluster core SE of the centre at $t=0$ as depicted in Fig.~\ref{fig:orbit_fiducial}. {The two fiducial orbits have pericentre distances of 100 and 400 kpc, we refer to them as the "close" and the "distant" fiducial runs.}
\item {The  simulation steps that match the observations best are 0.75 Gyr and  $1\Gyr$ after pericentre passage for the "close" and the "distant" orbit, respectively.}
\item Our line-of-sight is close to perpendicular to the orbital plane.
\end{itemize}

In the following subsections, we give the reasons why we regard this configuration as the most likely.

%**********************
\subsection{General features and merger geometry} 
Along with the observed X-ray image, brightness residual maps and temperature map, Fig.~\ref{fig:A496_maps} also displays the corresponding maps for our fiducial simulations, using the same colour scales.  We have rotated our simulated maps by $15\degree$ clockwise to match the observed orientation of features. In this configuration, the fiducial runs reproduce all qualitative sloshing  features \textit{simultaneously}: the spiral pattern of the CFs and the brightness excess, the orientation of the brightness excess spiral, the large-scale brightness excess towards the N-NW, the strongest large-scale brightness deficit towards the E, and the  appearance of the temperature map.

%**********
\subsubsection{Spiral morphology and line-of-sight}
The spiral-like arrangement of the brightness excess and the CFs is a typical signature of gas sloshing seen along a line-of-sight (LOS) approximately perpendicular to the orbital plane of the subcluster. A LOS about parallel to the orbital plane would result in concentric arcs on opposite sides of the cluster core (see Fig.~\ref{fig:fiducial-LOSs}), which is not the case here.

%**********
\subsubsection{Orientation on the sky}
The orientation of the subcluster orbit on the sky can be constrained by two properties:
\begin{itemize}
\item the orientation of the brightness excess spiral, if its full extent is known,
\item and the orientation of the large-scale brightness asymmetry induced by sloshing.
\end{itemize}

%¤¤¤
\paragraph{Fiducial orientation:}
In A496, the full extent of the brightness excess spiral is covered by the XMM-Newton observation. In the configuration described above, our fiducial simulation reproduces both properties simultaneously.  Thus, we derive a subcluster orbit from the SW to the N-NW with a pericentre SE of the cluster core, as sketched in Fig.~\ref{fig:orbit_fiducial}.  

From their sloshing simulations for the Virgo cluster, R11 predicted  the existence of a large-scale asymmetry as observed here. In Virgo, this asymmetry could not be confirmed observationally, because the data did not extend far enough. Moreover, Virgo is still a very dynamic cluster and the sloshing-related asymmetry may never be disentangled from other perturbations. Here, in A496, we can for the first time confirm the connection between gas sloshing and the large-scale asymmetry, which is found in simulations and  observations at the same position.

%¤¤¤
\paragraph{Uncertainty of orbit orientation due to preferred direction of sloshing in elliptical cluster:} \label{sec:orbitorientation}
The elongation of an elliptical  cluster introduces a preferred  direction of sloshing along the axis of elongation, introducing an uncertainty in the orbit orientation. In order to assess this uncertainty, we have tested different orbit orientations in the $xy$-plane. 
%
%FFFFFFFFFF
\begin{figure}
\centering\includegraphics[width=0.35\textwidth,angle=-30]{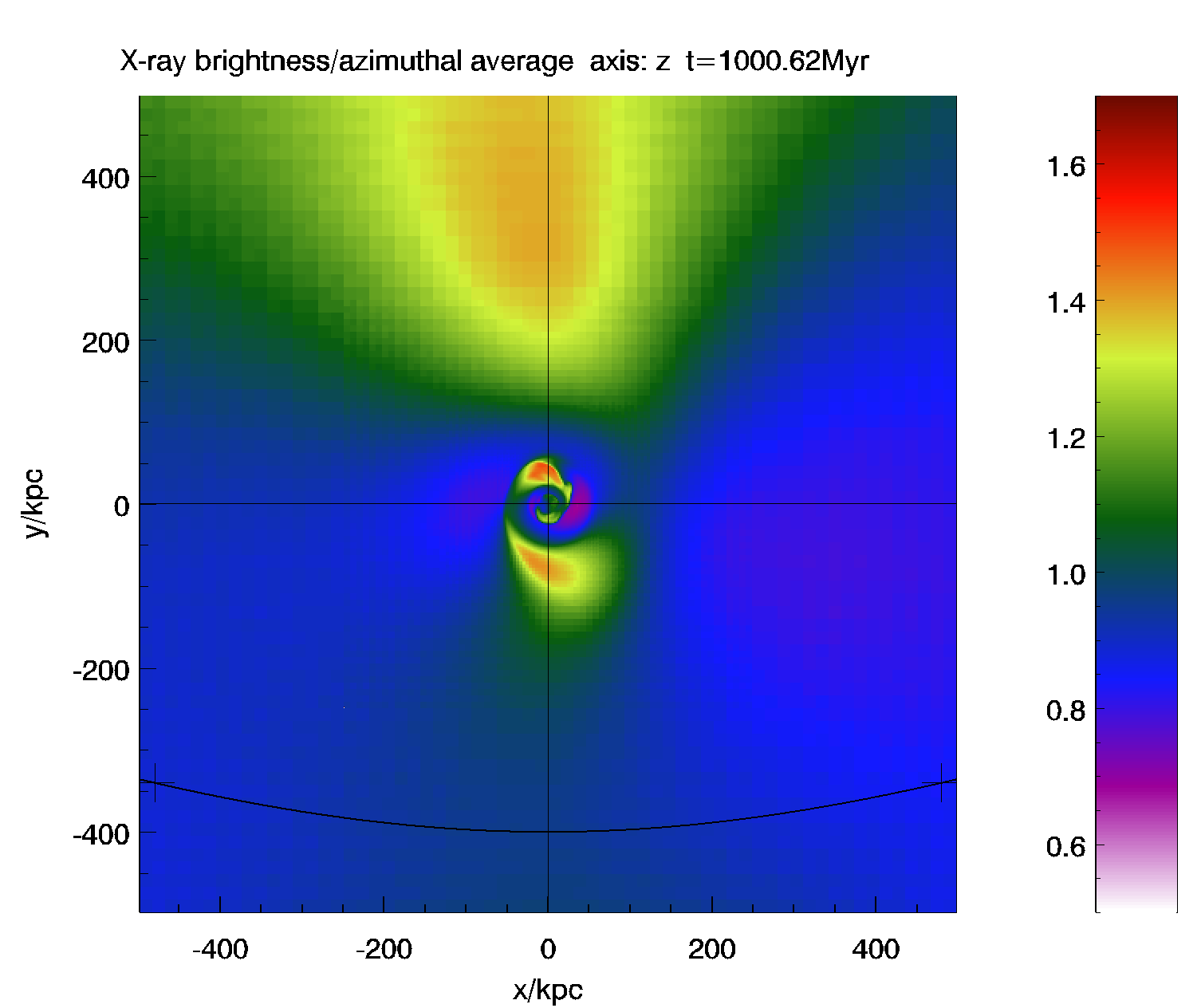}\vspace{-1cm}\newline
\centering\includegraphics[width=0.35\textwidth]{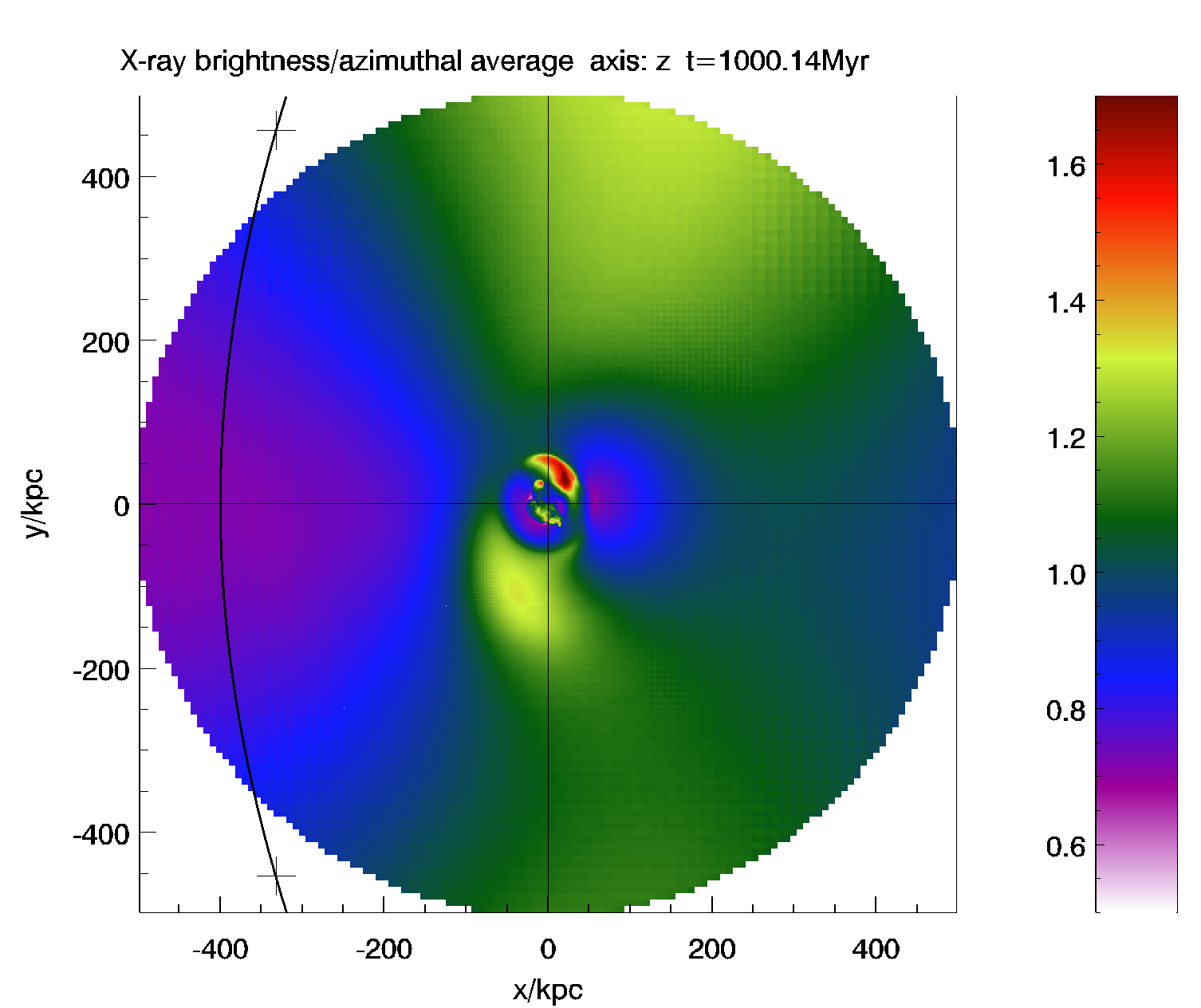}
\caption{Large-scale brightness residual maps. In the top panel, the orbit is parallel to the $x$-axis, in the bottom panel parallel to the $y$-axis. The simulations are identical to the fiducial run otherwise. We have oriented the panels such that they match the observations. While the morphology and positions of the innermost northern and southern cold front are very similar to the observation, the large-scale residuals are not. The morphology of the outer southern cold front differs from the observation in both cases,  especially in the bottom panel. The brightness excess in the N-NW is similar in the top panel, but the observed E-W asymmetry is missing. The latter is present in the bottom panel, but the N-NW excess is much too weak.}
\label{fig:parallel_x_y}
\end{figure}
%FFFFFFFFFF
%
The outcome of this test is demonstrated in Fig.~\ref{fig:parallel_x_y}. The morphology and positions of the innermost northern and southern CF are very similar to the observation in both cases, reflecting the preference for the  direction of sloshing. The large-scale residuals, however, differ. The morphology of the {outer southern CF} differs from the observation in both cases,  especially when the orbit is parallel to the $y$-axis. The brightness excess in the N-NW is similar to the observed one when the orbit is parallel to the $x$-axis, but then the observed E-W asymmetry is missing. The latter is present when the orbit is parallel to the $y$-axis, but in this case the N-NW excess is much too weak. If we rotate the orbit even further clockwise such that the subcluster finally moves towards the N-NW, the outer large-scale brightness excess flips to the S, which  disagrees with the observation. Hence, the large-scale structure clearly favours an orbit orientation comparable to our fiducial case.

%¤¤¤
\paragraph{Higher age combined with orbit rotated by 180$\degree$:} \label{sec:rotateorbit}
If we had data on only the central $\sim 200\Kpc$, there would be another degree of freedom. We could rotate our simulated maps by $180\degree$ and wait another couple of 100 Myr. Then the major CF, now in the S, would have moved further south to replace the observed outer southern CF.  The secondary, now northern CF would have moved out far enough to mimic the  CF observed in the N, and a tertiary front in about the right position in the S would match the observed inner southern CF. Even though the overall morphology of these fronts would be somewhat more disturbed than observed, at $t=1.8\Gyr$ the secondary and tertiary front would simultaneously be at the correct radii. The major, now outer southern CF would fall 50 kpc short of the radius of the observed position. In this situation, we would not know the full extent of the central brightness excess spiral, nor the large-scale brightness excess, and thus would not be able to unambiguously infer the orbit orientation. However, our knowledge of the brightness distribution outside $200 \Kpc$ clearly argues against this configuration and we can disregard it.

%***********************
\subsection{CF radii and age} \label{sec:radii_age}
%
%
%FFFFFFFFFF
\begin{figure}
\includegraphics[width=0.48\textwidth]{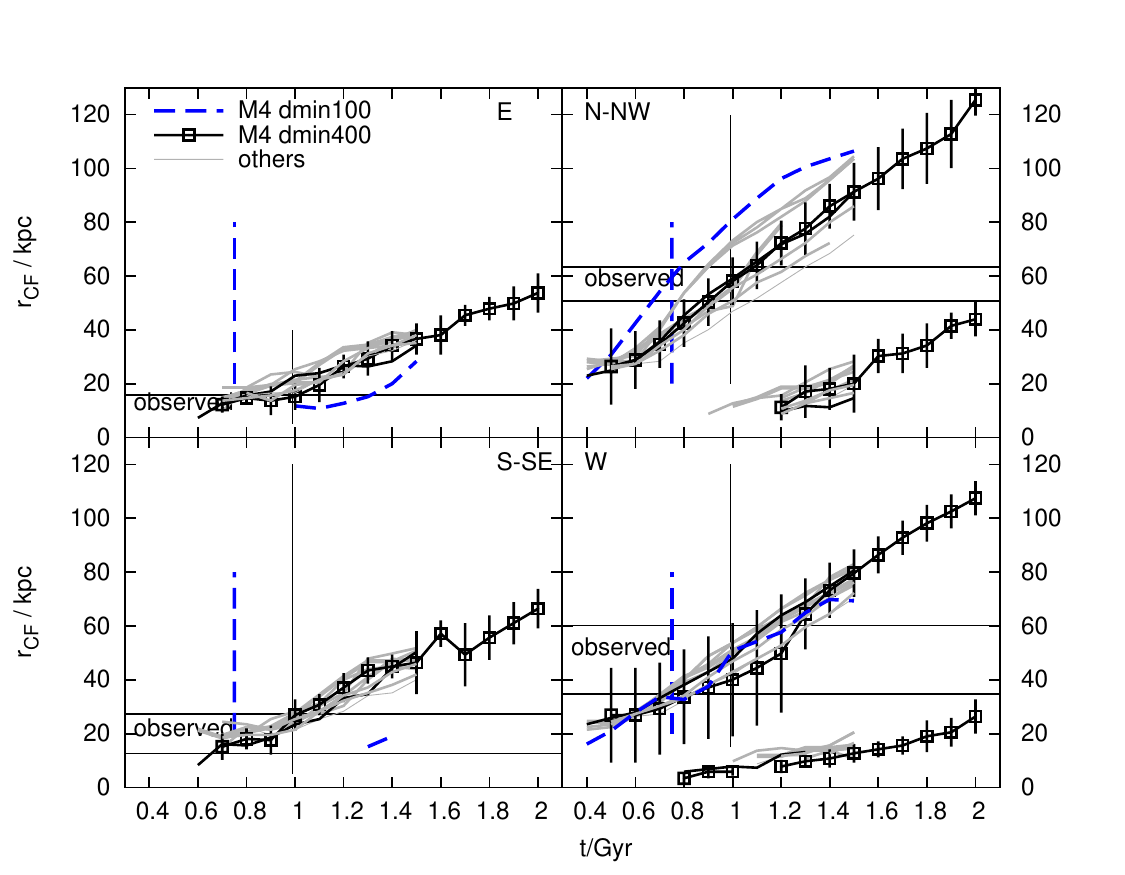}
\caption{Cold front radii as a function of time. Cold front radii are derived from directional profiles in the sectors indicated in the top  of each panel. The thin black horizontal lines mark the radial range of the observed cold fronts. The thin  vertical  line marks the inferred cold front age {for the "distant" fiducial run, the dashed vertical line for the "close" fiducial run.} We show the results of all simulation runs in light grey lines and highlight the fiducial {"distant" run  (black) and the fiducial "close" run (blue dashed).} In the latter case, the fiducial subcluster passes the cluster core at a distance of only 100 kpc and considerably distorts the cool core. This is also the strongest impact case studied. Otherwise, the cold front radii depend mostly on time and only moderately on subcluster and orbit characteristics. The error bars indicate the width of the fronts in the synthetic observations.}
\label{fig:evol_CFradii}
\end{figure}
%FFFFFFFFFF
%
%
With time, all CFs move outwards. This outward motion is mainly governed by the underlying potential of the main cluster and depends only moderately on the properties of the subcluster and its orbit. This is demonstrated in  Fig.~\ref{fig:evol_CFradii}, where we plot the evolution of the CF radii towards different directions for several simulation runs. The same effect was already seen for the Virgo cluster (R11). Thus, the cluster-centric radii of the CFs are a good tracer of their age, i.e.~the time since the subcluster's pericentre passage. At a simulation time of 1 Gyr, our fiducial model {"distant"} \textit{simultaneously} reproduces the CF radii towards all directions within 5 kpc. Given that our model uses a static cluster potential, this is a remarkably good agreement with the observations. The CF ages derived from different simulations vary between  0.9 to 1.1 Gyr, and in each case the simulation simultaneously reproduces the CF radii towards all directions with the accuracy stated above.   The only exception is the case with the strongest impact, the  fiducial {"close" run, where} the  subcluster passes the cluster centre at a distance of only 100 kpc and significantly distorts the cool core. {Here the CFs move outwards somewhat faster, reaching the observed positions in the N, S, and W already at 0.75 Gyr. There is no CF towards the E.} 

 Our rigid potential method overestimates the CF age by about 200 Myr (see \citealt{Roediger2011fastslosh}), hence we conclude that the CFs in A496 are about {0.6 to 0.8} Gyr old. D07 estimated an age of about 0.5 Gyr from a qualitative comparison with the simulations of \citet{Ascasibar2006}, which were aimed at a more massive cluster.

%*************************************************
\subsection{Contrasts across the cold fronts and subcluster mass/size/orbit} \label{sec:mass_jumps}
Generally, the impact of more massive and/or more compact subclusters on orbits with smaller pericentre distances  leads to stronger contrasts in all quantities across the CFs. Using this trend,  R11 were able to constrain the subcluster mass within a factor of 2 for the case of the Virgo cluster. Attempting the same for A496, in Fig.~\ref{fig:A496_profs_ell} we compare directional profiles for X-ray brightness, projected temperature and metallicity for simulations and observations. 

We plot results from three representative simulations: the two fiducial runs and a low-mass case with a subcluster mass of only $10^{13}M\Sun$ and a pericentre distance of 100 kpc, which should leave a milder imprint. {As in the "close" fiducial run the CFs move out faster, we consider a timestep of $0.75\Gyr$ for this case instead of  $1\Gyr$ used in the other two cases. }

The two weaker impact cases achieve an excellent agreement  with the observations within the inner 130 kpc. {For the fiducial "close" run the temperature contrasts are slightly too large.} As discussed in Sect.~\ref{sec:CFradii_obs}, the situation in A496 is  complicated by the fact that the CFs reside still inside the cool core of the cluster, where the CF jumps are in the same directions as  the intrinsic gradients of the cluster, and thus blend with the general cluster profile. Consequently, the simulations show a strong degree of degeneracy. 

Only the regions outside the cool core  towards the S and the E, i.e.~in the region of the {outermost southern CF}, show distinguishable results. Here, subclusters with masses below $2\times 10^{13}M\Sun$ leave a negligible imprint and are incompatible with the observations. The best match in temperature and metallicity across the {outer southern CF} is achieved with the {fiducial "close" run. In the fiducial "distant" case}, the temperature and metallicity gradients are not quite strong enough, and the truth is presumably in between these cases. Thus, our comparison suggests a subcluster mass of 2 to $4\times 10^{13}M\Sun$. {For comparison, the inner 150 to 200 kpc of A496 contain $4\times 10^{13}M\Sun$,  and the cluster mass within 1 Mpc is $2\times 10^{14}M\Sun$.} A more precise constraint of the subcluster and orbit properties requires a full hydrodynamics+N-body treatment in the simulations and tighter observational constraints on temperature profiles and contrasts across the fronts.

%% file: discussion.tex
%**********************
\section{Discussion} \label{sec:discussion}
%**********************
%
%
%***********************
\subsection{Variations in initial cluster model} 
% 
%***********************
\subsubsection{Spherical models for A496} 
Spherical cluster models for A496 reproduce the observed profiles for all quantities inside 150 kpc as well as  the elliptical ones. Our estimate for the CF age marginally increases by 100 Myr at most. Outside 150 kpc, the spherical  simulations slightly over-predict the X-ray brightness in E and W direction because they neglect the intrinsic ellipticity.

\begin{figure}
\includegraphics[width=0.4\textwidth]{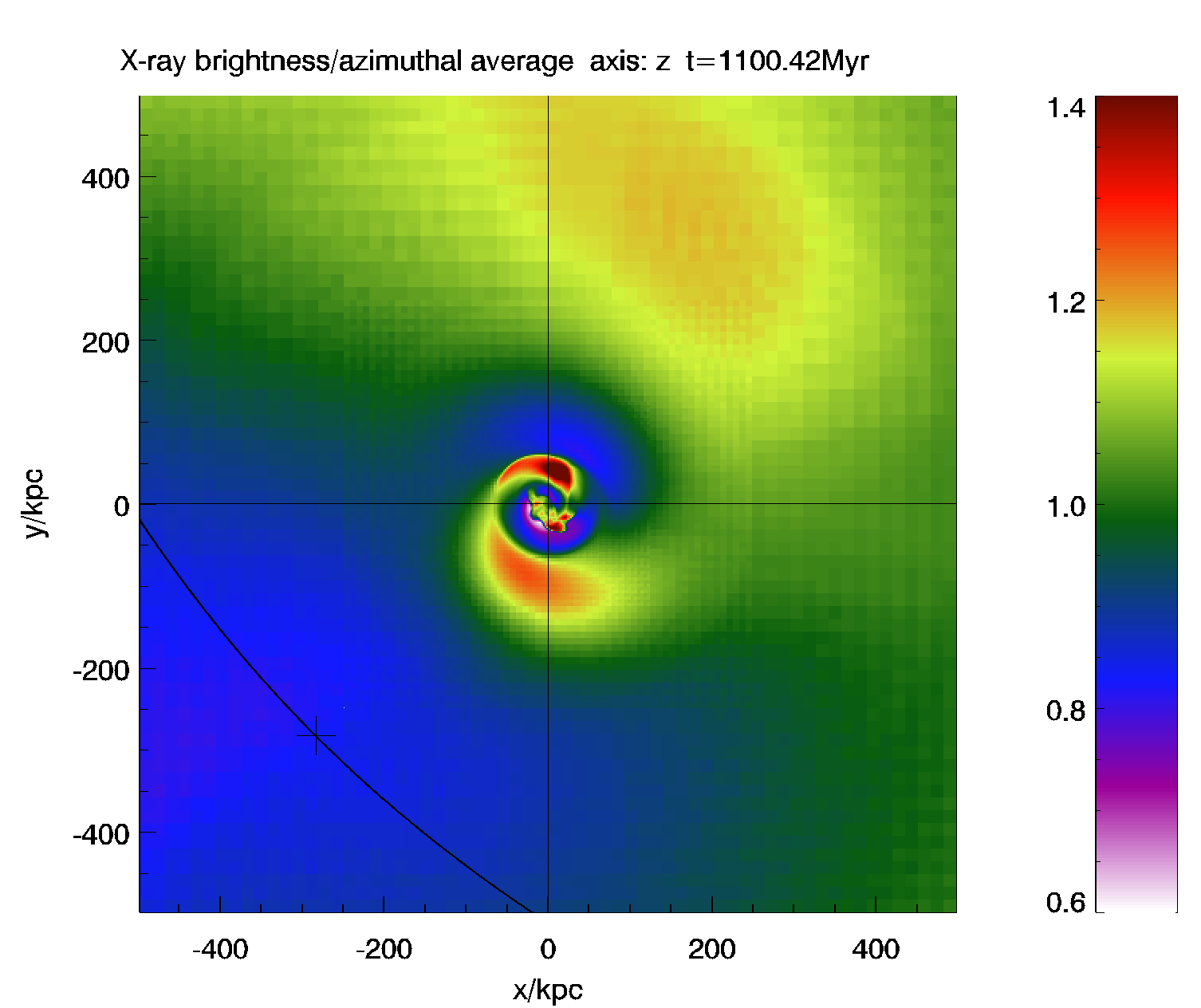}
\caption{X-ray brightness residual map for the initially spherical cluster. The sloshing spiral is more regular. The fiducial subcluster  moved on a diagonal orbit (see black line). }
\label{fig:A496_sph}
\end{figure}

The brightness excess spiral is, however, less elongated compared to the elliptical cluster and much more regular (Fig.~\ref{fig:A496_sph}). With a diagonal SW-NE orbit orientation (see black line in Fig.~\ref{fig:A496_sph}), these models also reproduce the large-scale brightness excess towards the N-NW, but the outer brightness deficit is located towards the SE. This difference to the elliptical models arises because the elliptical structure alone  leads to a characteristic brightness enhancement towards N and S and deficits towards E and W. 
%
%FFFFFFFFFF
%FFFFFFFFFF
%%

Overall, the elliptical cluster model results in a better match.

%***************************************
\subsubsection{Influence of initial temperature profiles} 
From the observations, the temperature profile of A496 is not well-constrained. Hence we tested several possible fits to the observations (see Fig.~\ref{fig:iniprofs}). Our results are  not sensitive to the details of the  temperature profile. The best match is achieved with profile 2, which is also the basis of our elliptical cluster model.

%****************
\subsection{The nature of the outer southern cold front}
{
In our simulations the outer southern CF never is a true discontinuity, while the observations suggest that it is. This is the only aspect the simulations do not reproduce. In the stronger merger simulation, the gradients across it are steeper in all quantities, but it is still continuous. R11 found the same structure in their Virgo sloshing simulations and named it cold fan to distinguish from a proper cold front. We verified that this is not an artefact of the rigid potential approximation employed here (\citealt{Roediger2011fastslosh}). A different orientation between the line-of-sight and the orbital plane would not make the feature appear sharper. }

{
Alternatively, we could alter our fiducial merger geometry and history in order achieve a true outer southern CF, but we would sacrifice the good match we achieve in all other properties:  After about 4.5 Gyr after pericentre passage, the now inner southern CF would have moved out far enough to reach the position of the observed outer front and could replace it. At this stage, the northern CF would have reached a distance of about 270 kpc, where no CF is observed. The same is true for the western CF. An inclination of 77$\degree$ between the orbital plane and the line-of-sight would shorten the apparent distances of the CFs to the observed values, but only along one axis. At this strong inclination, the clear spiral pattern would be absent. The second option is to consider the stage at 1.8 Gyr after pericentre passage and rotate the merger geometry by 180$\degree$, such that the previously outer northern front replaces the outer southern one, and the previously inner southern one replaces the outer northern one. As discussed in Sect.~\ref{sec:rotateorbit}, this configuration does neither match the overall brightness distribution, especially at large scales,  nor all CF radii simultaneously. Thus, our fiducial configuration is the best match. 
}

{
 The single minor merger scenario we suggest  for the history of A496 is a very simple one, and an almost complete match between predicted and observed features can be regarded as a success. Very likely the history of A496 has been more complex,  e.g.~there may be remnant bulk flows from an earlier merger. The interaction of outer southern cold front/fan with this remnant perturbation may transfer the cold fan into a sharp cold front.
}

%***********************
\subsection{Identifying the responsible subcluster}
\citet{Flin2006} analysed the galaxy distribution in A496 out to large radii and  detected a subcluster or galaxy group towards the N-NW  at the outer edge of their field-of-view. They applied their substructure detection method to a large number of clusters, and their general description of the method implies that each cluster is studied out to 1.5 Mpc from the cluster centre. Specifically for A496, they compare their result to the ones of \citet{Durret2000}. This comparison reveals that in the case of A496 the  field-of-view was larger, and that the subcluster in question is about 55 arcmin or 2.1 Mpc from the cluster centre towards the N-NW. We have marked the subcluster position in Fig.~\ref{fig:orbit_fiducial} along with the predicted position of our simulated subcluster, i.e.~at  $1\Gyr$ after pericentre passage. 

Our prediction of the current subcluster position requires two more considerations. First, the rigid potential method slightly over-estimates the CF ages. Secondly, the employed subcluster orbit is a simple test mass orbit, which does not include dynamical friction. While this is accurate enough for the orbit prior to pericentre passage, dynamical friction will slow down the subcluster significantly after pericentre distance (see \citealt{Roediger2011fastslosh}). Both considerations require that we expect the subcluster at a much smaller distance to the cluster centre than naively expected in our first estimate, i.e., around 1 Mpc towards the N-NE. While there is some intrinsic uncertainty in the direction of the orbit due to the ellipticity of the cluster, in Sect.~\ref{sec:orbitorientation} we have excluded orbits which bring the subcluster towards the N-NW. Furthermore, dynamical friction makes it unlikely that a subcluster that passed the cluster centre only about 1 Gyr ago or less can already have reached the observed distance of the Flin-Krywult subcluster. {These reasons argue against  this subcluster being responsible for the sloshing signatures in A496. }

{The only other subcluster detected towards the N is LDCE0308 about 1.3 Mpc to the N-NW. While its distance is favourable, our simulations clearly favour a position towards N-NE instead of N-NW. Furthermore, this subcluster is detected in the 2MASS survey only and probably of low mass. }

This leaves the identity of the subcluster  still an open question, because in the region where we expect the subcluster, there is no obvious galaxy concentration. However, it may be impossible to ever identify the responsible subcluster due to the effect of tidal forces. During the pericentre passage, a subcluster is tidally compressed. A few 100 Myr afterwards, however, it suffers substantial mass loss due to tidal decompression or tidal stripping, and may well be dispersed and not recognisable as a compact structure anymore.  

Extrapolating from the simulations of \citet{Ascasibar2006}, D07 speculated that the disturber should be rather massive with about $8\times 10^{13} M\Sun$. Our simulations show that such a massive disturber very likely causes  too strong an impact and destroys the cool core, but a less massive on of $4\times 10^{13} M\Sun$ is sufficient.

%***********************
\subsection{Origin of the disturbed inner cold fronts} \label{sec:KH}
{
In high-resolution sloshing simulations without viscosity or magnetic fields, the sloshing cold fronts tend to be subject to the Kelvin-Helmholtz instability (KHI) (e.g.~R11, \citealt{ZuHone2011}). The KHI can be suppressed by magnetic fields aligned with the fronts (\citealt{ZuHone2011}), and the fronts also appear smoother in a viscous ICM (\citealt{ZuHone2010}). While in many clusters the CFs indeed appear as smooth arcs free from signs of instabilities, in A496 they  show kink-like deviations from a smooth arc on length scales of about 20 kpc. In our  "close" fiducial run, we reproduce KHI-induced distortions in the inner CFs  that resemble the observed ones, also the double CF towards the W. The length scale of the simulated distortions is even independent of resolution (Fig.~\ref{fig:res_maps2}), indicating that the complex interplay of the curvature of the shearing layer and its outward motion set the instability length scale instead of the numerical viscosity. The \emph{presence} of KHI puts an upper limit on the magnetic field strength in this cluster, because a  mean tangential magnetic field, $B$, at the CFs stronger than (\citealt{Vikhlinin2002})
%--------------
\begin{eqnarray}
\frac{B^2}{\mu_0} &>& 0.5 \gamma \mathrm{Ma}^2 \frac{p\ICM}{(1+T\Cold/T\Hot)} \\
&=& (11\,\mu\mathrm{G})^2 \left(\frac{\mathrm{Ma}}{0.5}\right)^2 
\left(\frac{p\ICM}{2.8\times 10^{-2} \KeV\ccm}\right) \times \nonumber \\
&&\left(\frac{1+T\Cold/T\Hot}{1.5}\right)^{-1} \nonumber\\
&\approx&  0.1  p\ICM \left(\frac{\mathrm{Ma}}{0.5}\right)^2  \left(\frac{1+T\Cold/T\Hot}{1.5}\right)^{-1}
\end{eqnarray}
%========
should suppress the KHI, where $p\ICM$ is the ICM pressure, $\mathrm{Ma}$ is the Mach number of the shear flow, and $T\Hot$ and $T\Cold$ are the temperatures at the warmer and colder side of the discontinuity, respectively.  \citet{ZuHone2011} showed that gas sloshing typically amplifies the magnetic fields at the CFs by up to an order of magnitude, implying  initial field strengths of below a few $\mu$G or magnetic pressure below several hundredths in A496, which is within observational limits for galaxy clusters in general. 
}

{We note that the presence of substructure in CFs is not restricted to A496. Also the CFs in the centres of NGC 7618 and UGC 12491, a pair of merging galaxy groups (\citealt{Kraft2006}), show significant substructure in the form of kinks and wings (Kraft et al., in prep.). The same is true for the leading edges of M89 (NGC 4552, \citealt{Machacek2006a}) and NGC 4472 (M49, \citealt{Kraft2011}), two Virgo ellipticals moving through the Virgo cluster. Thus, the presence of KHI at the CFs in some clusters and their absence in others may provide a sensitive probe for magnetic field strengths.}

{
Both, the presence of the instabilities in the inner CFs and the stronger contrast across the outer southern one favour the fiducial "close" run, but the cleaner match for the inner CF positions and contrasts across them favours the fiducial "distant" one. 
A distinction between both requires better observational constraints on the cluster  potential and including the evolution of the DM components of both clusters in the simulation. Especially in the close encounter case, the central potential of A496 will deform somewhat, which will influence the exact positions of the CFs.
}

%***********************
\subsection{Reconstructing the mass distribution in a  sloshing cluster}
\citet{Markevitch2001} pointed out that the presence of the CF discontinuities in ICM density and temperature profiles can cause flawed results in cluster mass estimates. If the density and temperature profiles across a CF are used to derive the cluster mass profile assuming hydrostatic equilibrium, the resulting mass profile will have an unphysical discontinuity at the CF radius. {\citet{Keshet2010} discussed that the tangential flows associated with sloshing naturally lead to centrifugal accelerations that add to the radial force balance.}

\begin{figure}
\centering\includegraphics[width=0.39\textwidth]{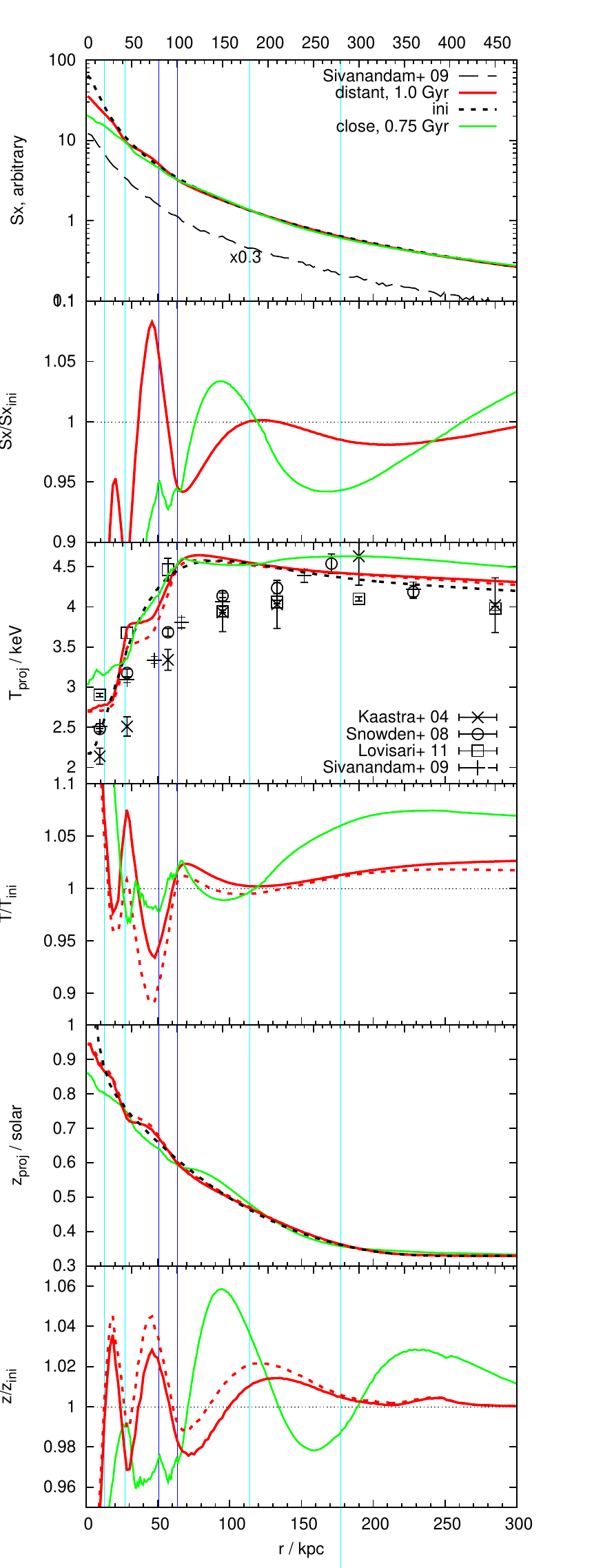}
\caption{
Evolution of fully azimuthally averaged profiles of X-ray brightness (top), projected temperature (third) and metallicity (fifth), in our {two fiducial models (see legend).} The second, fourth and sixth panel show the relative difference to the initial state. The vertical double lines mark the cold fronts towards the N (dark blue) and S (light blue).  For the temperature and metallicity in the fiducial "distant" run, we show the results of two averaging approaches: either, we first derive a synthetic map and average azimuthally over annuli in this map (solid lines). Alternatively, we calculate the integrals in Eqn.~\ref{eqn:mazzotta} not only along the LOS for each pixel, but also over annuli, and perform the division in Eqn.~\ref{eqn:mazzotta} afterwards (dashed lines). This corresponds to first  binning the observed data over annuli into cumulative spectra and deriving the temperature and metallicity from them. The latter method leads to stronger imprints of the cold fronts in the profiles. The observed fully averaged profiles are also plotted here for comparison, but  show no trace of the cold fronts.}
\label{fig:evol_av}
\end{figure}

Here, we investigate to what extent the underlying cluster mass can be inferred  from fully azimuthally averaged profiles. To this end, in Fig.~\ref{fig:evol_av} we plot the evolution of fully azimuthally averaged (i.e.~over $360\degree$) profiles for X-ray brightness, projected temperature and metallicity.
%
%FFFFFFFFFF
%FFFFFFFFFF
% 

Except for the central heating and brightness decrease, which are due to our neglect of radiative cooling, sloshing modifies the azimuthally averaged profiles by less than 10 percent. The {"close" fiducial case, the distortions of the averaged profiles are weaker in the centre, but stronger at large radii. The inner cold fronts leave clearer traces in the "distant" fiducial run.} Here, the modification is systematic: inside each CF,  brightness and metallicity are enhanced and the temperature decreases, while outside each CF the opposite is true. In order to detect this effect observationally, both, the brightness and the temperature profile need to be known with at least 10 kpc resolution and with an accuracy of better than 10 percent.  Given the scatter in the available data, the distortion of the  profiles due to sloshing does not seem to introduce an observable difference and will  not be a major error source for cluster mass profiles. The observed fully azimuthally averaged profiles for A496 show no trace of the CFs. 
 
Fig.~\ref{fig:evol_av} also shows that the accuracy of the averaged temperature profile will depend on its derivation. Dividing the X-ray data into annular rings and fitting the cumulative spectrum of each ring with a single temperature results in a stronger imprint of the CF in the profile (dotted coloured lines). Instead, profiles could be derived in different directions and averaged azimuthally afterwards (solid coloured lines), which leads to a smaller  deviation. Thus, for the purpose of mass reconstruction, the latter approach should be taken, while the former one will make CFs easier to detect already in averaged profiles.

%***********************
\subsection{Metal redistribution}
In the bottom two panels of Fig.~\ref{fig:evol_av}, we demonstrate that the metal redistribution introduced by gas sloshing is mainly an oscillation around the original metallicity profile. Hence, sloshing does not truly broaden the metallicity distribution. The same features were found in the Virgo cluster (R11).

%% file: appendix.tex
%***********************
\section{Comparison of observational datasets} \label{sec:compare_datasets}
%***********************
%
%TTTTTTTTTTTTT
\begin{table*}
\caption{Observational datasets for A496 used in this work.}
\begin{center}
\begin{tabular}{|l|l|l|l|}
\hline
Reference & instrument & observing date & net exposure time  in ks\\
&&&(for XMM-Newton:\\
&&&(MOS1+MOS2+pn)\\
\hline
\citet{Dupke2003} & \textsl{Chandra}, ACIS-S3 &  10/2001 &   8.7  \\
\citet{Dupke2007} & \textsl{Chandra}, ACIS-S3 & 07/2004 & 60  \\
\citet{Tanaka2006} & \textsl{XMM-Newton} & 02/2001 & 14.3 + 14.0 + 7.4 \\
%\citet{Ghizzardi2010} & \textsl{XMM-Newton}  &  &  \\
\citet{Lovisari2011}, \citet{Ghizzardi2010} & \textsl{XMM-Newton} & 08/2007, 02/2008  & (59+61) + (59+60) + (42+46) \\
%\citet{Ghizzardi2010} &   &  &  \\
\hline
\end{tabular}
\end{center}
\label{tab:datasets}
\end{table*}
%TTTTTTTTTTT

%
%FFFFFFFFFF
\begin{figure*}
\includegraphics[width=0.49\textwidth]{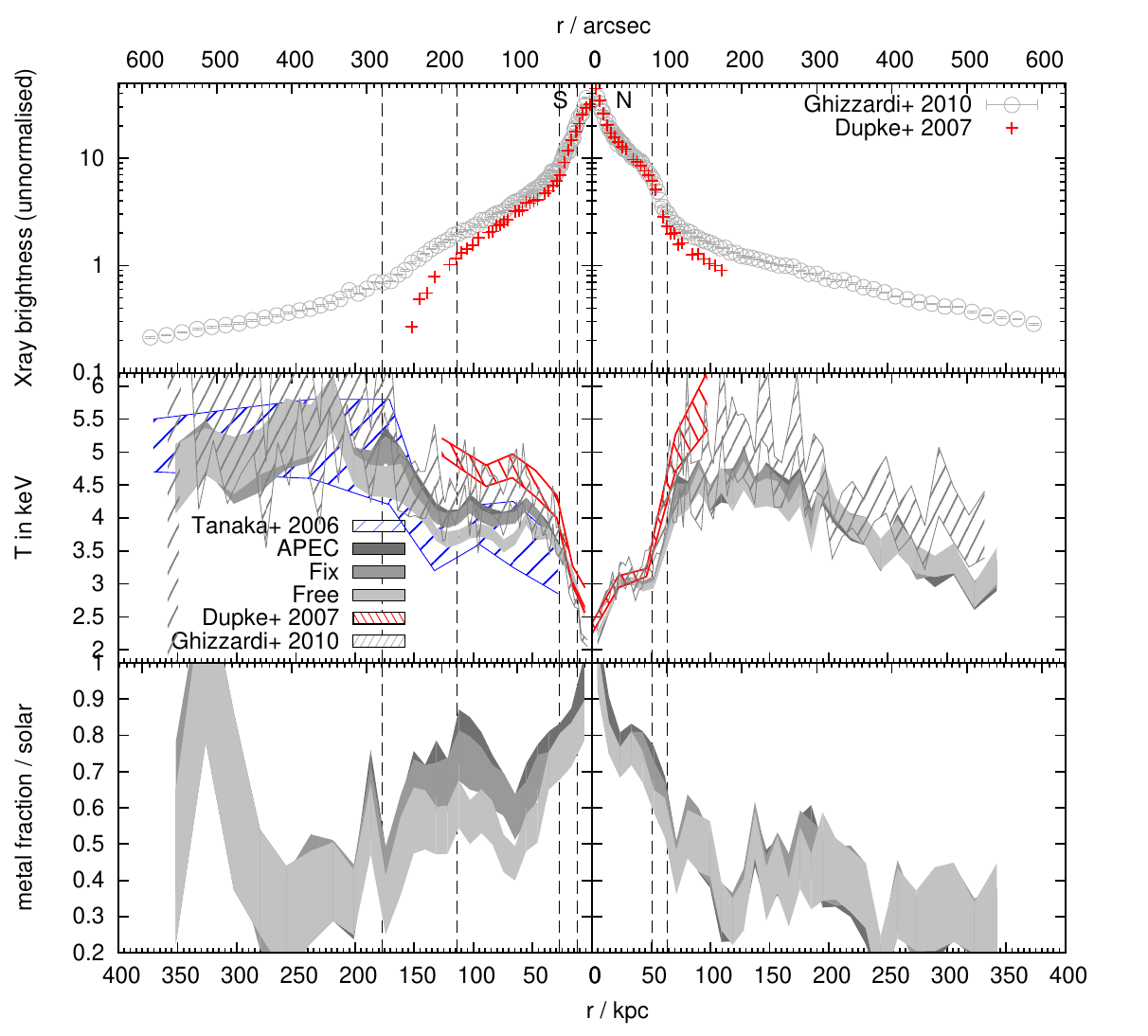}
\includegraphics[width=0.49\textwidth]{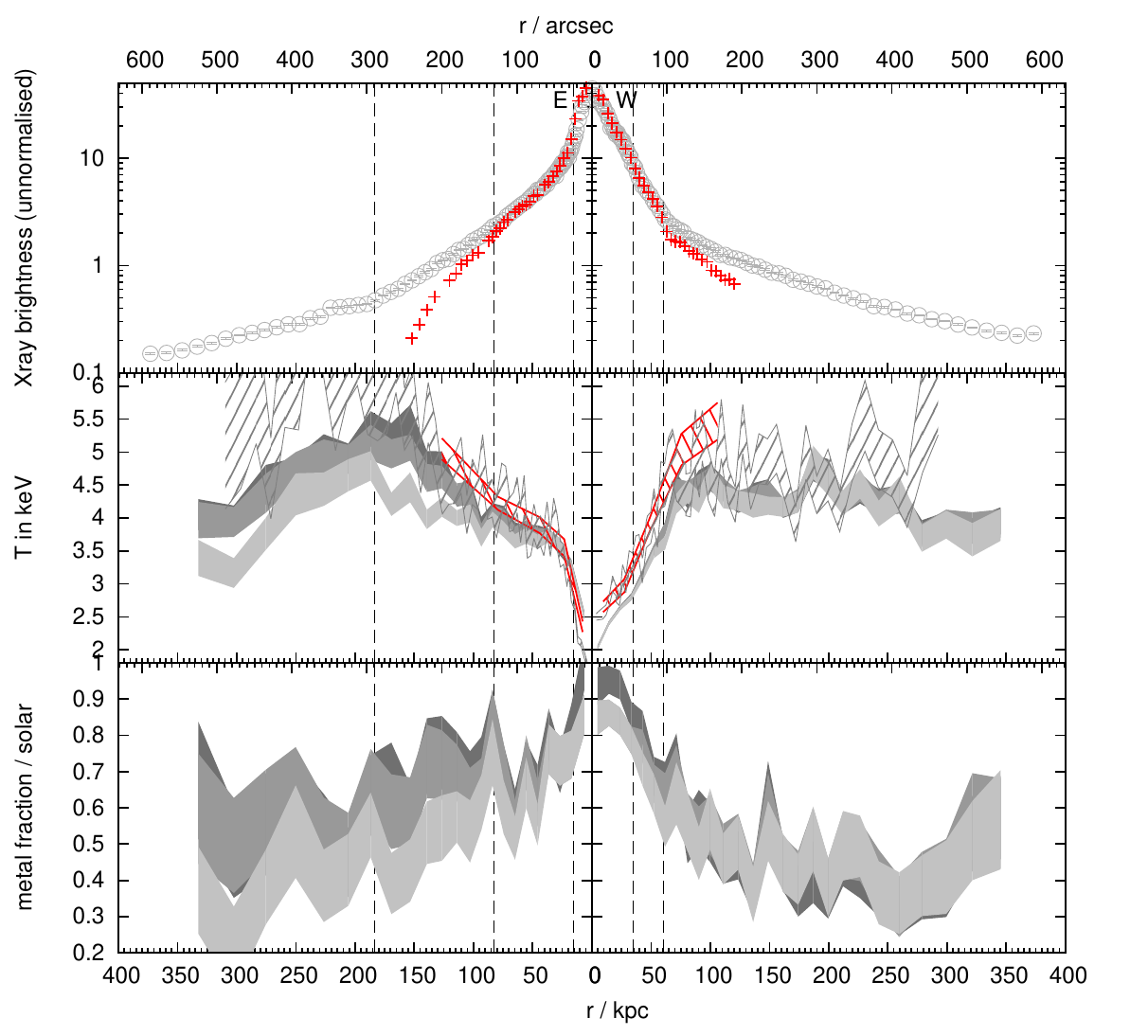}
\caption{Comparison of observed  profiles of X-ray brightness (top panels), projected temperature (middle panels) and metallicity (bottom panels) towards N and S (left figure), and E and W (right figure), for different datasets, see legend and Appendix~\ref{sec:compare_datasets}. For the orientation and range of the sectors, see Fig.~\ref{fig:sectors}. The dashed vertical double lines mark the position of the cold fronts  as stated in Table~\ref{tab:CFradii}.}
\label{fig:A496_profs_observed}
\end{figure*}
%FFFFFFFFFF
%

Detailed observations of A496 have been presented by several authors. Here, we use the datasets listed in Table~\ref{tab:datasets}. In Fig.~\ref{fig:A496_profs_observed} we compare directional profiles for X-ray brightness, projected temperature and metallicity. From the XMM archival data used by G10 and L11, we extract spectra from concentric annuli to determine the projected temperature, metallicity and iron abundance profiles. First we fit the data with an absorbed APEC model in the 0.4-10 keV band with the hydrogen column density, $n_H$, fixed at the galactic value to obtain temperature and metallicity (labelled "APEC"). Subsequently, the iron abundance has been determined using a VAPEC model in the same energy band with $n_H$ first fixed (labelled "Fix") at the galactic value and then left free to vary (labelled "Free"). During the fitting procedure the abundances of O, Mg, Si, S, Ar, Ni were left  free to vary. The ill-constrained  hydrogen column density towards A496  results in the systematic scatter in temperature and metallicity (or iron abundance) between the different assumptions and also the difference to the G10 profiles.  

The X-ray brightness profiles from the D07 Chandra data fall significantly below the ones from the G10 XMM-Newton data outside $\sim 70\Kpc$, which is due to different methods of background subtraction. In the Chandra observation, the whole field-of-view contained cluster emission, leading to an over-subtraction of the background. Therefore, for comparisons to simulations, we  use the X-ray brightness profiles from G10. The simulated temperature and metallicity profiles will be compared to the ones of G10 and our three recalculations, where we regard the range of observational results collected here as representative for their systematic error.

%iiiiiiiiiiiiiiiii
\input{method}

%********************
\section{Other lines-of-sight} \label{sec:LOS}
%********************
%
%FFFFFFFFFFFFFFF
\begin{figure}
\includegraphics[trim=30 -25 275 30,clip,height=4.2cm]{Figs/A496_ell2/M4A100_I400MAX3_45_HR/proj_Excess_z_size2_0200}
\includegraphics[trim=1300 -25 100 30,clip,height=4.2cm]{Figs/A496_ell2/M4A100_I400MAX3_45_HR/proj_Excess_z_size2_0200}
\includegraphics[trim=30 0 215 115,clip,angle=90,height=4.2cm]{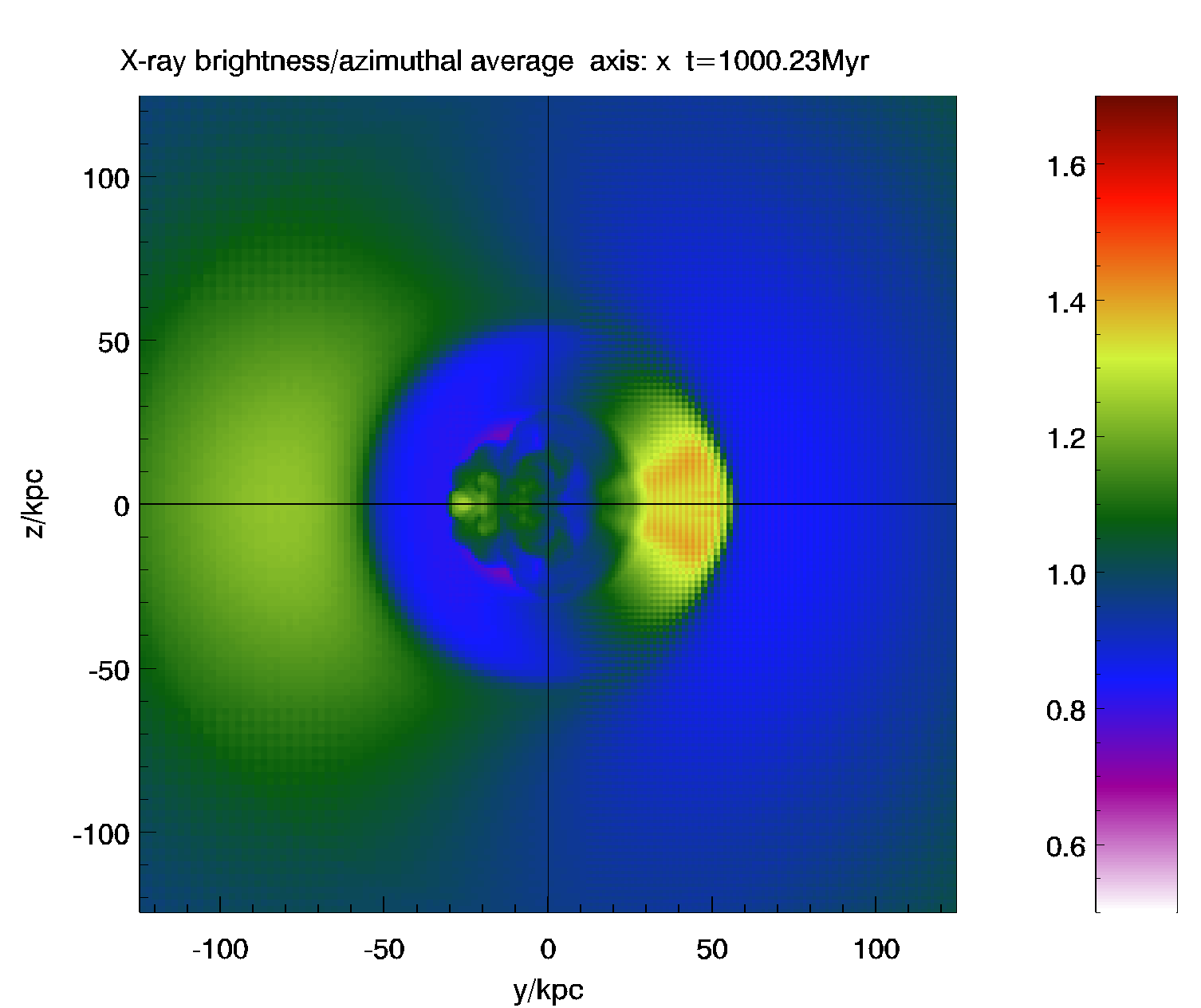}
\newline
\includegraphics[trim=30 -25 275 115,clip,angle=0,height=4.cm]{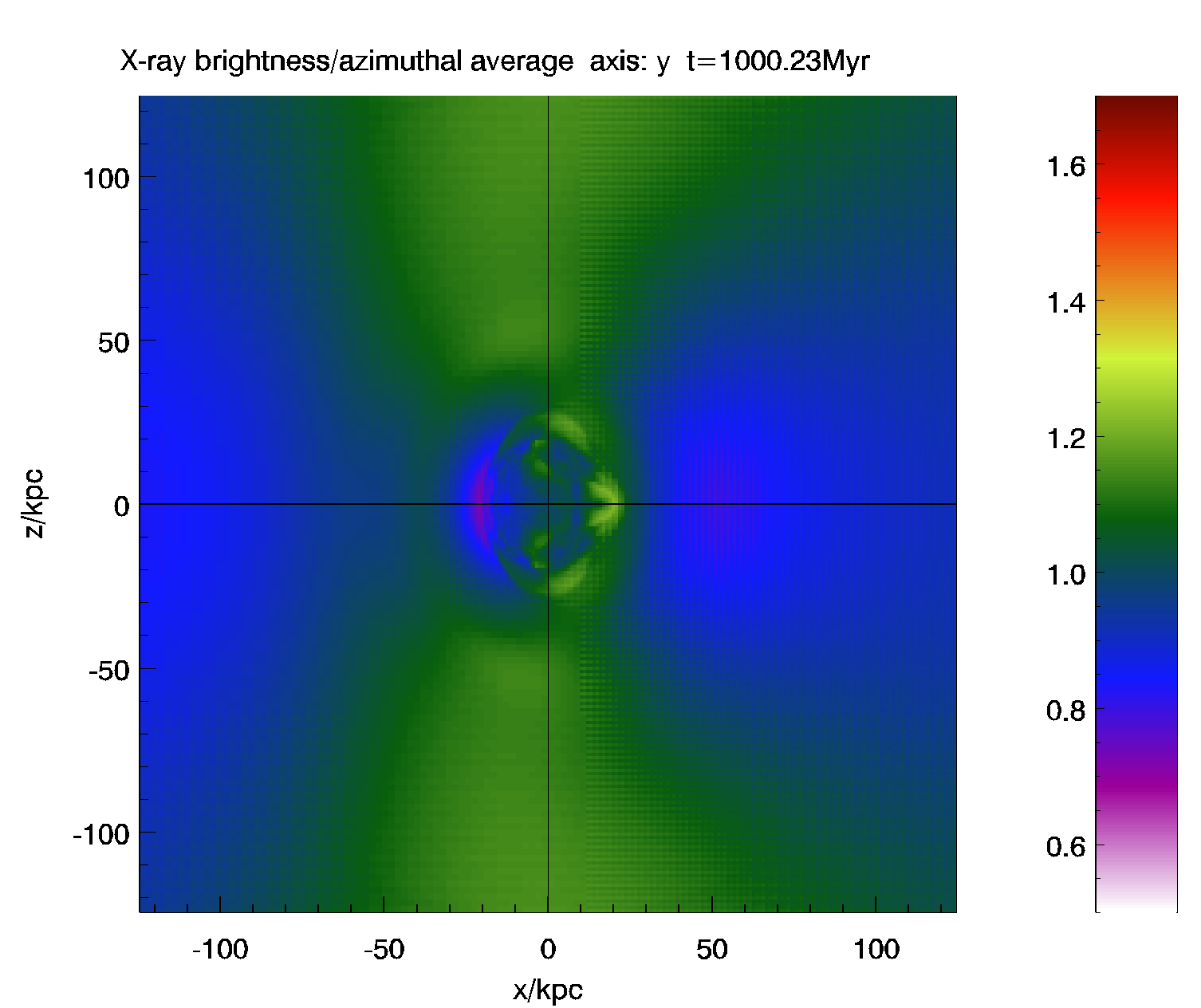}
\hspace{-0.3cm}
\includegraphics[trim=100 0 80 0,clip,height=4.2cm,origin=t,angle=15]{Figs/residuals_smooth8_ncut150num_xoff0_yoff0}
\caption{Brightness residual maps along different LOSs: synthetic images for fiducial run at $t=1\Gyr$ along three grid axes. The elliptical cluster is elongated along the $y$-axis, the subcluster orbit is in the $xy$-plane, approximately from SW over SE to N-NE. Top left: along the $z$-axis, which is perpendicular to the orbital plane. Bottom left: along the $y$-axis. Top right: along the $x$-axis.  Bottom right: residual map derived from the Chandra image of D07.}
\label{fig:fiducial-LOSs}
\end{figure}
%FFFFFFFFFFF

In Fig.~\ref{fig:fiducial-LOSs} we show brightness residual maps along different LOSs.  The CFs can be detected in all cases. The spiral-like arrangement of brightness residuals and CFs is only seen along the LOS perpendicular to the subcluster orbit. For LOSs parallel to the orbital plane, CFs and brightness residuals take the appearance of staggered arcs. Hence, we conclude that in A496 we are looking approximately perpendicular onto the orbital plane of the subcluster.

%\clearpage

%********************
\section{Resolution test} \label{sec:resolution}
%********************

%FFFFFFFFFF
\begin{figure*}
\includegraphics[trim=420 310 560 400,clip,height=4.5cm]{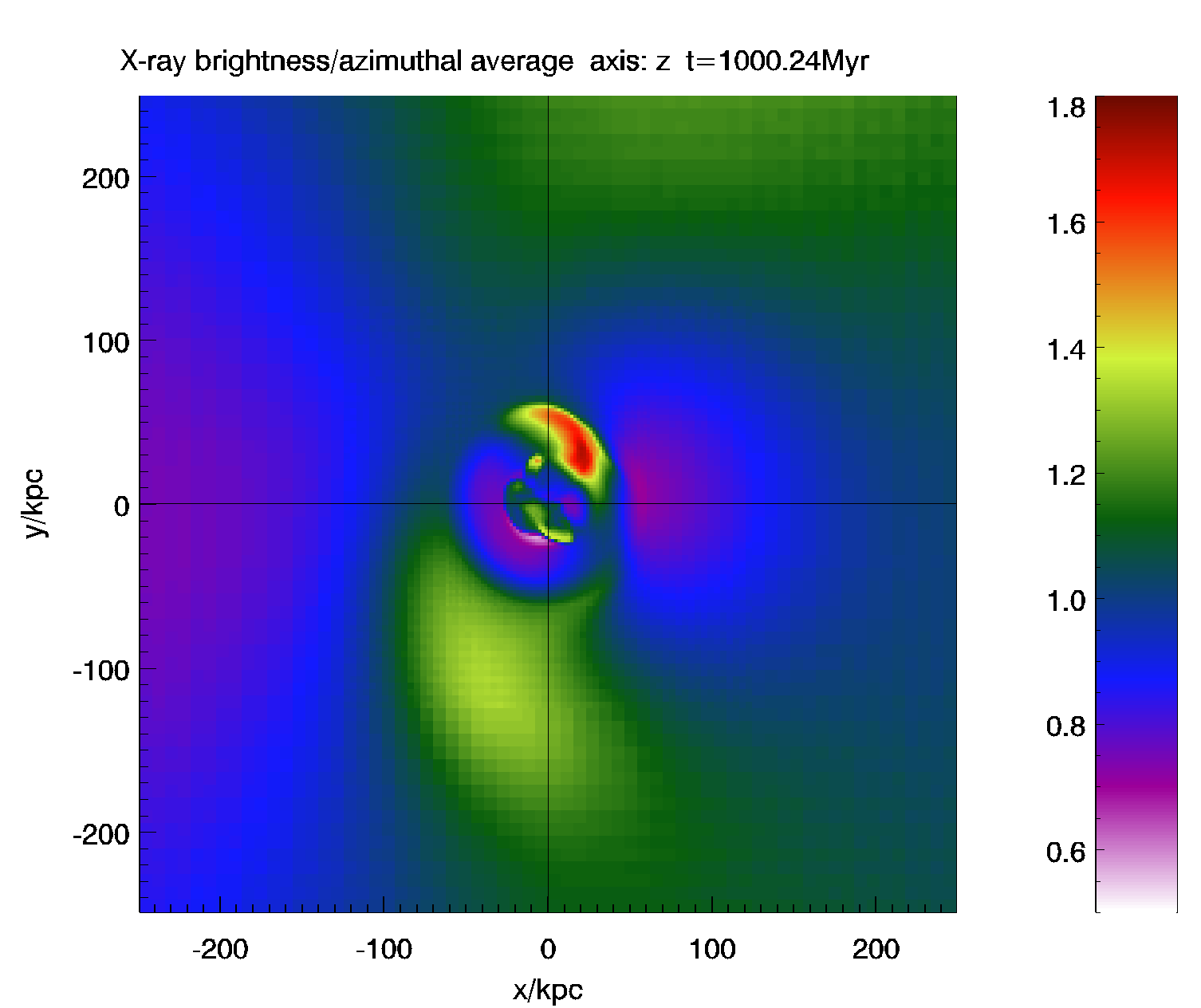}
\includegraphics[trim=0 0 200 200,clip,height=4.5cm]{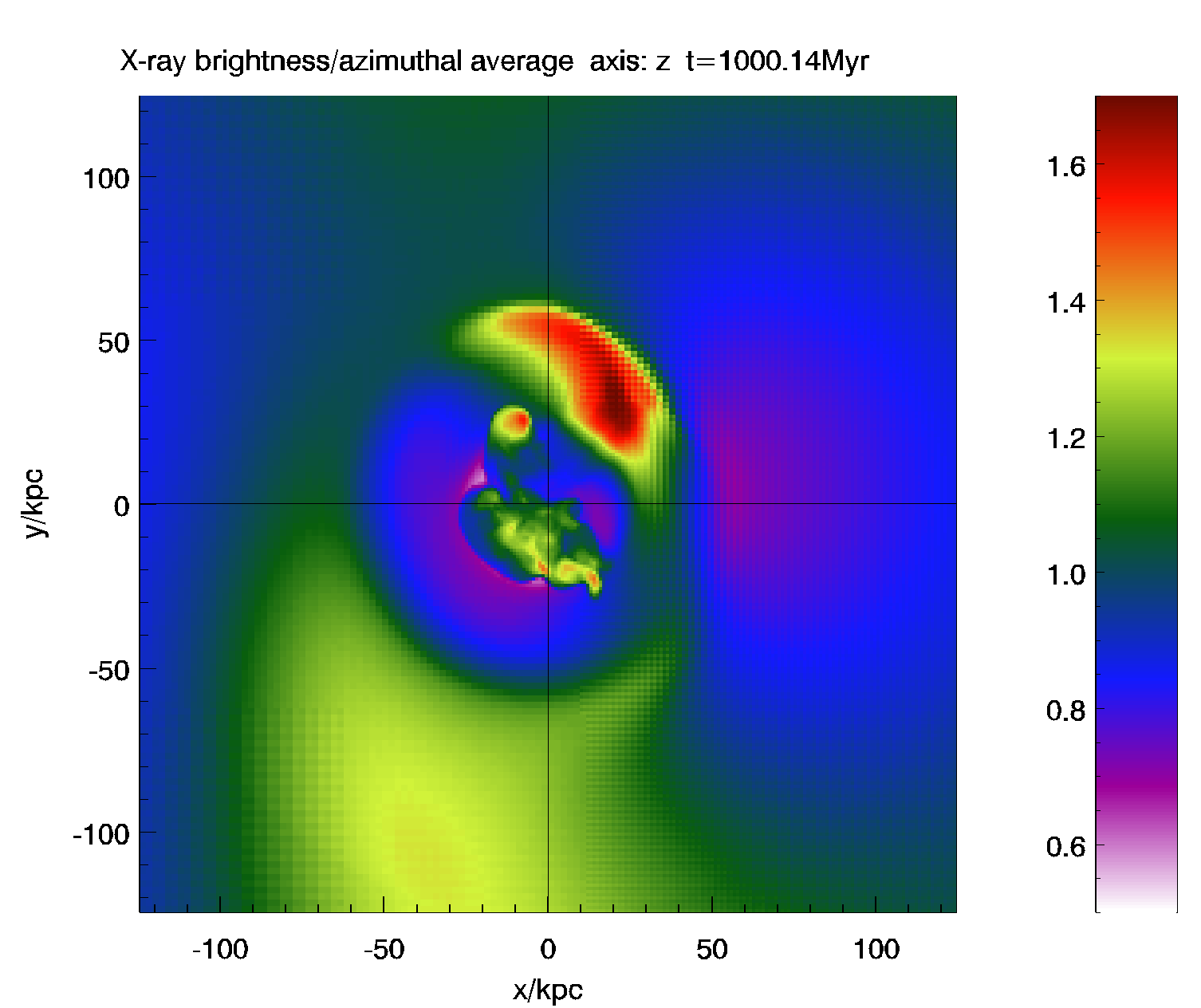}
\includegraphics[trim=0 0 0 200,clip,height=4.5cm]{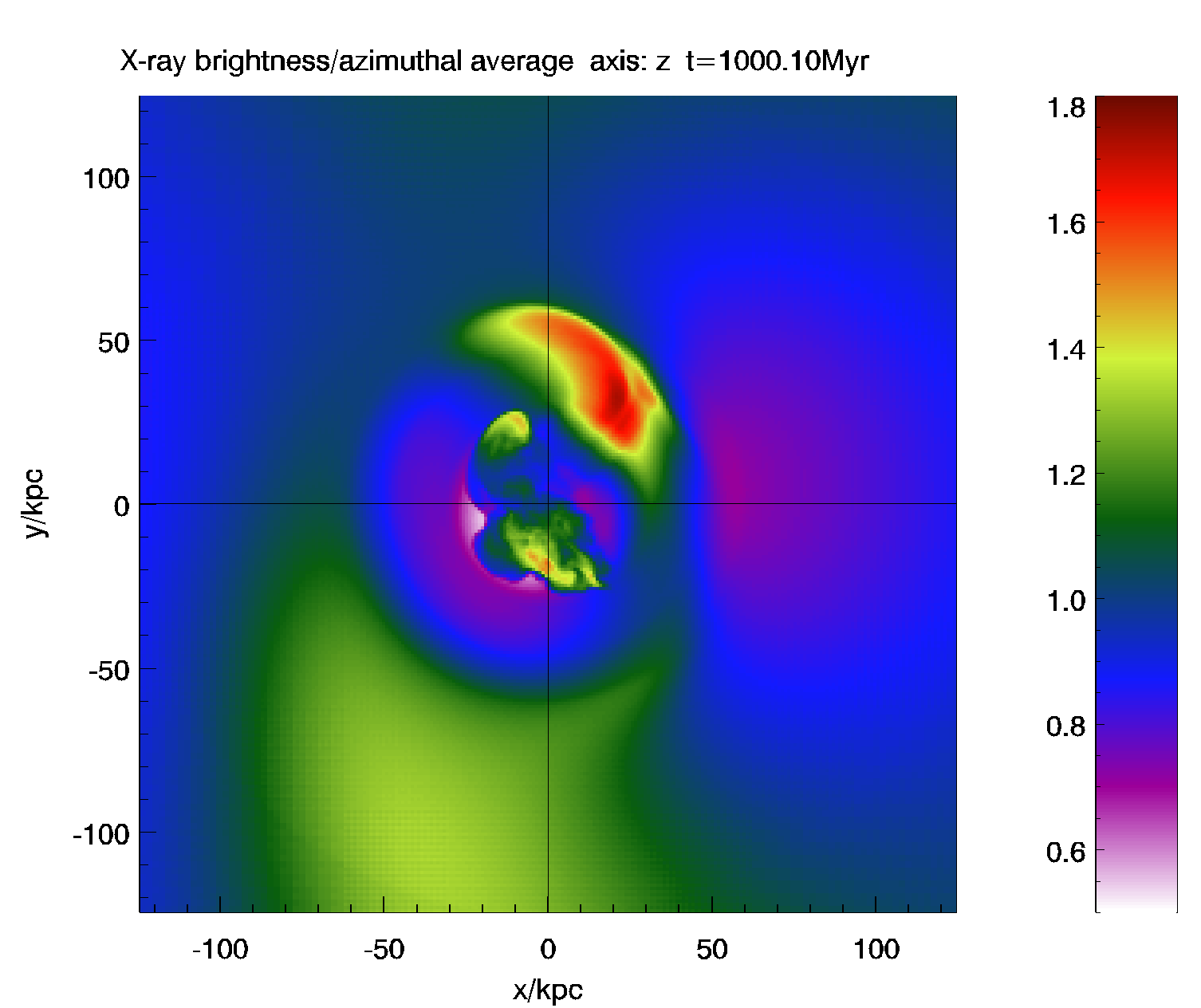}
\caption{Resolution test: brightness residual maps for our fiducial subcluster on orbit parallel to $y$-axis  at $t=1\Gyr$. The resolution increases by a factor of 2 from left to right between neighbouring images.}
\label{fig:res_maps}
\end{figure*}
%FFFFFFFFFF

%FFFFFFFFFF
\begin{figure}
\includegraphics[width=0.45\textwidth]{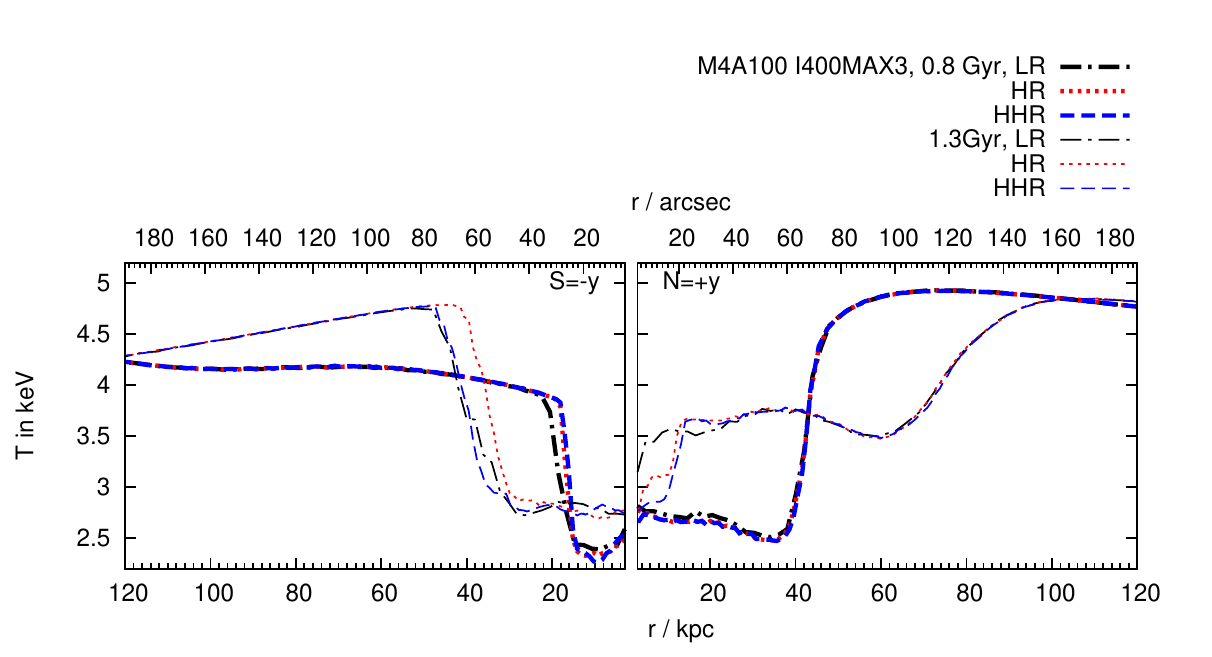}
\includegraphics[width=0.45\textwidth]{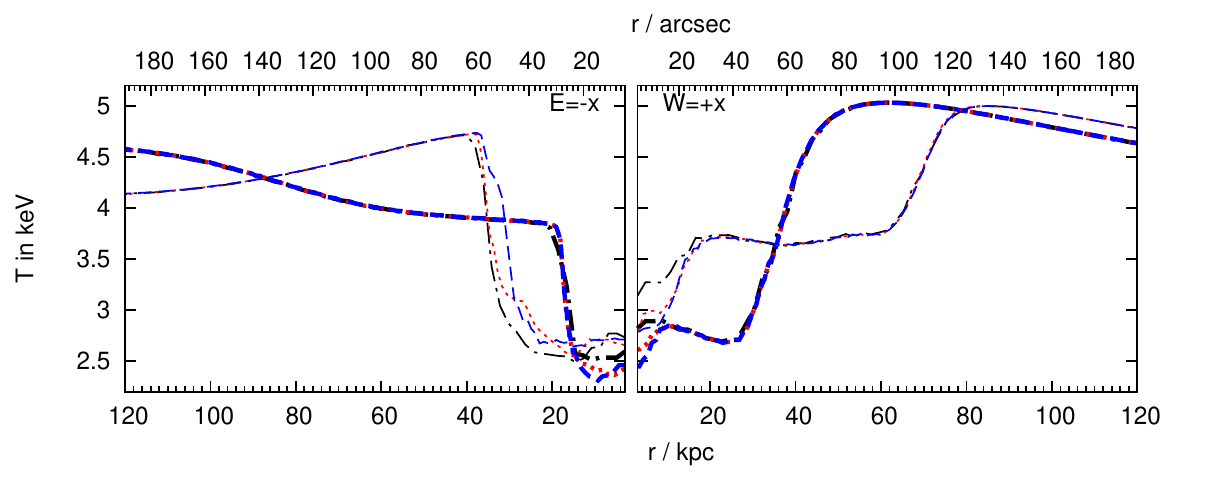}
\caption{Resolution test for same case as in Fig.~\ref{fig:res_maps}: projected directional temperature profiles  for three different resolutions at 0.8 Gyr and 1.3 Gyr after the pericentre passage. The resolution increases by a factor 2 from LR to  HR and again to HHR.}
\label{fig:res_profiles}
\end{figure}
%FFFFFFFFFF

\begin{figure*}
\includegraphics[trim=420 310 560 400,clip,height=4.5cm]{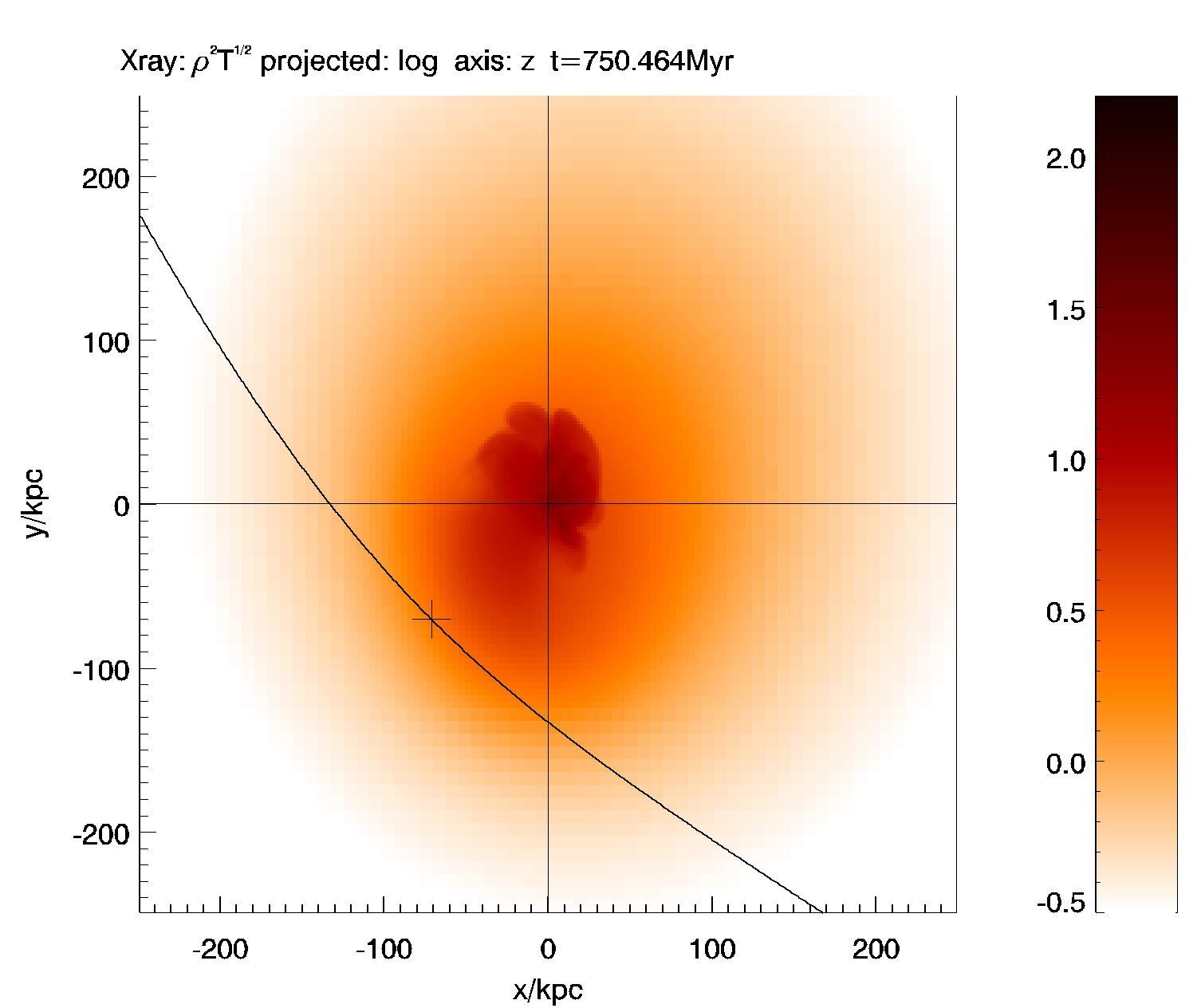}
\includegraphics[trim=0 0 200 200,clip,height=4.5cm]{Figs/A496_ell2/M4A100_I100MAX3_45_HR/proj_Xray_z_size2_0175.png}
\includegraphics[trim=0 0 0 200,clip,height=4.5cm]{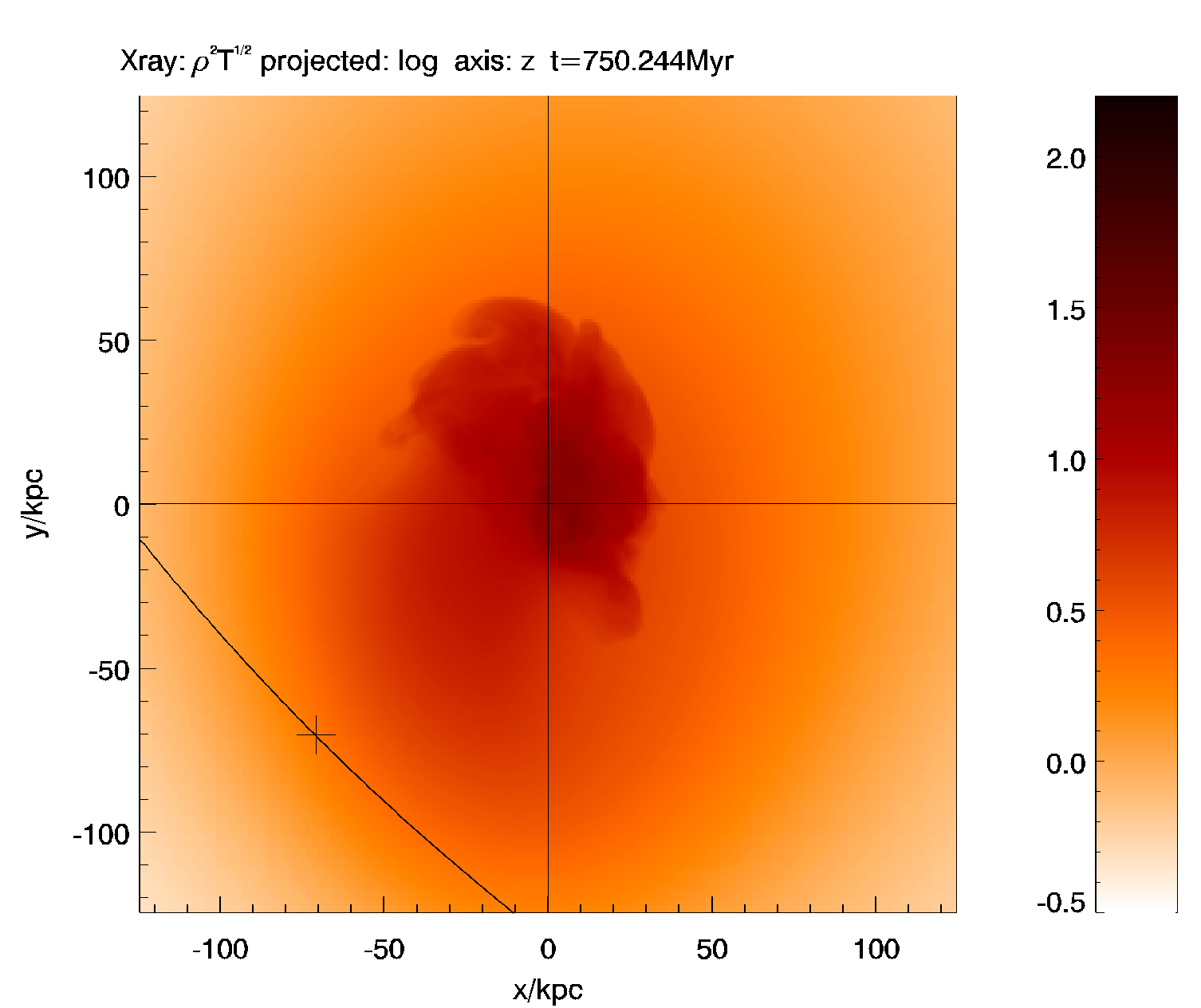}
\includegraphics[trim=420 310 560 400,clip,height=4.5cm]{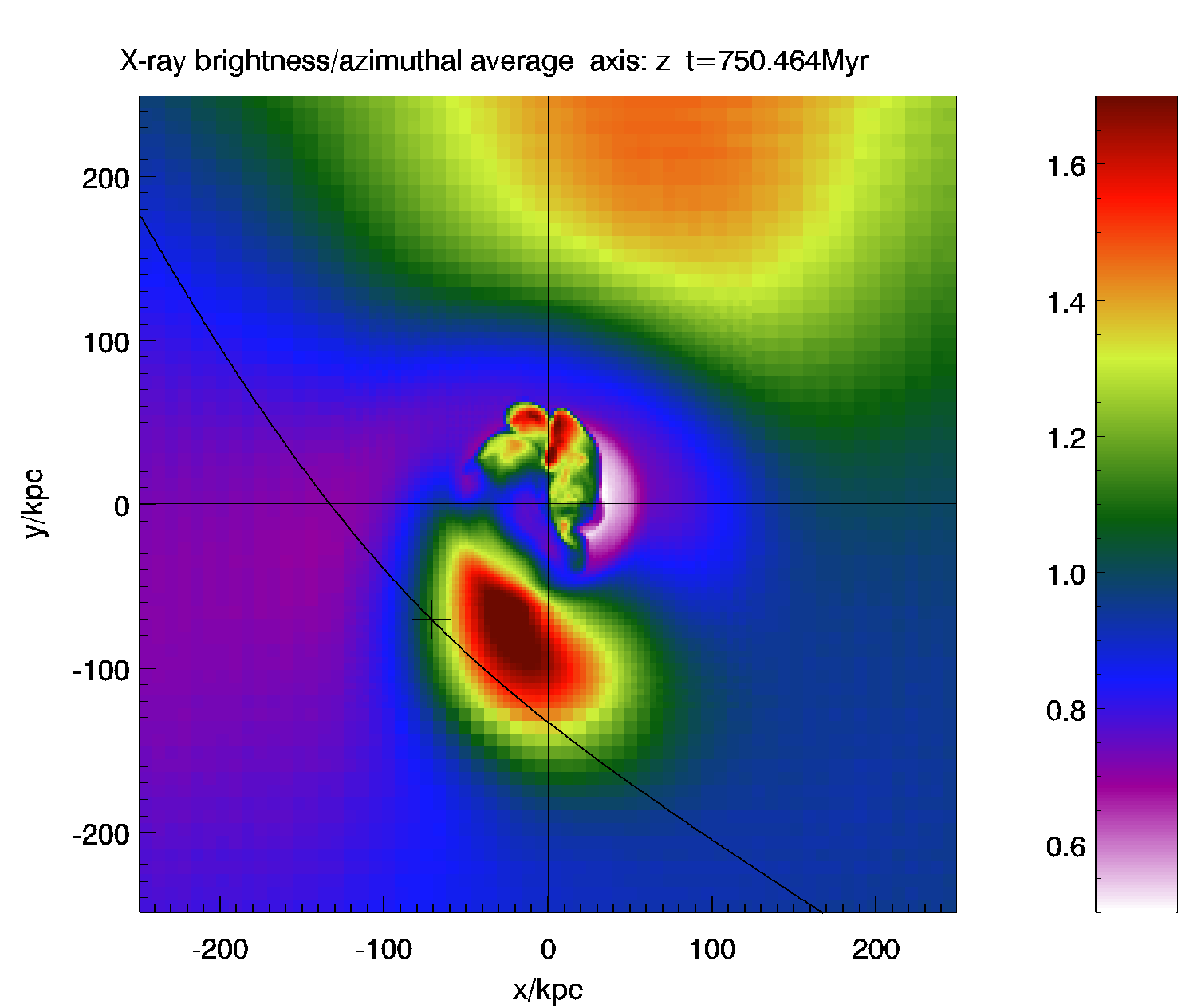}
\includegraphics[trim=0 0 200 200,clip,height=4.5cm]{Figs/A496_ell2/M4A100_I100MAX3_45_HR/proj_Excess_z_size2_0175.png}
\includegraphics[trim=0 0 0 200,clip,height=4.5cm]{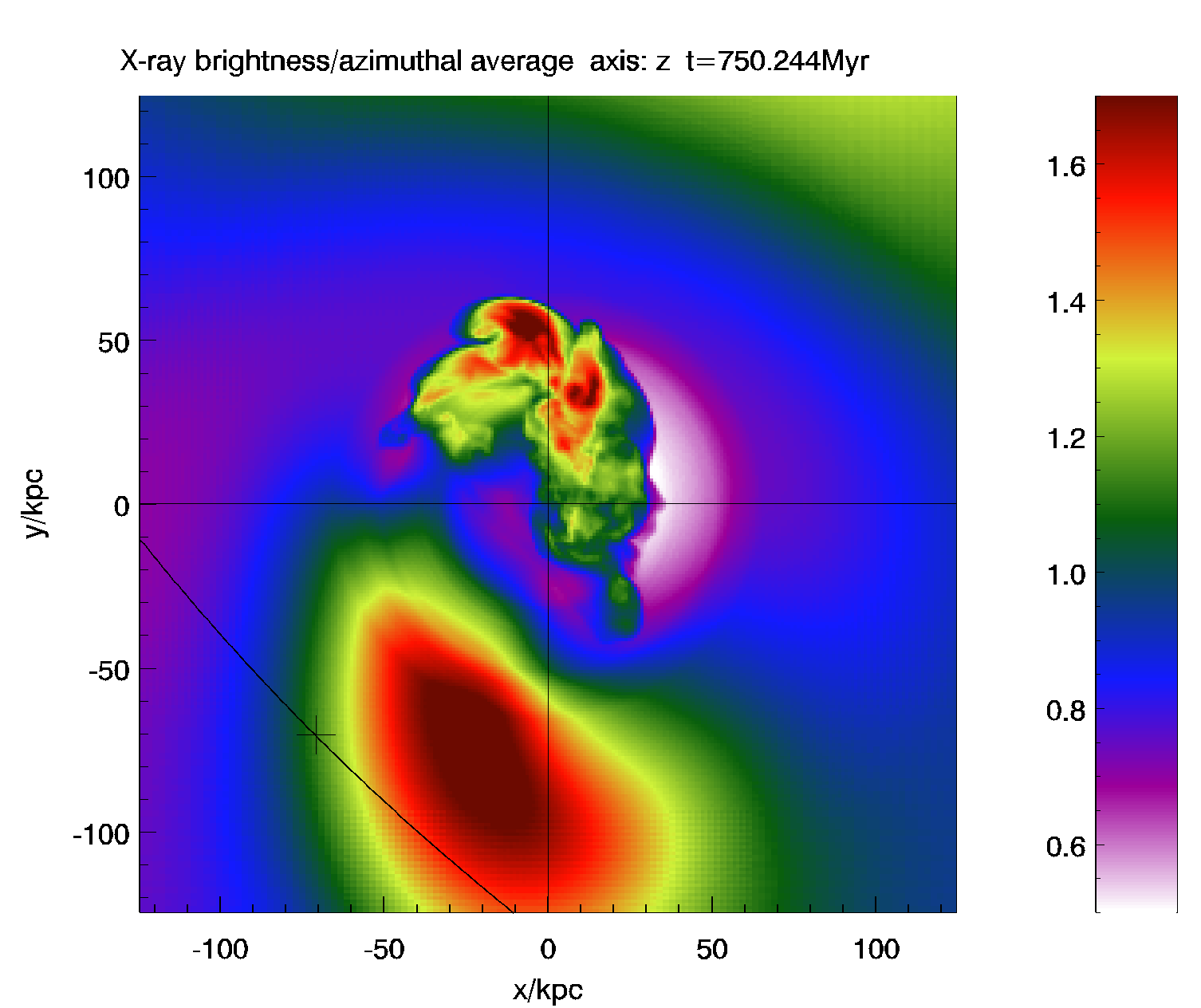}
\caption{
{
Resolution test: brightness residual maps for our fiducial run "close" with pericentre passage at 100 kpc. The resolution increases by a factor of 2 from left to right between neighbouring images. In the rhs image, the northern CF is resolved with 1 kpc. The distortions of the inner CFs are largely independent of resolution.
}
}
\label{fig:res_maps2}
\end{figure*}

We have performed resolution tests for several runs by varying the resolution by a factor of 4. In Fig.~\ref{fig:res_maps} we show examples of  brightness residual maps for our the run with the fiducial subcluster on an orbit parallel to the $y$-axis {with pericentre distance 400 kpc},  for three different resolutions.
In Fig.~\ref{fig:res_profiles}, we show projected  temperature profiles for the same case at two different epochs. Only at late times and only in the inner 30 kpc, these profiles differ slightly for the low-resolution case. 
{In Fig.~\ref{fig:res_maps2} we show X-ray and brightness residual maps for the fiducial "close" run. The distortions of the inner CFs are largely independent of resolution.}

\newpage

%FFFFFFFFFF
%FFFFFFFFFF

%% file: method.tex
%*************
\section{Simulation method and initial model} \label{sec:method}
We follow the approach of R11  and perform  idealised simulations of a minor merger between a main cluster, here A496, and a smaller, gas-free subcluster. The ICM gas physics are described by the hydrodynamical equations. Additionally, the ICM is subject to the gravitational forces due to the main cluster and the subcluster. These are modelled by the adapted rigid potential approximation described  and verified by \citet{Roediger2011fastslosh}. Instead of modelling the DM distributions  by the N-body method, we assume static potentials for both clusters, which speeds up the simulations considerably. The simulations are run in the rest frame of the main cluster. As this is not an inertial frame, the resulting inertial accelerations are taken into account. We concentrate on minor mergers, where the subcluster is significantly less massive than the main cluster. A major merger would  cause additional structure, e.g., destroy the cool core, which is not the case for A496. In A496, only in the inner $\sim 15\Kpc$  the cooling time is shorter than 1 Gyr. Hence, in the outer regions of interest for our analysis, the cooling time is long and we neglect radiative cooling in our simulations. Finally, we  only model the first core passage of the subcluster and the subsequent gas sloshing.

%*************
\subsection{Code}

Our simulations use the FLASH code (version 3.2, \citealt{Dubey2009}).  FLASH is a modular block-structured AMR code, parallelised using the Message Passing Interface (MPI) library. It solves the Riemann problem on a Cartesian grid using the Piecewise-Parabolic Method (PPM). The simulations are performed in 3D and all boundaries are reflecting. We use a large simulation grid of size $4\times 3\times 3\Mpc^3$ to ensure that no reflected waves reach the cluster centre during our simulation. In most simulations, we resolve the inner 32 kpc with $\Delta x=2\Kpc$ and the inner 128 kpc with $\Delta x=4\Kpc$. The fiducial run uses a twice as good resolution.  We performed resolution tests for several cases by rerunning with a twice and a four times as good resolution. In Appendix~\ref{sec:resolution} we show that none of our results depend on resolution.

%*******************
\subsection{Model for A496}

The main cluster A496 is modelled either as a spherical or triaxial ellipsoidal cluster. 

%**********
\subsubsection{Spherical model}\label{sec:maincluster}

Given the ICM density and temperature profile for A496 and assuming  hydrostatic equilibrium, we calculate the gravitational acceleration due to the underlying main cluster potential as a function of radius. When the cluster is allowed to evolve for 2 Gyr in isolation, no sloshing or other modification of the ICM distribution is observed.

The ICM density and temperature profile are fitted to observational data, which is summarised in  Figure~\ref{fig:iniprofs}:
%
%FFFFFFFFFF
\begin{figure}
\centering\includegraphics[width=0.49\textwidth]{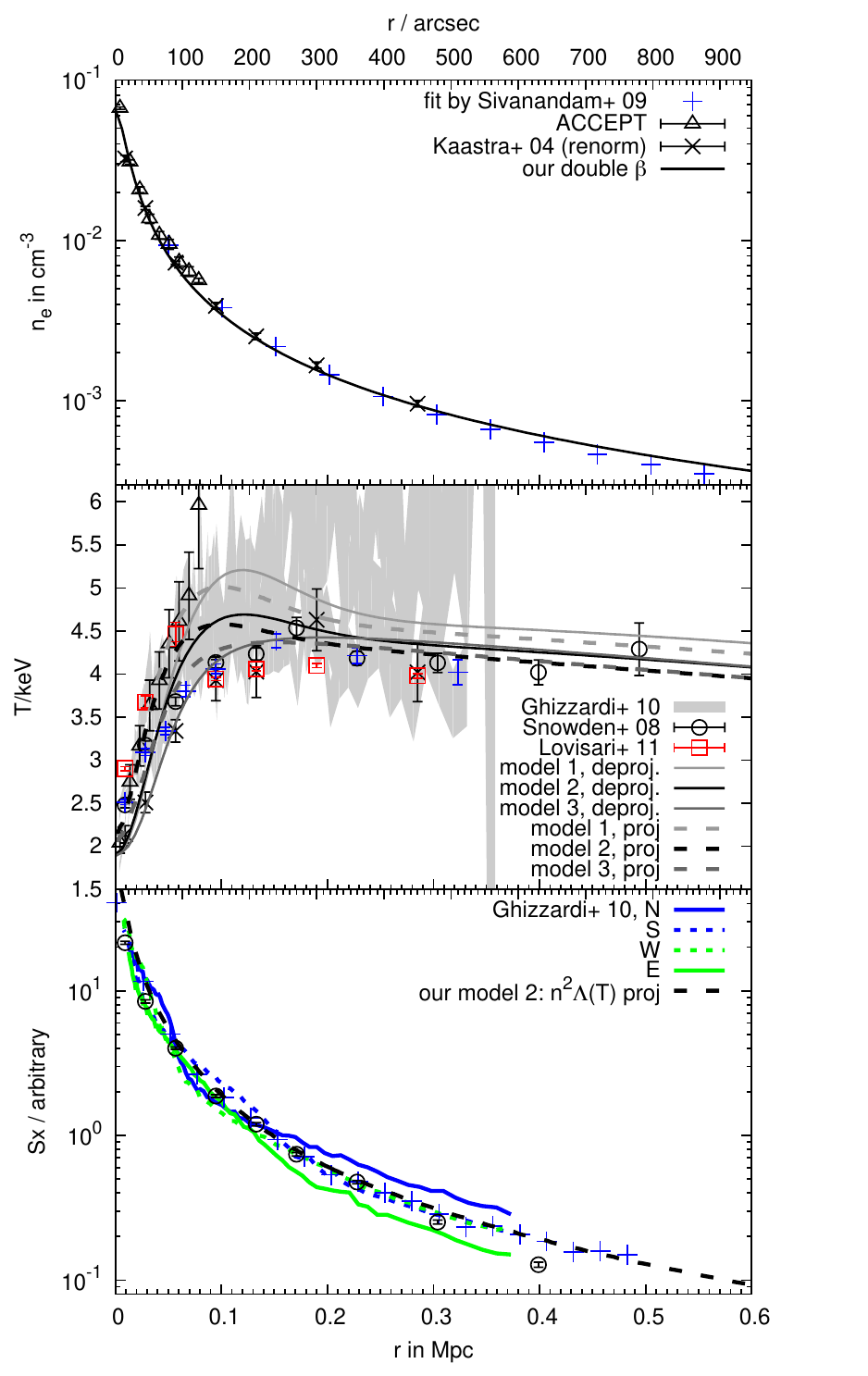}
\caption{Initial profiles of our model cluster and comparison to observations. The panels from top to  bottom are electron density, temperature, are X-ray surface brightness. Black and grey solid lines show our deprojected model profiles, while dashed black and grey lines show projected quantities from our model. Coloured lines, shaded areas and data points are observational data from several works, see legends. Lines/symbols appearing in several panels are explained in the panel where they appear first.   For details see Sect.~\ref{sec:maincluster}.}
\label{fig:iniprofs}
\end{figure}
%FFFFFFFFFF
%
the first panel shows the radial electron density profile, $n_e(r)$, from the ACCEPT sample (\citealt{Cavagnolo2009}), from \citet{Kaastra2004} (converted from proton density), and the single-$\beta$ fit given by \citet{Sivanandam2009}.
In the second panel, the grey-shaded regions display projected temperature profiles, $T\Proj(r)$, towards N, S, SE, W, E from G10. The data points show azimuthally averaged projected temperature profiles from the ACCEPT sample and from  \citet{Kaastra2004}, \citet{Snowden2008}, L11, and  \citet{Sivanandam2009}.
The third panel shows a collection of surface brightness profiles towards N, S, W, and E (D07,  G10), and the azimuthally averaged profiles from \citet{Snowden2008} and \citet{Sivanandam2009}. 

We describe the density profile of our model cluster by a double-$\beta$ profile. For the temperature, we utilise the analytical formula
%-----------
\begin{eqnarray}
T_1(r) &=& \left( m(r-r_0) + T_0 \right) \times \mathrm{step}(r-r_0) \nonumber \\
&&\times\, \mathrm{drop}(r-r_0)\;\;\textrm{with} \label{eqn:temp} \\
\mathrm{step}(x) &=&  y_l +  \frac{ 1- y_l}{1 + \exp\left[ -(x-r_s)/a_s \right] }   \nonumber\\
\mathrm{drop}(x) &=& \frac{1 + \left( x/r_d \right)^n }{D + \left( x/r_d \right)^n} , \nonumber
\end{eqnarray}
%======
which describes a linearly decreasing temperature in the outer part, a central decrease and  an intermediate temperature enhancement. Given the scatter in the temperature data we test three different temperature profiles for the model cluster. Our parameters for the density and temperature profiles are listed in Table~\ref{tab:Cluster_parameters}.
%
%TTTTTTTTT
\begin{table}
\caption{ICM parameters for the cluster model. The fiducial model is based on temperature profile 2.}
\begin{center}
\begin{tabular}{|l|c|c|c}
\hline
 \multicolumn{4}{l}{\textbf{Density: double $\beta$ profile}} \\
 \multicolumn{2}{l}{core radii $r_{1,2}/\Kpc$: }& 10 & 16\\
 \multicolumn{2}{l}{core densities $\rho_{01,02}/(\gccm)$:}  & $5.64 \cdot 10^{-26}$ & $6.6\cdot 10^{-26}$\\
 \multicolumn{2}{l}{$\beta_{1,2}$: }& 1 & 0.42\\ 
\vspace{1ex}\\
 \multicolumn{4}{l}{\textbf{Temperature:  see Eqn.~\ref{eqn:temp}}}\\
                    & profile 1 & profile 2& profile 3 \\
 $m/(\K \PC^{-1})$: & \multicolumn{3}{c}{ -11.6}\\
$T_{0}/(10^7\K)$: & 5 &  4.7 & 4.7\\
 $r_0/\Kpc$: & 0 & 0 & -10\\
$y_l$                   &   1.6 & 1.3 & 1 \\
$r_s/\Kpc$: & \multicolumn{3}{c}{12}\\
$a_s/\Kpc$: & \multicolumn{3}{c}{45}\\
$r_d/\Kpc$: & 33 & 29.6 & 45\\
$D$                & 4.1 & 3.15 & 2.5 \\
$n$              & 2 & 2 & 3\\
\vspace{1ex}\\
 \multicolumn{4}{l}{\textbf{Metal density:  Hernquist profile, see Eqn.~\ref{eqn:hernq}}}\\
\multicolumn{2}{l}{scale radius $r_{H}/\Kpc$: }& \multicolumn{2}{l}{300}\\ 
\multicolumn{2}{l}{core density $\rho_{H}/(\gccm)$: }& \multicolumn{2}{l}{$2.6\cdot 10^{-29}$}\\ 
\multicolumn{2}{l}{metallicity floor / solar: }& \multicolumn{2}{l}{0.33}\\ 
\hline
\end{tabular}
\end{center}
\label{tab:Cluster_parameters}
\end{table}
%TTTTTTTTTTTT

The black solid line in the top panel of Fig.~\ref{fig:iniprofs} displays our deprojected electron density profile. The thick dashed black  line in the third panel displays the X-ray brightness profile of our model cluster, derived by projecting $n_e^2 \Lambda(T,z)$, where $\Lambda(T,z)$ is the metallicity-dependent cooling function of \citet{Sutherland1993}.  Our density profile differs slightly from the fit given by \citet{Sivanandam2009}, but results in a better fit to the X-ray surface brightness profiles. We experimented with the Sivanandam density profile, with both the original single-$\beta$ version and with an additional central core. The resulting initial X-ray brightness profile runs along the upper range of the observed profiles shown in panel 3 of Fig.~\ref{fig:iniprofs}. After sloshing, the X-ray profiles end up being too broad in all directions, hence we prefer our fit. 

The black and grey solid lines in the second panel of Fig.~\ref{fig:iniprofs} display our deprojected temperature profiles, while the black and grey dashed lines present the corresponding projected temperature profiles. These are derived following \citet{Mazzotta2004} by calculating a weighted average of the temperature along each LOS using the weights
%-----------
\begin{equation}
W=n^2/T^{3/4}. \label{eqn:mazzotta}
\end{equation}
%======

%The remaining three panels in Fig.~\ref{fig:iniprofs} display the dynamical properties of our model cluster: the radial gravitational acceleration (panel 4), the cumulative mass (panel 5), and the rotational velocity. We give the results for all three temperature models.

%FFFFFFFFFF
\begin{figure}
\centering\includegraphics[width=0.45\textwidth]{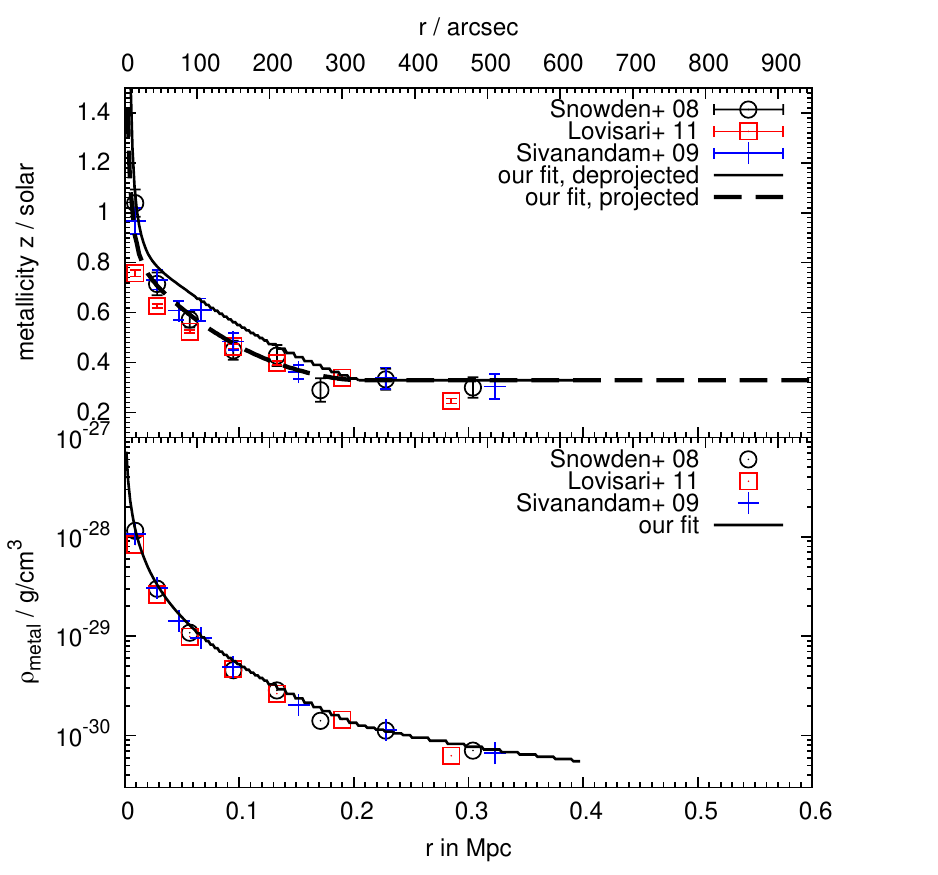}
\caption{Radial profiles of metallicity (top) and metal density (bottom): we show our projected and deprojected model and observational data, see legends. Correctly speaking, the data from \citet{Lovisari2011} display the iron abundance instead of metallicity, leading to the  differences near the centre.}
\label{fig:iniprofs_met}
\end{figure}
%FFFFFFFFFF
%
The top panel of Fig.~\ref{fig:iniprofs_met} demonstrates the observed, azimuthally averaged metallicity in A496 from \citet{Snowden2008}, L11 (iron abundance instead of metallicity), and  \citet{Sivanandam2009}.  Together with the ICM density profile, this translates into a metal density profile (bottom panel), which we fit with a Hernquist  profile (\citealt{Hernquist1990}),
%-----------
\begin{equation}
\rho\Fe (r) = \frac{\rho_H}{2\pi} \frac{r_H}{r}\frac{1}{\left( r/r_H +1 \right)^3} . \label{eqn:hernq}
\end{equation}
%======
Furthermore, we impose a metallicity floor at 0.33 solar. The resulting metal density profile is shown in the bottom panel of Fig.~\ref{fig:iniprofs_met}.  The parameters for the core density, $\rho_H$, and scale radius, $r_H$, are given in Table~\ref{tab:Cluster_parameters}.  The resulting deprojected and projected, emission-weighted metallicity is shown in the top panel in comparison to observational data.

%**********
\subsubsection{Elliptical model}\label{sec:maincluster_ell}
A496 is slightly elliptical along the N-NW to S-SE direction (T06, \citealt{Lagana2010}). This feature can also be derived from a close inspection of the observed directional surface brightness profiles in the third panel of  Fig.~\ref{fig:iniprofs}. The average brightness along the E-W axis declines somewhat faster than the one along the N-S axis, which indicates an elongation  along the N-S axis. 

The spherical  cluster described in Sect.~\ref{sec:maincluster} can be made triaxial while maintaining hydrostatic equilibrium by replacing the spherical radius, $r=\sqrt{x^2 + y^2 + z^2}$, with the elliptical radius,
%-----------
\begin{equation}
r_e = \sqrt{ \frac{x^2}{a^2} + \frac{y^2}{b^2} +\frac{z^2}{c^2} }, \label{eqn:r_ell}
\end{equation}
%======
in the double-$\beta$ profile for the ICM density and the temperature profile in Eqn.~\ref{eqn:temp}. We use $a=1.2$ and $b=c=1$, which makes the cluster elongated along the $y$-axis. For our elliptical cluster we use temperature profile 2. 
In Fig.~\ref{fig:profs_xray}, we compare the observed and modelled  surface brightness profiles for separate directions. 

%**************************
\subsection{Subcluster and orbits}  \label{sec:method_orbits}
We test several combinations of subcluster masses, scale radii and orbits. Once we know the main cluster's potential, we calculate the orbit of a test mass moving through this potential. During the course of the simulation, the subcluster potential is shifted along its orbit through the main cluster. The time normalisation is chosen such that pericentre passage happens at $t=0$. At the start of each simulation, the subcluster is placed at its orbit 1 Gyr prior to pericentre distance. We stop the simulations at $t\approx 1.5 \Gyr$ after core passage, shortly after a good match with the observed CFs is reached. This moment is well before the second core passage of the subcluster.

The gravitational potential of our subclusters is described by a Hernquist halo (\citealt{Hernquist1990}). We vary its mass and scale radius between $0.5$ and $4\times 10^{13}M\Sun$ and between 50 and $200\Kpc$, respectively (see Table~\ref{tab:runs}). A mass of $4\times 10^{13}M\Sun$ equals the mass of the inner 150 to 200 kpc of A496. In our cluster model, A496 contains $2\times 10^{14}M\Sun$ within 1 Mpc, thus our simulations are for mass ratios above 5. At the upper end of our mass range, we are restricted by the condition that we want to model minor mergers only.  The lower end of the mass range is determined such that the subcluster still leaves an imprint that is comparable to the observations.   
%TTTTTTTT
\begin{table}
\caption{List of simulation runs, stating the subcluster mass, scale radius, pericentre and apocentre of its orbit.}
\begin{center}
\begin{tabular}{llll}
$M\Sub/10^{13}M\Sun$  &  $a\Sub/\Kpc$  &  $d\Min/\Kpc$  & $d\Max/\Mpc$ \\
\hline
0.5 &   50 & 100 & 3 \\
1   &    50 & 100 & 3 \\
1   &  100 & 100 & 3 \\
2   &  100 & 100 & 3 \\
2   &  100 & 100 & 10 \\
4   &  100 & 100 & 3 \\
2   &  100 & 400 & 3 \\
4   &  100 & 400 & 3 \\
4   &  200 & 400 & 3 \\
4   &  200 & 400 & 10 \\
\hline
\end{tabular}
\end{center}
\label{tab:runs}
\end{table}%
%TTTTTTT

We assume our subclusters to be gas-free. R11 have shown that in the Virgo cluster a subcluster sufficiently massive to cause the observed CFs   can be completely ram-pressure stripped before pericentre passage. The ram pressure in A496 in general exceeds the one in Virgo by at least a factor of 2, and we study the same subcluster mass range here. Thus, also A496 should be able to ram-pressure strip the subclusters under consideration. Also the observations do not show any sign of gas ram pressure stripped  from a companion subcluster.

We test  pericentre distances of the subcluster orbit of 100 and 400 kpc, and apocentre distances of  3 and 10 Mpc, leading to different velocities during pericentre passage. Furthermore, we vary the orientation of the orbits as described in Sect.~\ref{sec:orbitorientation}.

%***************************
\subsection{Synthetic observations} \label{sec:mock_obs}
From our simulations, we calculate synthetic maps and radial profiles. For the X-ray brightness, we project $n^2 \Lambda(T,z)$ along the LOS, where $\Lambda(T,z)$ is the metallicity-dependent cooling function of \citet{Sutherland1993}.

\label{sec:residual_center}

%FFFFFFFFFF
\begin{figure}
\includegraphics[trim=200 200 120 80,clip,width=0.23\textwidth]{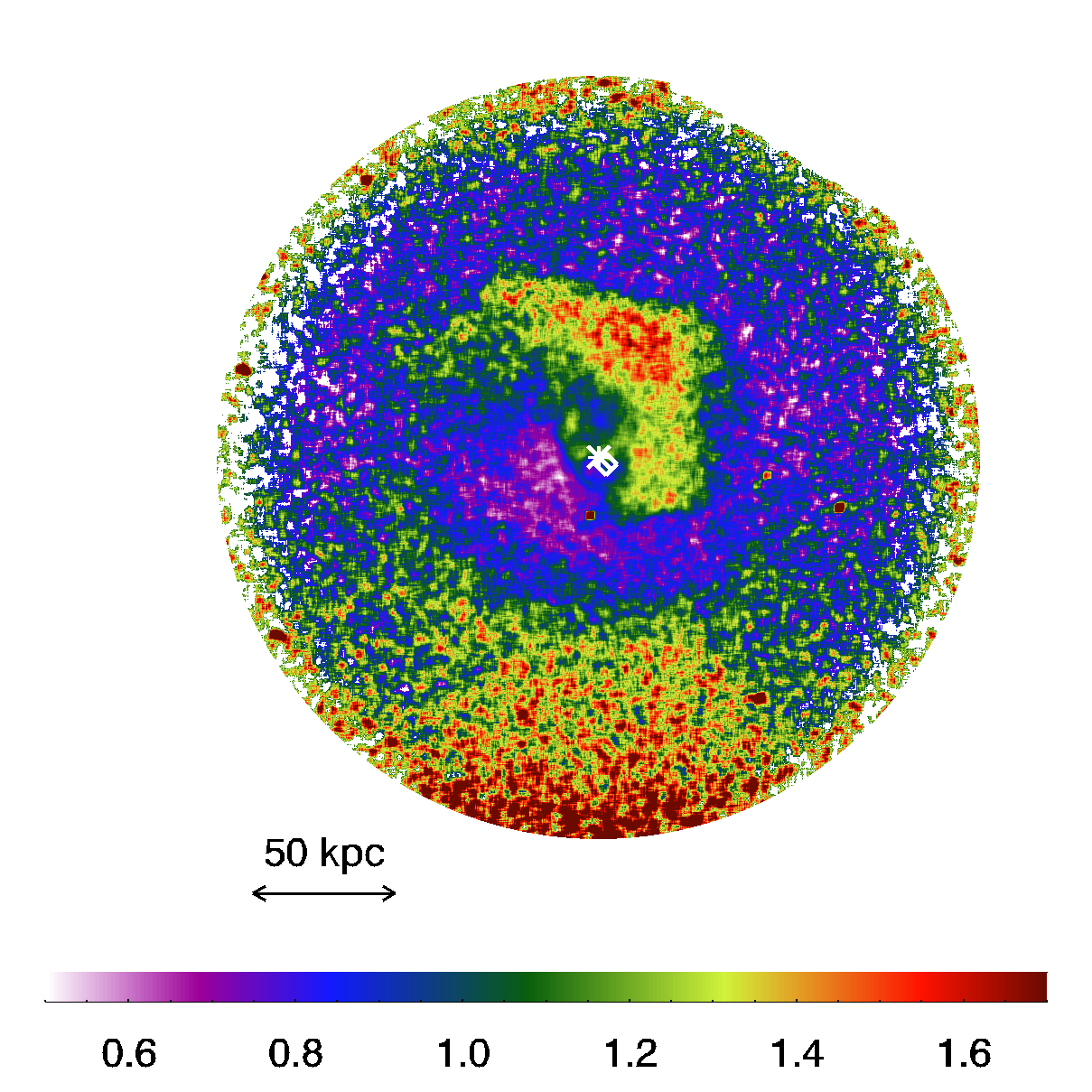}
\includegraphics[trim=200 200 120 80,clip,width=0.23\textwidth]{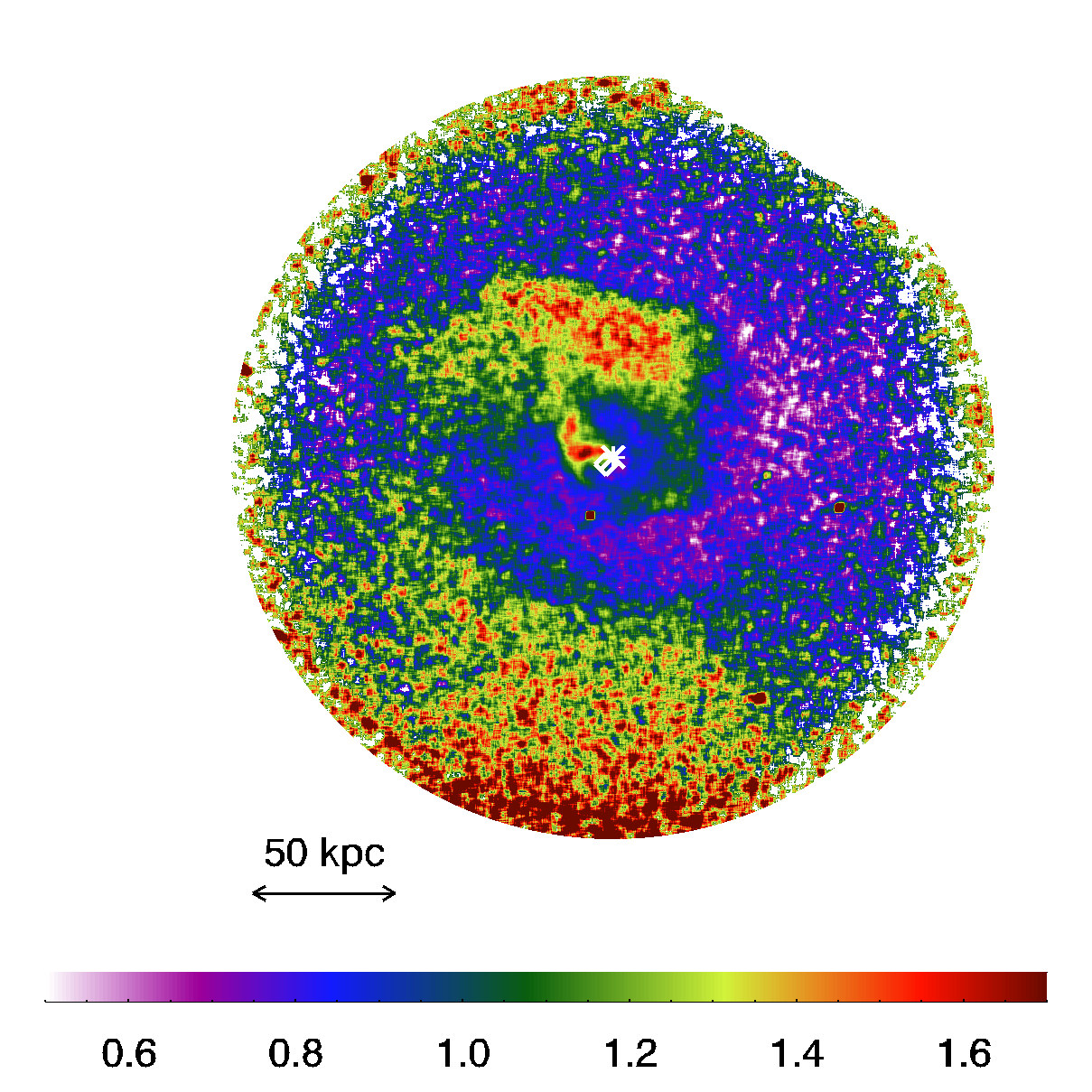}
\includegraphics[trim=200 290 120 80,clip,width=0.23\textwidth]{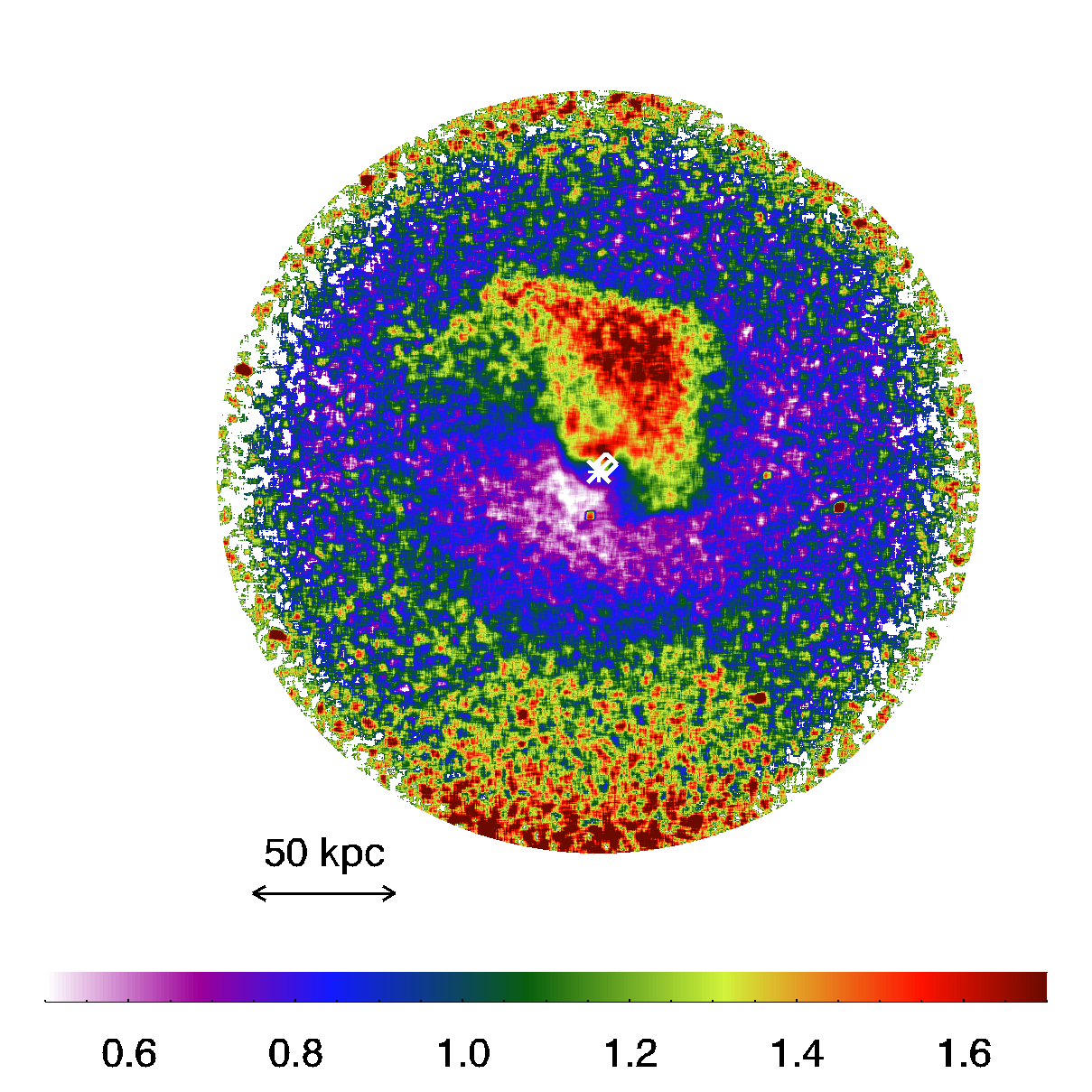}
\includegraphics[trim=200 290 120 80,clip,width=0.23\textwidth]{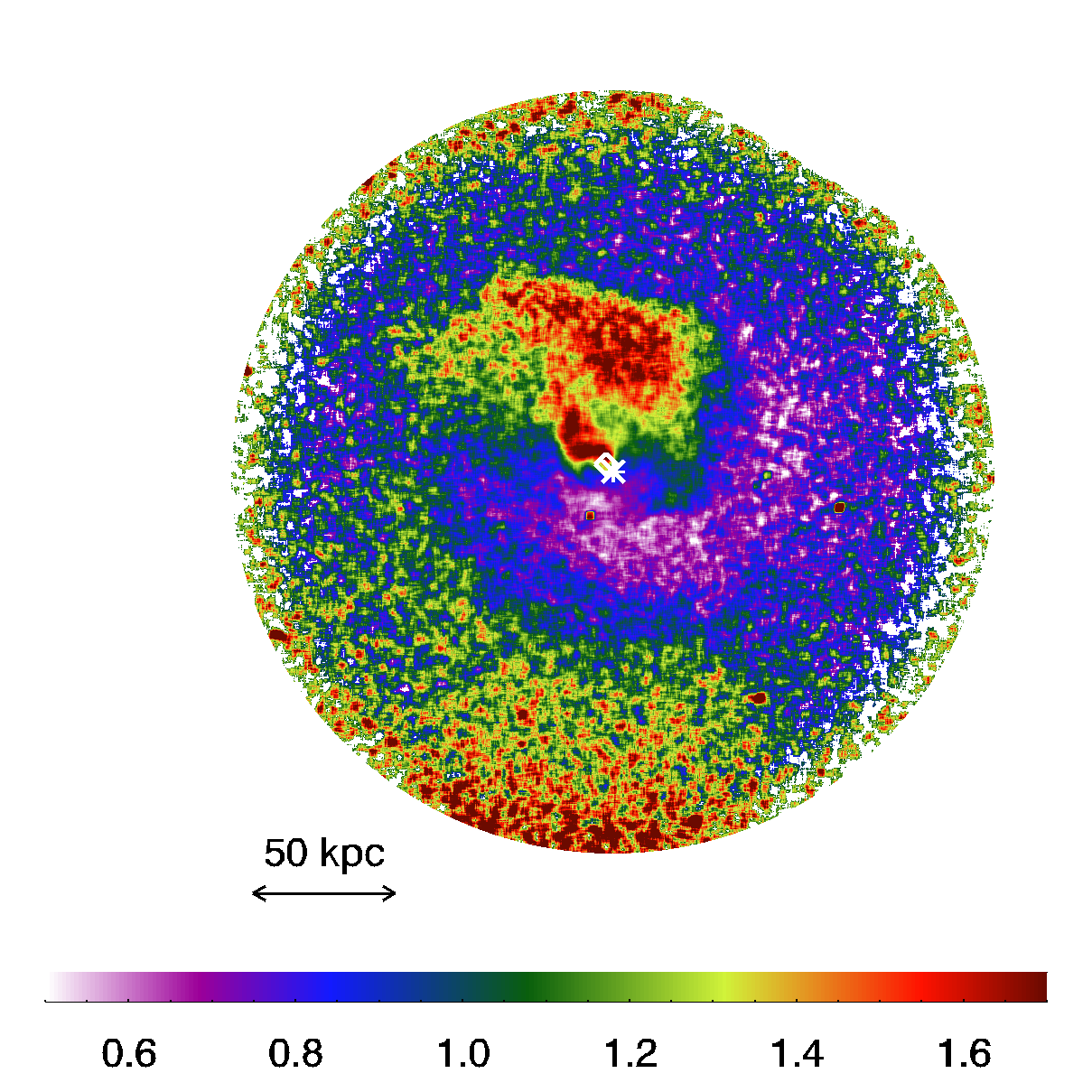}
\caption{Dependence of brightness residual maps on centre determination. The diamond marks the centre of the BCG in A496, the cross marks the centre used for the residual map. It is offset by 5.6 arcsec along the diagonals. The colour scale is the same as in Fig.~\ref{fig:A496_maps}.}
\label{fig:generate_residuals}
\end{figure}
%FFFFFFFFFF
%
For each X-ray image, we calculate a brightness residual map by dividing  the image by its azimuthal average, thus highlighting deviations from circular symmetry. This procedure requires the identification of the cluster centre around which the azimuthal average is taken. The details of the resulting residual map, especially near the cluster centre, depend somewhat on the positioning of the cluster centre, as shown in Fig.~\ref{fig:generate_residuals}. However, the overall structure of the residual map including the spiral-shaped brightness excess, the northern CF and the brightness excess towards the S is detected robustly.

For the projected temperature, we employ the weighting scheme of \citet{Mazzotta2004} as explained above (Eqn.~\ref{eqn:mazzotta}). The projected metallicity is the emission-weighted metallicity averaged along each LOS. 

\begin{figure}
\includegraphics[width=0.49\textwidth]{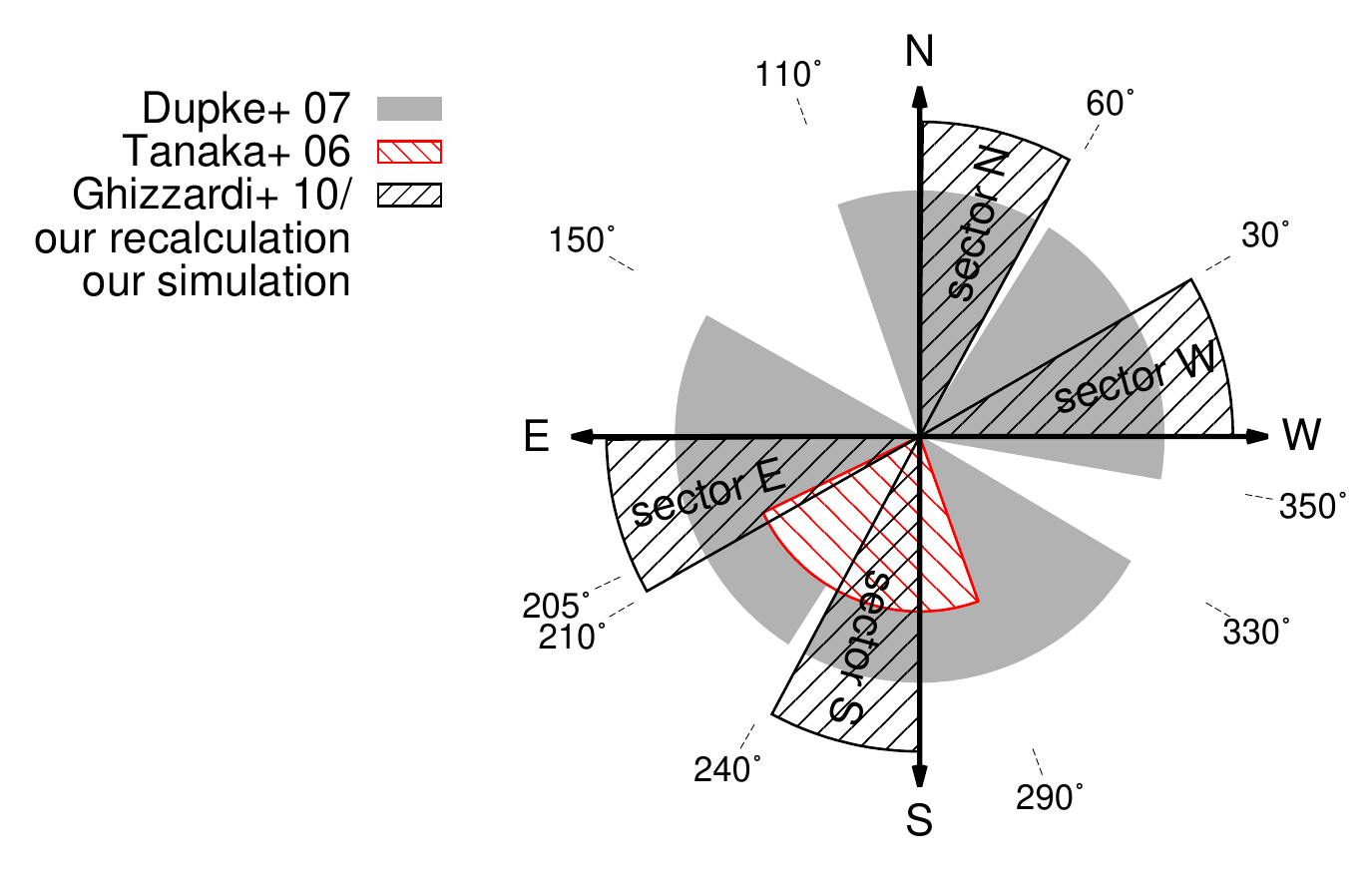}
\caption{Directional profiles are averaged over the sectors marked in this graph. See legend for different datasets. For our work, we consistently use the sectors marked with "sector N/W/S/E".}
\label{fig:sectors}
\end{figure}

From these synthetic maps, we calculate directional profiles which are averaged over a $30\degree$ azimuthal range. We use the same azimuthal sectors as in G10 and as shown in Fig.~\ref{fig:sectors}.
%FFFFFFFFFF
%FFFFFFFFFF